\documentclass{aa}
\usepackage{import}
\usepackage{booktabs}
\usepackage{longtable}
\usepackage{graphicx}
\usepackage[varg]{txfonts}
\usepackage{hyperref}
\usepackage{siunitx}
\usepackage{float}
\usepackage{caption}
\usepackage{subcaption}
\usepackage{mathtools}
\usepackage{textgreek}
\usepackage{supertabular}
\usepackage{multirow}

\usepackage{natbib}
\bibpunct{(}{)}{;}{a}{}{,} 

\makeatletter
\renewcommand*\aa@pageof{, page \thepage{} of \pageref*{LastPage}}
\makeatother

\hypersetup{
    colorlinks=true,
    linkcolor=blue,
    filecolor=magenta,      
    urlcolor=blue,
    citecolor=blue,
    pdfpagemode=FullScreen,
}

\begin{document} 

\titlerunning{The powerful lens galaxy cluster PLCK G287.0+32.9}
\title{The powerful lens galaxy cluster PLCK G287.0+32.9 ($\theta_{\rm{E}} \sim \qty{43}{\arcsec}$)\thanks{Tables C.1, D.1 and E.1 are only available in electronic form at the CDS via anonymous ftp to \url{cdsarc.u-strasbg.fr} (\url{130.79.128.5}) or via \url{http://cdsweb.u-strasbg.fr/cgi-bin/qcat?J/A+A/}}}
\subtitle{Redshift catalog and new lens model using MUSE observations}

\authorrunning{D'Addona et al. 2023}
\author{
    M. D'Addona\inst{\ref{unisa},\ref{oacn}},
    A. Mercurio\inst{\ref{unisa},\ref{oacn},\ref{infn-unisa}},
    P. Rosati\inst{\ref{oas},\ref{unife}},
    C. Grillo\inst{\ref{unimi},\ref{inafmilano}},
    G. Caminha\inst{\ref{tum}, \ref{mpa}},
    A. Acebron\inst{\ref{unimi},\ref{inafmilano}},
    G. Angora\inst{\ref{oacn},\ref{unife}},
    P. Bergamini\inst{\ref{oas}, \ref{unimi}},
    V. Bozza\inst{\ref{unisa},\ref{oacn},\ref{infn-unisa}},   
    G. Granata\inst{\ref{unimi},\ref{inafmilano}},
    M. Annunziatella\inst{\ref{madrid}},
    A. Gargiulo\inst{\ref{inafmilano}},
    R. Gobat\inst{\ref{pucv}},
    P. Tozzi\inst{\ref{oaa}},
    M. Girardi\inst{\ref{unitr},\ref{oat}},
    M. Lombardi\inst{\ref{unimi}},
    M. Meneghetti\inst{\ref{oas}},
    P. Schipani\inst{\ref{oacn}},
    L. Tortorelli\inst{\ref{munich},\ref{zurich}},
    E. Vanzella\inst{\ref{oas}},
}

\institute{
    Università di Salerno, Dipartimento di Fisica "E.R. Caianiello",
    Via Giovanni Paolo II 132, I-84084 Fisciano (SA), Italy \label{unisa} %
    \and
    INAF – Osservatorio Astronomico di Capodimonte, Via Moiariello 16, I-80131 Napoli, Italy \label{oacn} %
    \and
    INFN – Gruppo Collegato di Salerno - Sezione di Napoli,  Dipartimento di Fisica "E.R. Caianiello", Università di Salerno, via Giovanni Paolo II, 132 - I-84084 Fisciano (SA), Italy. \label{infn-unisa} %
    \and
    INAF – OAS, Osservatorio di Astrofisica e Scienza dello Spazio di Bologna, via Gobetti 93/3, I-40129 Bologna, Italy \label{oas} %
    \and
    Dipartimento di Fisica e Scienze della Terra, Università degli Studi di Ferrara, via Saragat 1, I-44122 Ferrara, Italy \label{unife} %
    \and
    Dipartimento di Fisica, Università degli Studi di Milano, via Celoria 16, I-20133 Milano, Italy
    \label{unimi} %
    \and
    INAF - IASF Milano, via Corti 12, I-20133 Milano, Italy \label{inafmilano} %
    \and
    Technical University of Munich, TUM School of Natural Sciences, Department of Physics, James-Franck-Str 1, 85748 Garching, Germany \label{tum} %
    \and
    Max-Planck-Institut für Astrophysik, Karl-Schwarzschild-Str. 1, 85748 Garching, Germany \label{mpa} %
    \and
    Centro de Astrobiología, Instituto Nacional de Técnica Aeroespacial, Ctra de Torrejón a Ajalvir, km 4, 28850 Torrejón de Ardoz, Madrid, Spain \label{madrid} %
    \and
    Instituto de Física, Pontificia Universidad Católica de Valparaiso, Avda. Universidad 330, Placilla, Valparaíso, Chile \label{pucv} %
    \and
    INAF – OAA, Osservatorio Astrofisico di Arcetri, Largo Enrico Fermi 5, I-50125 Firenze, Italy \label{oaa} %
    \and
    Universit\`a degli Studi di Trieste, Dipartimento di Fisica - Sezione di Astronomia, via Tiepolo 11, I-34143 Trieste, Italy \label{unitr}
    \and INAF - Osservatorio Astronomico di Trieste, via Tiepolo 11, I-34143 Trieste, Italy \label{oat} %
    \and
    Faculty of Physics, University Observatory, Ludwig-Maximilians-Universität München, Munich, Germany \label{munich}
    \and
    Institute for Particle Physics and Astrophysics, ETH Zürich, Zürich, Switzerland \label{zurich}
}

\date{Received 21 December 2023; accepted 28 January 2024}

 
\abstract
{}
{We present a new high-precision strong-lensing model of PLCK G287.0$+$32.9, a massive lens galaxy cluster at $z=0.383$, with the aim of obtaining an accurate estimation of its effective Einstein radius and total mass distribution. We also present a spectroscopic catalog containing accurate redshift measurements for close to 500 objects up to redshift $z = 6$, including multiply lensed sources and cluster member galaxies.}
{We exploited high-quality spectroscopic data from the Multi Unit Spectroscopic Explorer (MUSE), covering a central $3~\rm{arcmin}^2$ region of the cluster. We supplemented the spectroscopic catalog by including redshift measurements from VIsible MultiObject Spectrograph (VIMOS) and DEep Imaging Multi-Object Spectrograph (DEIMOS). We identified 129 spectroscopic cluster member galaxies with redshift values of $0.360 \leq z \leq 0.405$, and $m_{\rm{F160W}} \leq 21$. We complemented this galaxy cluster member sample with 24 photometric members identified with a convolutional neural network (CNN) approach. We also identified 114 multiple images from 28 background sources, of which 84 images from 16 sources are new and the remaining ones have already been identified in previous works. From these, we extracted a "golden sample" of 47 secure multiple images and used them, together with the selected cluster member, to build and optimize several strong-lensing models with the software lenstool.} 
{The best-fitting lens model shows a root mean square (RMS) separation value between the predicted and observed positions of the multiple images of $\ang[angle-symbol-over-decimal]{;;0.75}$. Using its predictive power, we found three new multiple images and we confirm the configuration of three systems of multiple images that were not used for the optimization of the model. For a source at a redshift of $z_s = 2$, we found a cluster with an Einstein radius of $\theta_{\rm{E}} = 43.4\arcsec \pm 0.1\arcsec$. This value is in agreement with previous estimates and corresponds to a total mass enclosed in the critical curve of $\rm{M_{E}} = {3.33}_{-0.07}^{+0.02} \times{ 10^{14} \rm{M}_{\odot}}$.}
{The combined application of ancillary Hubble Space Telescope (HST) imaging, VIMOS and DEIMOS data, and the new MUSE spectroscopic observations allowed us to build a new lens model of the galaxy cluster PLCK G287.0+32.9, with an improvement in terms of reconstructing the observed positions of the multiple images of a factor of $2.5$ with respect to previous models. The derived total mass distribution confirms this cluster to be a very prominent gravitational lens, with an effective Einstein radius of $\theta_{\rm{E}} \sim {\ang{;;43}}$. We were also able to construct an extensive spectroscopic catalog containing 490 objects, of which 153 are bright cluster members with $m_{\rm{F160W}} \leq 21$, and 114 are multiple images.}


\maketitle


\section{Introduction}
\label{sec:introduction}

Galaxy clusters are the largest gravitationally bound structures known in the Universe. They are not only valuable laboratories in the investigation of galaxy formation and evolution processes, but given they can act as giant gravitational lenses, they also are precious tools for investigating distant and faint objects (\citealt{Vanzella2016,Vanzella2017}). For example, \cite{Vanzella2021} was able to peer into the internal structure of galaxies at high-redshift lensed galaxies and unveil star-forming complexes matching the scales of bound star clusters. The outcome of the strong-lensing models can be used to constrain cosmological models, such as the standard lambda cold dark matter cosmology (\textLambda-CDM, \citealt{WMAP5,WMAP7,Planck2015,Planck2018}), and validate their predictions: for example, the analysis of the time delays between the multiple images of variable strongly lensed sources \citep{Refsdal1966}, such as supernovae \citep{Kelly2015,Rodney2021,Gobar2017} or quasars \citep{Inada2012,Onguri2013,Sharon2017,Coubrin2018,Bonvin2018,Acebron2022}, has been used to measure the value of the Hubble constant, $\rm{H}_0$ \citep{Suyu2017,Tewes2013,Grillo2018,Birrer2019,Sluse2019,Wong2020}; statistical analyses of the total mass distributions derived from lens models and the predictions of cosmological simulations have indicated that the hierarchical formation of the large-scale structure may not be compatible with the \textLambda-CDM model \citep{Giocoli2008,Giocoli2010,Wang2020,Meneghetti2020,Meneghetti2022,Meneghetti2023}; analyses of the distribution of the effective Einstein radius, $\theta_{\rm{E}}$, the radius of the area enclosed within the critical curve of infinite magnification if it were a circle, has in recent years led to a tension between observations \citep{Zitrin2012} and theoretical predictions based on \textLambda-CDM \citep{Oguri2009,Hennawi2007a}.\\

Even if this tension has been partially scaled down, the study of the observed distribution of Einstein radii plays an important role in testing the validity of cosmological models \citep{Waizmann2012}. In fact, since the Einstein radius is strictly linked to the lens total mass, due to the shape of the universal mass function \citep{Tinker2008}, $\theta_{\rm{E}}$ is expected to be on the order of tens of arcseconds (for a lensed source at a redshift of $z_{\rm{s}} \sim 2$), with clusters with larger Einstein radii becoming rarer and rarer \citep{Richard2010, Onguri2012, Zitrin2015, Sharon2020}. To date, only a handful of clusters characterized by $\theta_{\rm{E}} \geq \qty{40}{\arcsec}$ are known to date \citep{Zitrin2009, Acebron2020,Broadhurst2005,Cerny2018}. Therefore, the number of these massive clusters can enforce strong constraints on structure formation and evolution models.\\

Our study focuses on one of these rare gems: the galaxy cluster PLCK G287.0+32.9 (PLCK-G287 hereafter), a powerful gravitational lens at a redshift of $z_c = 0.383$ and the second-most significant Sunyaev-Zel'dovich detection from the Planck catalog \citep{Planck2011}. Using radio observations from the Karl Guthe Jansky Very Large Array (VLA) and the Giant Metrewave Radio Telescope (GMRT), along with X-ray observations from XMM-Newton, \cite{Bagchi2011} found corroborating evidence for PLCK-G287 to be defined as a massive post-merger system ($\rm{M}_{\rm{500}} = 1.5 \times 10^{15}~\rm{M}_{\odot}$) with a very complex structure. This has also been confirmed by weak-lensing analyses carried out by \cite{Gruen2014} and \cite{Finner2017} and a study from \cite{Bonafede2014} that is also based on GMRT observations. More recently, \cite{Zitrin2017}  built the first strong-lensing model for this cluster by using the light-traces-mass method (LTM, \citealt{Zitrin2009b}), estimating an effective Einstein radius of $\theta_{\rm{E}} = \qty{42}{\arcsec}$ with a total uncertainty of $\sim 10\%$. At the highly non-linear tail of the probability distribution of Einstein radii, this large uncertainty can make the difference between a mild outlier and a peculiar object that can be used to challenge the predictions of the \textLambda-CDM model. \cite{Zitrin2017} also identified $60$ candidate multiple images of $20$ background systems using the values of their photometric redshift estimates, but none of them was spectroscopically confirmed. Of these multiple images, only $35$ of them, from $10$ background sources, were confirmed by the model.\\

Observations from instruments such as the Multi-Unit Spectroscopic Explorer (MUSE) spectrograph \citep{MuseREF2010} of the European Southern Observatory (ESO) Very Large Telescope (VLT) have been shown to be a real game-changer that allowed the identification of large samples of secure multiple images and cluster members with accurate redshift measurements (e.g., \citealt{Cerny2018,Mahler2018,Lagattuta2019,Mercurio2021,Bergamini2023b,Granata2023}), and leading to the flourishing of high-precision strong-lensing models of galaxy clusters. In this context, we present a new strong-lensing model and a catalog of redshifts for the galaxy cluster PLCK-G287.\\

The structure of this paper is as follows: in Sect. \ref{sec:data}, we briefly describe the data and data analysis used to identify multiple images and cluster members used for the strong-lensing modeling described in Sect. \ref{sec:model}. In Sect. \ref{sec:results}, we present our results, followed by a discussion and a comparison with the findings of previous works from the literature in Sect. \ref{sec:summary}. In Appendices \ref{appendix:multiple_images}, \ref{appendix:cluster_members}, and \ref{appendix:z_spec_catalog} we report the identified multiple images, the identified cluster members, and the spectroscopic catalog furnished with a list of most prominent emission and absorption features identified for each object. In this work, we assume a \textLambda-CDM cosmology with $\Omega_{m0} = 0.3$, $\Omega_{\Lambda0} = 0.7$ and $\rm{H}_{\rm{0}} = 70~\text{km}~\text{s}^{-1}~\text{Mpc}^{-1}$, for which $\qty{1}{\arcsec} \simeq 5.23~\text{kpc}$ at the redshift of the cluster PLCK-G287 ($z_c=0.383$).

\begin{figure*}[hbtp]
    \centering
    \includegraphics[width=1.0\textwidth]{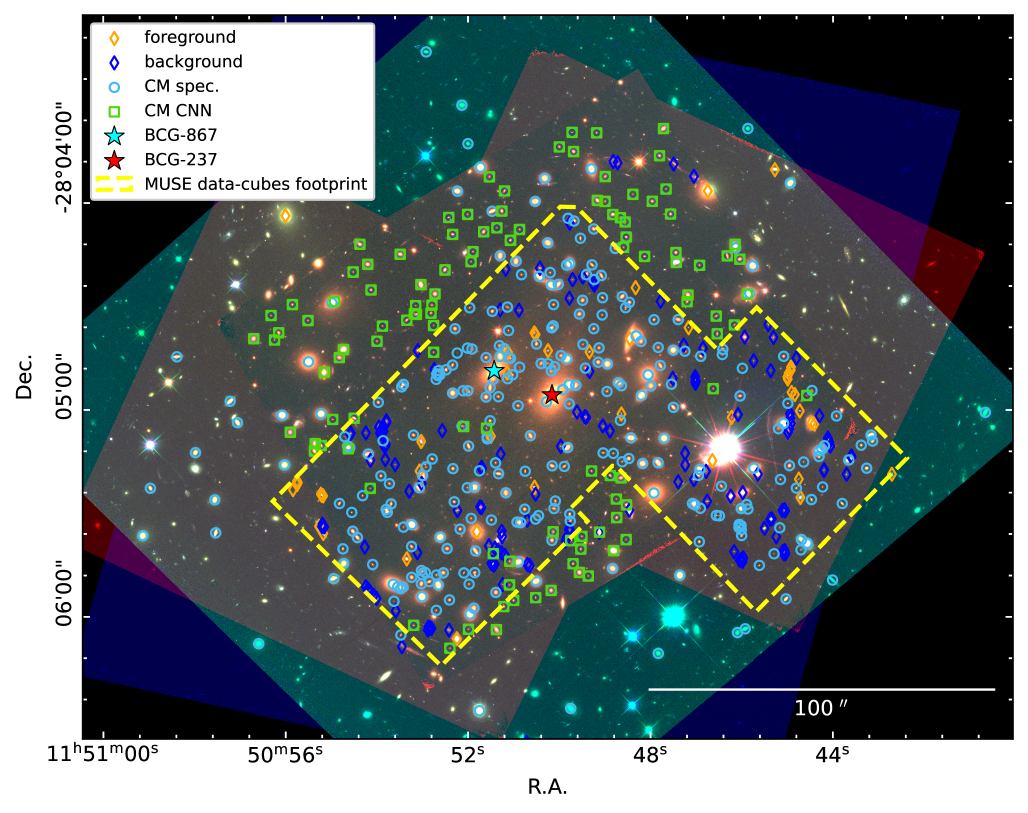}
    \includegraphics[width=1.0\textwidth]{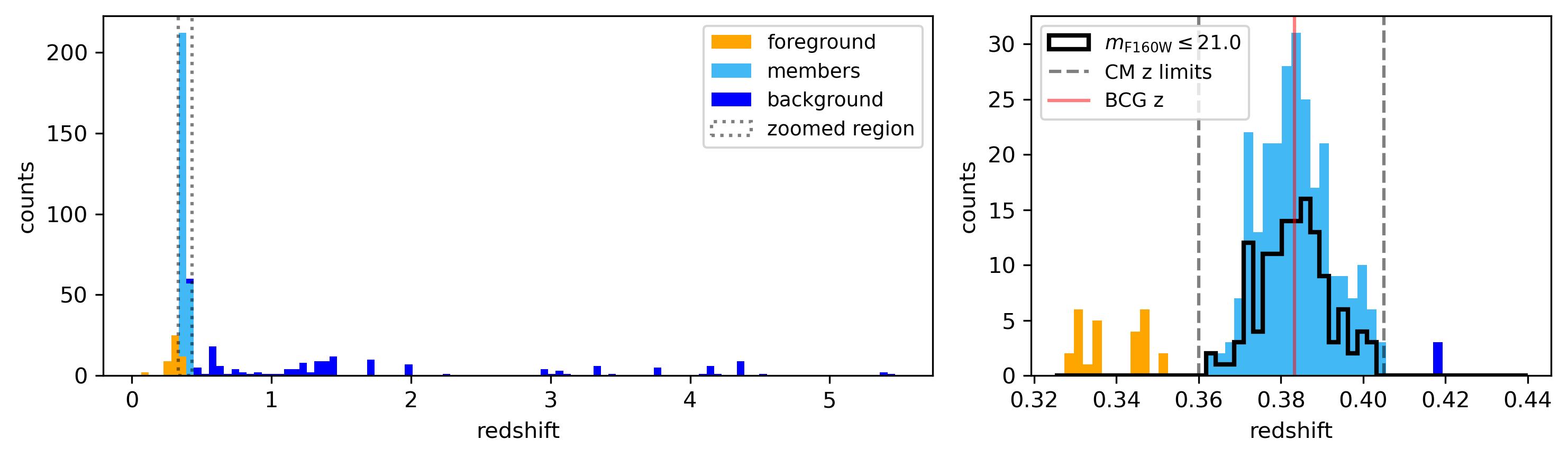}
    \caption[]{RGB image of the galaxy cluster PLCK-G287 obtained by combining the HST images with the software fits2rgb \protect\footnotemark (Red: F105W, F110W F125W, F140W F160W. Green: F814W. Blue: F435W, F475W, F606W). The yellow dashed line indicates the footprint of the MUSE spectral data cubes. The cyan circles and green squares indicate the spectroscopic and CNN-identified cluster members, respectively (see Sect. \ref{sub:model.members}), while the orange and blue diamonds indicate, respectively, the spectroscopic foreground and background objects. The two BCGs that have been modeled independently of the other members are indicated by the red and cyan stars. The redshift estimates for the objects outside the MUSE footprint come from VIMOS and DEIMOS observations (see Sect. \ref{sub:data.redshift}). Histograms show the redshift distribution of objects in the HST footprint: the one on the right is a zooming that shows the redshift of the BCG ($z = 0.383$, solid red line) and the thresholds used to select cluster members ($0.360 \leq z \leq 0.405$, dashed lines). The solid black line shows the distribution of members with F160W Kron magnitude values equal or lower than 21.}
    \label{fig:cluter_overview}
\end{figure*}
\footnotetext{fits2rgb: \href{https://github.com/mauritiusdadd/fits2rgb}{https://github.com/mauritiusdadd/fits2rgb}. This is a simple python3 script, based on the astropy python package \citep{Astropy2022}, that can merge several monochromatic FITS images into an RGB one.}

\section{Data}
\label{sec:data}


\subsection{Photometric data}
PLCK-G287 has been observed by the Hubble Space Telescope (HST) in the framework of the Reionization Lensing Cluster Survey (RELICS) survey (P.I.: Dan Coe, \citealt{RELICS}, program ID 14096) using both the Advanced Camera for Surveys (ACS) and the Wide Field Camera 3 (WFC3). Previous observations made with HST in cycle 23 (P.I.: Seitz, program ID 14165, \citealt{Seitz2016}) were also integrated into the RELICS data, for a total of three orbits for the ACS filters group and two orbits for the WFC3 one. We used publicly available images in the two resolutions of $\qty{0.03}{\arcsec}$ (30 mas) and $\qty{0.06}{\arcsec}$ (60 mas) per pixel. They cover an area, centered on the brightest cluster galaxy (BCG), with a radius  of $\sim\qty{1.7}{\arcmin}$ in the optical bands F435W, F475W, F606W, and F814W; and of $\sim\qty{1.0}{\arcmin}$ in the infrared bands (IR) F105W, F110W, F125W, F140W, and F160W. Each science image is also supplemented with a weightmap image. For the first group of filters, the total exposure time varies from a minimum of $711~\rm{s}$ for band F125W to a maximum of $11447~\rm{s}$ for band F110W, while for the second group it ranges from a minimum of $2125~\rm{s}$ for band F435W to a maximum of $4680~\rm{s}$ for band F814W. Using a subset of 28 non-saturated stars, we measured the full width at half maximum (FWHM) of the point spread function (PSF) as a range that goes from a minimum of $\ang[angle-symbol-over-decimal]{;;0.11} \pm \ang[angle-symbol-over-decimal]{;;0.02}$ in the band F814W to a maximum of $\ang[angle-symbol-over-decimal]{;;0.22} \pm \ang[angle-symbol-over-decimal]{;;0.02}$ in the band F160W (see Table \ref{tab:hst_psf}).\\

\begin{table}
    \caption{\label{tab:hst_psf}PSF FWHM measured from the HST images.}
    \resizebox{\columnwidth}{!}{%
    \begin{tabular}{cc|cc}
        FILTER &  PSF FWHM $[\arcsec]$& FILTER & PSF FWHM $[\arcsec]$\\
        \hline\hline
        F435W & $0.12 \pm 0.02$  & F105W & $0.20 \pm 0.02$ \\
        F475W & $0.13 \pm 0.03$  & F110W & $0.21 \pm 0.03$ \\  
        F606W & $0.12 \pm 0.02$  & F125W & $0.21 \pm 0.03$ \\  
        F814W & $0.11 \pm 0.02$  & F140W & $0.21 \pm 0.02$ \\
              &                  & F160W & $0.22 \pm 0.02$ \\
        \hline
    \end{tabular}
    }
    \tablefoot{PSF FWHM measured from the HST images taken by the two instruments ACS (left) and WFC3 (right) with different filters (see Sect. \ref{sec:data})}
\end{table}

We used the 60 mas HST images and their corresponding weightmaps to extract sources and compute the photometry. For each band, we ran Sextractor \citep{Sextractor} in dual-image mode, using the F814W image as the detection image. We used this band for the detection because of the lower PSF FWHM and higher exposure time and, secondly, because a large portion of the emission from the continuum of the spectrum of member galaxies falls in this filter (from $\sim 5000~\text{\AA}$ to $\sim 7000~\text{\AA}$ at the redshift of the cluster). We obtained a catalog that contains the ICRS R.A. and Dec. and pixel coordinates of the detected sources (indicated respectively with ALPHA\_J2000, DELTA\_J2000, X\_IMAGE and Y\_IMAGE); their isophotal semi-major and semi-minor axis and rotation angle (A\_IMAGE, B\_IMAGE and THETA\_IMAGE); the peak surface brightness (MU\_MAX); the stellarity index (CLASS\_STAR); the isophotal flux and magnitude (FLUX\_ISO and MAG\_ISO); and the fluxes and magnitudes computed at 14 fixed apertures ranging from 2 to 160 pixels (FLUX\_APERX\_n and MAG\_APER\_n), corresponding to a range between $\ang[angle-symbol-over-decimal]{;;0.12}$ and $\ang[angle-symbol-over-decimal]{;;9.6}$.\\

We tuned the extraction parameters to minimize the spurious detections due to noise and we use a semi-automated procedure to mask star spikes and identify potential spurious detection, which is described in Appendix \ref{appendix:start_spike}. After this cleaning process, the photometric catalog contains a total of 3168 objects. Note that for some extended objects (i.e., lensed galaxies and arcs) single bright features, such as star formation clumps, appear as distinct objects.

\subsection{Spectroscopic data and redshift estimation}
\label{sub:data.redshift}
This study exploits new, extensive, and high-quality spectroscopic data derived from observations made with MUSE and the Adaptive Optics Facility (AOF) working in wide field mode \citep{MUSE-AOa,MUSE-AOb} on three nights in 2019 March-May (P.I.: Amata Mercurio, ESO program 0102.A-0640(A)). These observations produced three spectral data cubes corresponding to three pointings covering the center of the galaxy cluster (see Fig. \ref{fig:cluter_overview}) for a total area of $\sim 3~\rm{arcmin}^2$  with a spatial resolution of \ang[angle-symbol-over-decimal]{;;0.2} and with a spectral resolution of $1.25~\text{\AA}~\text{pixel}^{-1}$ in the vacuum wavelength range from $4700~\text{\AA}$ to $9350~\text{\AA}$, with a gap between $5805~\text{\AA}$ and $5967~\text{\AA}$ due to the emission generated by the guiding laser of the AOF. All the wavelengths in this work are referred to vacuum. We process and merge the spectral data cubes following the prescriptions of \cite{Caminha2019}, using the reduction pipeline version 2.8.3 \citep{Weilbacher2020A}. The final data cube has an exposure time on target of 3.1 hours in two pointings and 3.8 hours in the westernmost pointing. The PSF measured from stars on the pseudo-white images, obtained by staking the data cube along the spectral axis, exhibits $\rm{FWHM} \approx \ang[angle-symbol-over-decimal]{;;0.50} - \ang[angle-symbol-over-decimal]{;;0.55}$ across all pointings.\\

For each object in the photometric catalog that falls into the data cube footprint, we extract the spectrum with the program python-specex\footnote{python-specex: a python package and a set of programs we developed to handle spectroscopic data-cubes and to extract 1D spectra from them. More information can be found at \href{https://github.com/mauritiusdadd/python-specex} {https://github.com/mauritiusdadd/python-specex} \citep{DAddonaSpecex}.} using a circular aperture of radius $\qty{0.4}{\arcsec}$. We visually inspect and determine the redshift ($z$) of each spectrum using the program EZ \citep{EZ2010}. A quality flag QF is assigned to each spectrum, that indicates the reliability of the redshift estimation \citep{Balestra2016,Caminha2016}: 1=insecure; 2=likely; 3=secure; 9=based on one emission line. We were able to determine the redshift for 531 objects, of which 432 have $\rm{QF} \geq 3$. This reliable spectroscopic sample contains 31 stars, 42 foreground objects ($z < 0.360$), 213 cluster members ($0.360 \leq z \leq 0.405$; see Sect. \ref{sub:model.members}) and 146 background objects ($z > 0.405$). Note that, as for the photometric catalog, clumps of extended objects are considered as distinct objects since this is useful for the identification of multiple images. For objects that are outside the MUSE field of view, we use redshift measurements from observations made with the VIsible MultiObject Spectrograph (VIMOS) on the VLT UT3 Telescope in service mode on four nights in 2015 February–March (P.I.: Mario Nonino, period 094.A-0529(B), 11 objects). We completed the spectroscopic catalog with redshifts measurements (47 objects) by \cite{Golovich2019a,Golovich2019b}, using observations made with the DEep Imaging Multi-Object Spectrograph (DEIMOS) on the Keck II telescope at the W. M. Keck Observatory on Maunakea (KECK) over the nights: 2013 January 26, 2014 July 14, 2014 September 5, 2013 December 3–5 (half nights), 2014 June 22–23, 2015 February 15, and 2015 December 13. With these redshift measurements, in the region covered by HST, we identified: 9 and 35 cluster member galaxies, 1 and 5 foreground objects, and 1 and 7 background objects from VIMOS and DEIMOS, respectively (see Table \ref{tab:zcat_stats}). The spectroscopic catalog is available at the CDS (see \ref{appendix:z_spec_catalog}).\\

\begin{table}
    \caption{\label{tab:zcat_stats}Distribution of the objects in the reliable spectroscopic sample.}
    \resizebox{\columnwidth}{!}{%
    \begin{tabular}{lrrrrr}
        $\rm{QF} \geq 3$ &  MUSE & VIMOS & DEIMOS & CNN \\
        \hline\hline\\
        Total & 432 & 11 & 47 & -- \\
        Stars & 31 & 0 & 0 & -- \\
        Foreground & 42 & 1 & 5 & -- \\
        Background & 146 & 1 & 7 & -- \\
        Cluster Members & 213 & 9 & 35 & 110 \\
        \hline\\[-0.5em]
        C.M. $m_{\rm{F160W}} \leq 21$ & 85 & 9 & 35 & 24 \\
    \end{tabular}
    }
    \tablefoot{For completeness, photometric cluster members identified by a CNN are also reported (see Sect. \ref{sub:model.members}).}
\end{table}

\section{Strong-lensing model}
\label{sec:model}
In this section, we describe the selection of cluster members and multiple images used to optimize the lens model. We used the software lenstool\footnote{lenstool: \href{https://projets.lam.fr/projects/lenstool}{https://projets.lam.fr/projects/lenstool}}\citep{Kneib1996,Jullo2007,Lenstool2009} to model the total mass distribution of the lens cluster following the prescriptions of previous works, such as \cite{Caminha2019}, \cite{Bergamini2021}, \cite{Acebron2022}, and \cite{Granata2023}. This pipeline takes advantage of Markov chain Monte Carlo (MCMC) to retrieve the best-fitting values of the parameters $\boldsymbol{\xi}$ of a parametric mass model, and their errors, by minimizing the chi-square function defined by Eq. (\ref{eq:lenstool_chi2})

\begin{equation}
    \label{eq:lenstool_chi2}
    \chi^2(\boldsymbol{\xi}) \coloneqq \sum_{j=1}^{N_{fam}}\sum_{i=1}^{N^j_{img}} \left( \frac{ \left\| \boldsymbol{x}^{obs}_{i,j} - \boldsymbol{x}^{pred}_{i,j}(\boldsymbol{\xi}) \right\|}{\Delta x_{i,j}} \right)^2,
\end{equation}

\noindent where $\boldsymbol{x}^{obs}_{i,j}$ and $\boldsymbol{x}^{pred}_{i,j}$ are the observed and predicted positions respectively, on the lens plane, of the i-th counter-image of the j-th family and $\Delta x_{i,j}$ is the corresponding uncertainty.
\subsection{Selection of cluster members}
\label{sub:model.members}
Guided by the distribution of the redshifts around the brightest cluster galaxy (BCG) with ID 237 (see Fig. \ref{fig:cluter_overview}), we identify 257 spectroscopically confirmed members in the redshift range [$0.360$,~$0.405$]. This corresponds to a peculiar velocity range of [$-5000$,$~+4700$] $\rm{km}~\rm{s}^{-1}$ in the cluster rest frame centered at $z_c = 0.383$. To maximize the completeness of the sample, we also include 110 photometric members that have been identified using the convolutional neural network developed by \cite{Angora2020}: for each object, this CNN takes in input a set of cut-outs extracted from the ACS and WCF3 images and returns a probability of the object being a cluster member. To test its performance on this cluster, we use the CNN to predict the cluster membership of all the MUSE spectroscopy members, resulting in a purity of $89\%$, a completeness of $88\%$, and an f1-score of $89\%$. Members that had a missing magnitude measurement in the band F160W have been inferred using a set of linear regressors (Appendix \ref{appendix:imputation}). Furthermore, for two bright members with IDs 3266 and 3266, that are near the very bright star in the western MUSE pointing and for which Sextractor was not able to reliably measure their photometry, we estimated their $m_{\rm{F106W}}$ values using astropy/photutils v1.8.0 \citep{Photutils180}.\\

\begin{figure}
    \centering
    \includegraphics[width=0.5\textwidth]{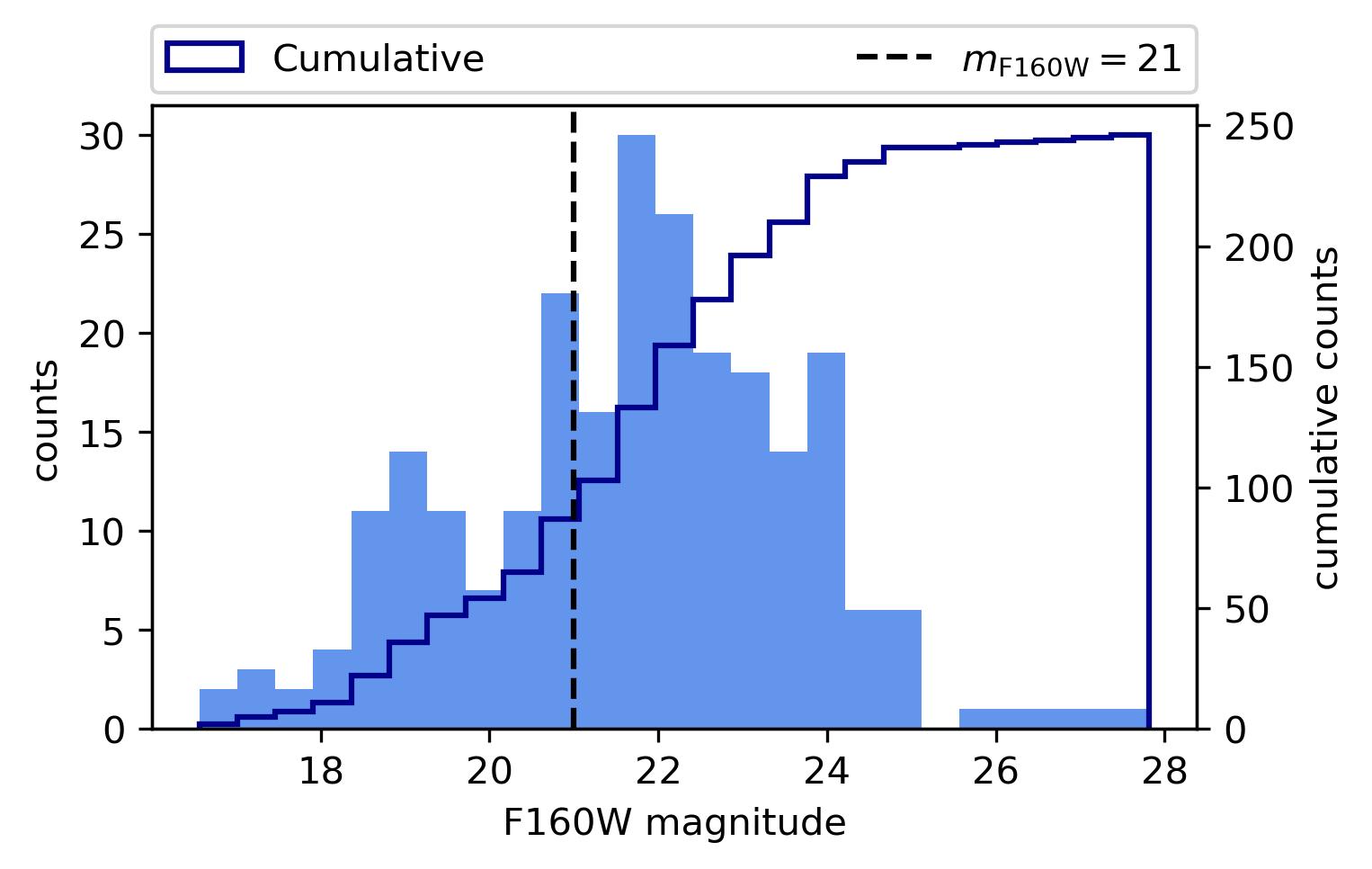}
    \caption{Histogram and cumulative distribution of the cluster members as a function of the Kron magnitudem,$m_{\rm{F106W}}$. The dashed line indicates the threshold used to select the members utilized in the strong-lensing model.}
    \label{fig:members_f160w_selection}
\end{figure}

Finally, to construct the cluster member sample for the lens model, we selected 129 spectroscopic members and 24 photometric ones for which $m_{\rm{F160W}} \leq 21$ (Fig. \ref{fig:members_f160w_selection}, Table \ref{tab:zcat_stats}), for a total of $N_{cm} = 153$ selected cluster members. We use this magnitude threshold to reduce the number of faint members and thus the computation time, after checking that it does not affect the goodness of our strong-lens modeling. This choice is also justified by the results by \cite{Bergamini2023b}, which shows that using a magnitude limit of 21 in the band F160W for the cluster member selection does not produce significant variation in the lens model metric. A similar study by \cite{Raney2021} also shows a similar behavior for different magnitude thresholds between 21 and 26 in the  F814W band. \\

\subsection{Multiple images}
\label{sub:model.multiple_images}
We identified multiple images by first searching for objects in the spectroscopic catalog having a similar redshift and then visually inspecting both the data-cube and the high-resolution 30 mas HST images. They are labeled using the format X.Yk, where X is an integer that identifies the background source system, Y is an integer that indicates any clearly identifiable substructure, such as star formation clumps, and k is a letter that differentiates among the different multiple images. A single background system may, in fact, have multiple clumps and substructures that can be used as lensed sources themselves. For this reason, we use the term "family" to indicate a set of multiple images of the same substructure. 

We also checked the multiple images reported in \cite{Zitrin2017} and spectroscopically confirmed 30 of them from 12 background sources, while others are too faint to be spectroscopically confirmed or they are outside the MUSE footprint. When possible, for confirmed multiple images, we used the same family number used in \cite{Zitrin2017}; namely, an image with the ID 7.1a in this work and one with the ID 7.2 in Zitrin's work are multiple images of the same background source. We were able to identify 114 multiple images of 28 background sources, of which 16 are newly identified multiply lensed sources, corresponding to 38 families in the redshift range $0.60 - 5.98$ (see Appendix \ref{appendix:multiple_images}). The position of each clump was refined by super-sampling a square cutout from the F814W image of size $\qty{1}{\arcsec} \times \qty{1}{\arcsec}$, centered at its initial position, and then taking the position of the peak nearest to cutout center. For HST-dark objects (mostly Lyman-\textalpha{} emitters), the cutout was extracted from a thin slice of the spectral data-cube, centered on the wavelength of the strongest emission line, which was stacked along the spectral axis. We used $\ang[angle-symbol-over-decimal]{;;0.25}$ as positional error for the multiple images identified in HST and $\ang[angle-symbol-over-decimal]{;;1}$ for those only visible in the MUSE data.\\

\begin{figure}
    \centering
    \includegraphics[width=0.5\textwidth]{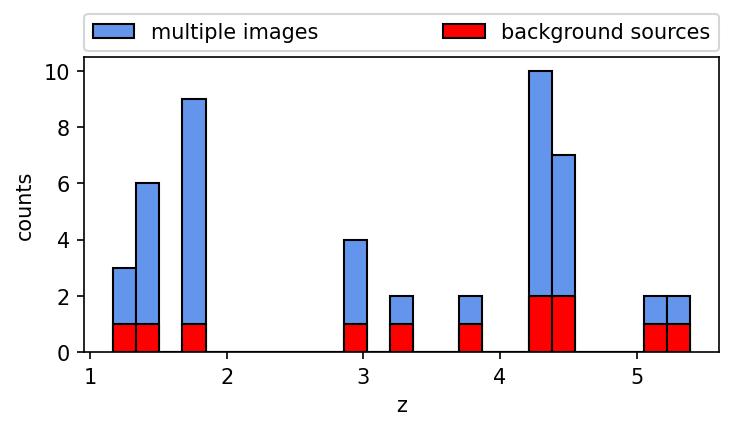}
    \caption{Redshift distribution of the multiple images (in blue) and of the corresponding background sources (in red) in the golden sample used in the strong-lensing model optimization.}
    \label{fig:lensed_images_hist_z}
\end{figure}

In order to reduce potential biases in the strong-lensing model optimization caused by uncertain or incorrect constraints \citep{Grillo2015}, we selected a reliable subset of multiple images that have a spectroscopic quality flag $\rm{QF} \geq 3$, for which there is no ambiguity in the identification of the clumps and that are not a galaxy-galaxy strong-lensing event (GGSL). This choice is due to the fact that the multiple images generated by these kinds of events could result in a very strong constraint on the mass of the galaxy that acts as a lens and could therefore introduce a possible unwanted bias in the scaling relations for the cluster members. This "golden sample" contains $N_{im}^{tot}=47$ multiple images of $N_{fam}=17$ families from 12 background sources, of which 15 multiple images of 6 background sources were previously identified in \cite{Zitrin2017}. This corresponds to a total number of observables of $N_{obs} = 2 \times (N_{im}^{tot} - N_{fam}) = 60$. It spans a wide range of redshift, from $z = 1.17$ to $z = 5.39$ (as shown in Fig. \ref{fig:lensed_images_hist_z}), and covers a large portion of the cluster core (see Fig. \ref{fig:model_absolute_mu} and Table C.1). The properties of the identified multiple images are summarized in Appendix \ref{appendix:multiple_images}.

\subsection{Lens model}
\label{sub:model.lens_model}

Following the path traced by other works to model the total mass distribution of galaxy clusters (see e.g., \citealt{Grillo2016,Caminha2017,Bergamini2021,Granata2022,Acebron2022b}), we decompose the total mass (or equivalently) the total gravitational potential,  $\phi_{tot}$, into several components, as follows (\ref{eq:mass_components}):
\begin{equation}
    \label{eq:mass_components}
    \phi_{tot} = \sum_{i=1}^{N_{cm}} \phi^{(cm)}_i + \sum_{j=1}^{N_{BCG}} \phi^{(BCG)}_{j} + \sum_{k=1}^{N_h} \phi^{(h)}_{k} + \phi_{\kappa,\gamma}
,\end{equation}

\noindent where the first term takes into account the contribution of the $N_{cm}$ cluster members, the second one is for the two BCGs (IDs 273 and 867, see Fig. \ref{fig:cluter_overview}), which are parameterized separately from the other members, the third one describes the contribution of the $N_h$ cluster-scale halos and the last one refers to a constant convergence or shear introduced by possible unaccounted lensing effects \citep{Acebron2022}.\\

Each cluster-scale halo potential, $\phi^{(h)}_{k}$, is represented by a dual pseudo-isothermal elliptical mass density (dPIE, \citealt{Limousin2005}) with seven free parameters: position $(x, y)$;  position angle ($\theta$); ellipticity ($e$); central velocity dispersion ($\sigma_0$ or, as equivalently adopted by lenstool, $\sigma_{LT} = \sqrt{2/3} \sigma_0$); core radius ($r_{\rm{core}})$; truncation radius ($r_{\rm{cut}}$). In order to reduce the number of free parameters \citep{Bergamini2023}, the value of the latter has been fixed to $\qty{2000}{\arcsec}$, a value that is large enough to be considered as infinite compared to the priors assumed for $r_{\rm{core}}$. We used singular dPIE profiles also for the potentials of the BCGs, $\phi^{(BCG)}_{j}$ (but fixing their central position), thereby introducing only four additional free parameters for each BCG. Cluster members, $\phi^{(cm)}_i$, are described with singular, circular dPIE profiles. In order to reduce the number of free parameters, the following scaling relations are used:

\begin{align}
    \label{eq:scaling_sigma}
    \sigma^{(cm)}_{LT,i} &= \sigma^{(ref)}_{LT} \left( \frac{L_i}{L_{ref}} \right)^{\alpha}, \\
    \label{eq:scaling_cut}
    r^{(cm)}_{cut,i} &=  r^{(ref)}_{cut} \left( \frac{L_i}{L_{ref}} \right)^{\beta_{cut}}, \\
    \label{eq:scaling_lm}
    \frac{M_i}{L_i} &\propto L^{\gamma}_i,
\end{align}

We measure the luminosities of the members with their Kron magnitude in band F160W, using as reference the magnitude $m^{(ref)}_{\rm{F160W}} = 16.563$ of the BCG-273. Following \cite{Bergamini2019}, we use $\gamma = 0.2$, which is consistent with the canonical fundamental plane \citep{Faber1989,Bender1992}, $\alpha = 0.35$ and $\beta_{cut} = \gamma - 2\alpha + 1 = 0.5$ . The two remaining free parameters $\sigma^{(ref)}_{LT}$ and $r^{(ref)}_{cut}$ are then computed for the reference luminosity, $L_{ref}$. Finally, the external convergence-shear potential has three parameters: convergence, shear, and position angle. The total number of free parameters of the model is thus: $N_{par} = 6N_h + 4N_{BCG} + 5,  $  which is then reduced to $N_{par} = 6N_h + 4N_{BCG} + 2$ if we do not include the shear component. The corresponding degrees of freedom is $N_{dof} = N_{con} - N_{par}$.\\ 

\begingroup
\setlength{\tabcolsep}{10pt}
\renewcommand{\arraystretch}{1.25} 
\begin{table}[t]
    \centering
    \caption{Model configurations with the associated reduced $\chi^2$ and $\Delta_{RMS}$ values.}
    \begin{tabular}{rccrr}
         $\phi_{\kappa,\gamma}$ & $N_h$ & $N_{BCG} $ & red. $\chi^2$ & $\Delta_{RMS} [\arcsec]$ \\
         \hline
         \hline
                              N &     1 &          2 &        151.11 & 3.44 \\
                              N &     2 &          2 &         32.90 & 1.74 \\
                              Y &     2 &          2 &         50.58 & 2.14 \\
                              N &     2 &          0 &         81.90 & 2.53 \\
                              N &     3 &          0 &         10.38 & 1.05 \\
                              N &     3 &          2 &          4.23 & 0.75 \\
                              N &     4 &          2 &          5.65 & 0.95 \\
        \hline
    \end{tabular}
    \tablefoot{The column $\phi_{\kappa,\gamma}$ indicates whether the model configuration includes (Y) or does not include (N) an external shear or convergence component; $N_h$ indicates the number of cluster scale halos; $N_{BCG}$ indicates the number of BCGs that are parameterized separately from the other cluster members; red. $\chi^2$ is the reduced $\chi^2$ of the optimized model; $\Delta_{RMS}$ is the quantity defined by Eq. (\ref{eq:delta_rms}).}
    \label{tab:model_configs}
\end{table}
\endgroup

We tested several models with different numbers of cluster-scale halos ($1 \leq N_h \leq 4$), with or without the shear term, and/or the two BCGs in the scaling relations. For each configuration, we computed the reduced $\chi^2$ and the $\Delta_{RMS}$ values, the latter is defined by (\ref{eq:delta_rms}) as the RMS of the separation between the observed position, $\boldsymbol{x}_{i}^{obs}$, and predicted one $\boldsymbol{x}_{i}^{pred} $ of the $N_{im}^{tot}$ multiple images:

\begin{equation}
    \label{eq:delta_rms}
    \Delta_{RMS} = \sqrt{\frac{1}{N_{im}^{tot}}\sum_{i=1}^{N_{im}^{tot}}{\lVert \boldsymbol{x}_{i}^{pred} - \boldsymbol{x}_{i}^{obs} \rVert}^2}
.\end{equation}

We find that the model that best reproduces the positions of the observed multiple images of the golden sample (and for which we describe the resulting properties in the following section) is the one featuring  three cluster-scale halos ($N_h = 3$), with the two BCGs out of the scaling relations ($N_{BCG} = 2$) and without the external convergence-shear component (see Table \ref{tab:model_configs}).\\

\section{Results}
\label{sec:results}

\begingroup
\renewcommand{\arraystretch}{2} 
\begin{table*}[t]
    \centering
     \caption{\label{tab:model_params} Input and optimized parameter values and assumed priors for the lens model.}
    \resizebox{\textwidth}{!}{%
    \begin{tabular}{|c|c|c|c|c|c|c|c|c|}
    \cline{3-9}
    \multicolumn{2}{c|}{} &\multicolumn{7}{|c|}{\textbf{Input parameter and assumed priors for the lens model}} \\\cline{3-9}
    \multicolumn{2}{c|}{} & \textbf{x}~$\mathbf{[\arcsec]}$ & \textbf{y}~$\mathbf{[\arcsec]}$ & $\boldsymbol{e}$ & $\boldsymbol{\theta}$ \textbf{[°]} & $\boldsymbol{\sigma} ~ \mathbf{[km~s^{-1}]}$ & $\boldsymbol{r_{\rm{core}}}~[\arcsec]$ & $\boldsymbol{r_{\rm{cut}}}~[\arcsec]$ \\
    \hline
    \multirow{3}{*}{\rotatebox{90}{\textbf{Cluster-scale}}} & $\boldsymbol{1^{st}}$ \textbf{Cluster halo} & $-30.0 \div ~30.0$ & $-30.0 \div ~30.0$ & $0.0 \div ~0.9$ & $0.0 \div ~180.0$ & $400 \div 2000$ & $0.5 \div 120.0$ & $2000.0$\\
    & $\boldsymbol{2^{nd}}$ \textbf{Cluster halo} & $-60.0 \div ~80.0$ & $-60.0 \div ~50.0$ & $0.0 \div ~0.9$ & $0.0 \div ~180.0$ & $200 \div 2000$ & $0.5 \div 120.0$ & $2000.0$\\
    & $\boldsymbol{3^{rd}}$ \textbf{Cluster halo} & $-60.0 \div ~50.0$ & $-60.0 \div ~50.0$ & $0.0 \div ~0.9$ & $0.0 \div ~180.0$ & $200 \div 2000$ & $0.5 \div 120.0$ & $2000.0$\\
    \hline
    \hline
    \multirow{3}{*}{\rotatebox{90}{\textbf{Subhalos}}}  & \textbf{BCG-273} & $0.0$ & $0.0$ & $0.0 \div ~0.9$ & $0.0 \div ~180.0$ & $150 \div 450$ & $0.001$ & $0.0 \div 120$\\
    & \textbf{BCG-867} & $-16.7$ & $7.0$ & $0.0 \div ~0.9$ & $0.0 \div ~180.0$ & $150 \div 450$ & $0.001$ & $0.0 \div 120$\\
    \cline{2-9}
     & \textbf{Scaling relations} & $\boldsymbol{N_{gal}} = 151$ & $\boldsymbol{m^{(ref)}_{\rm{F160W}}} = 16.563$ & $\boldsymbol{\alpha} = 0.35$ & $\boldsymbol{\sigma^{(ref)}_{LT}} = 50 \div 400$ & $\boldsymbol{\beta_{cut}} = 0.5$ & $\boldsymbol{r^{(ref)}_{cut}} = 1.0 \div 50.0$ & $\boldsymbol{\gamma} = 0.2$ \\
    \hline
    \end{tabular}
    }

    \bigskip

    \resizebox{\textwidth}{!}{%
    \begin{tabular}{|c|c|c|c|c|c|c|c|c|}
    \cline{3-9}
    \multicolumn{2}{c|}{} &\multicolumn{7}{|c|}{\textbf{Optimized output parameters for the lens model}} \\\cline{3-9}
    \multicolumn{2}{c|}{} & \textbf{x}~$\mathbf{[\arcsec]}$ & \textbf{y}~$\mathbf{[\arcsec]}$ & $\boldsymbol{e}$ & $\boldsymbol{\theta}$ \textbf{[°]} & $\boldsymbol{\sigma} ~ \mathbf{[km~s^{-1}]}$ & $\boldsymbol{r_{\rm{core}}}~[\arcsec]$ & $\boldsymbol{r_{\rm{cut}}}~[\arcsec]$ \\
    \hline
    \multirow{3}{*}{\rotatebox{90}{\textbf{Cluster-scale}}} & $\boldsymbol{1^{st}}$ \textbf{Cluster halo} & ${0.0}~_{-0.4}^{+0.5}$ & ${5.1}~_{-0.3}^{+0.5}$ & ${0.546}~_{-0.014}^{+0.009}$ & ${61.7}~_{-0.6}^{+1.3}$ & ${1330}~_{-20}^{+20}$ & ${29.1}~_{-0.7}^{+0.6}$ & $2000.0$\\
    & $\boldsymbol{2^{nd}}$ \textbf{Cluster halo} & ${61.0}~_{-2.0}^{+4.0}$ & ${-20.8}~_{-0.8}^{+1.4}$ & ${0.71}~_{-0.08}^{+0.09}$ & ${24.0}~_{-2.0}^{+2.0}$ & ${580}~_{-30}^{+40}$ & ${15.0}~_{-2.0}^{+3.0}$ & $2000.0$\\
    & $\boldsymbol{3^{rd}}$ \textbf{Cluster halo} & ${-51.0}~_{-1.0}^{+1.0}$ & ${-58.9}~_{-0.8}^{+1.7}$ & ${0.7}~_{-0.1}^{+0.1}$ & ${114.0}~_{-7.0}^{+8.0}$ & ${480}~_{-40}^{+60}$ & ${15.0}~_{-3.0}^{+3.0}$ & $2000.0$\\
    \hline
    \hline
    \multirow{3}{*}{\rotatebox{90}{\textbf{Subhalos}}} & \textbf{BCG-273} & 0 & 0 & ${0.73}~_{-0.03}^{+0.03}$ & ${119.0}~_{-5.0}^{+12.0}$ & ${334}~_{-8}^{+11}$ & $0.001$ & ${90.0}~_{-30.0}^{+20.0}$\\
    & \textbf{BCG-867} & $-16.7$ & $7.0$ & ${0.79}~_{-0.04}^{+0.03}$ & ${171.0}~_{-2.0}^{+1.0}$ & ${435}~_{-14}^{+6}$ & $0.001$ & ${100.0}~_{-20.0}^{+10.0}$\\
    \cline{2-9}
    & \textbf{Scaling relations} & $\boldsymbol{N_{gal}} = 151$ & $\boldsymbol{m^{(ref)}_{\rm{F160W}}} = 16.563$ & $\boldsymbol{\alpha} = 0.35$ & $\boldsymbol{\sigma^{(ref)}_{LT}} = {310.0}~_{-10.0}^{+10.0}$ & $\boldsymbol{\beta_{cut}} = 0.5$ & $\boldsymbol{r^{(ref)}_{cut}} = {16.0}~_{-2.0}^{+3.0}$ & $\boldsymbol{\gamma} = 0.2$ \\
    \hline
    \end{tabular}
    }
    \tablefoot{Input parameter values and assumed uniform priors (top panel) and median optimized parameter values with their 16th and 84th percentiles from the marginalized posterior distribution (bottom panel). Boundaries of uniform priors are separated by the $\div$ symbol; $N_{gal}$ is the number of cluster member galaxies optimized using the scaling relations (see Eqs. \ref{eq:scaling_sigma}, \ref{eq:scaling_cut}, and \ref{eq:scaling_lm}); $m^{(ref)}_{\rm{F160W}}$ is the magnitude in band F160W of BCG-273 and it has been used as reference for the scaling relations. We use the coordinates of the center of BCG-273 (ICRS $177.7090129, -28.0821343$) as the reference coordinates for the strong-lensing model. Note: for lenstool to compute the relative positions, the x and y coordinates are expressed in arcseconds with the x-axis directed toward the west (decreasing direction of R.A.) and the y-axis toward the north (increasing direction of Dec.).}
\end{table*}
\endgroup

\begin{figure}
    \centering
    \includegraphics[width=0.5\textwidth]{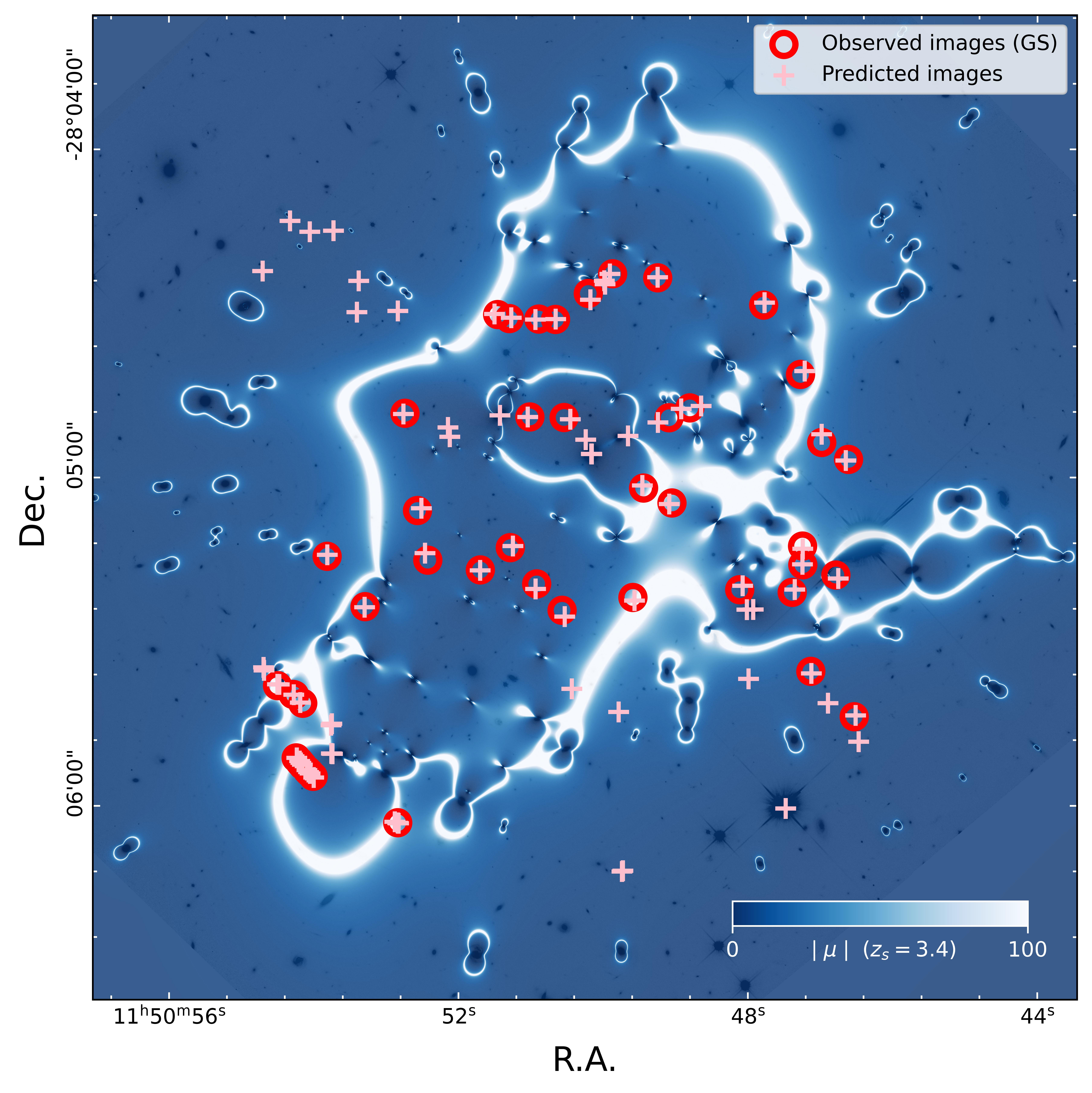}
    \caption{Absolute magnification, $\mid \mu \mid,$ map for images of a source at redshift of $z_s = 3.4$, overlaid onto the F814W image. The red circles indicate the position of the observed multiple images in the golden sample and pink crosses are the multiple images predicted with the optimized lens model.}
    \label{fig:model_absolute_mu}
\end{figure}

\begin{figure}
    \centering
    \includegraphics[width=0.5\textwidth]{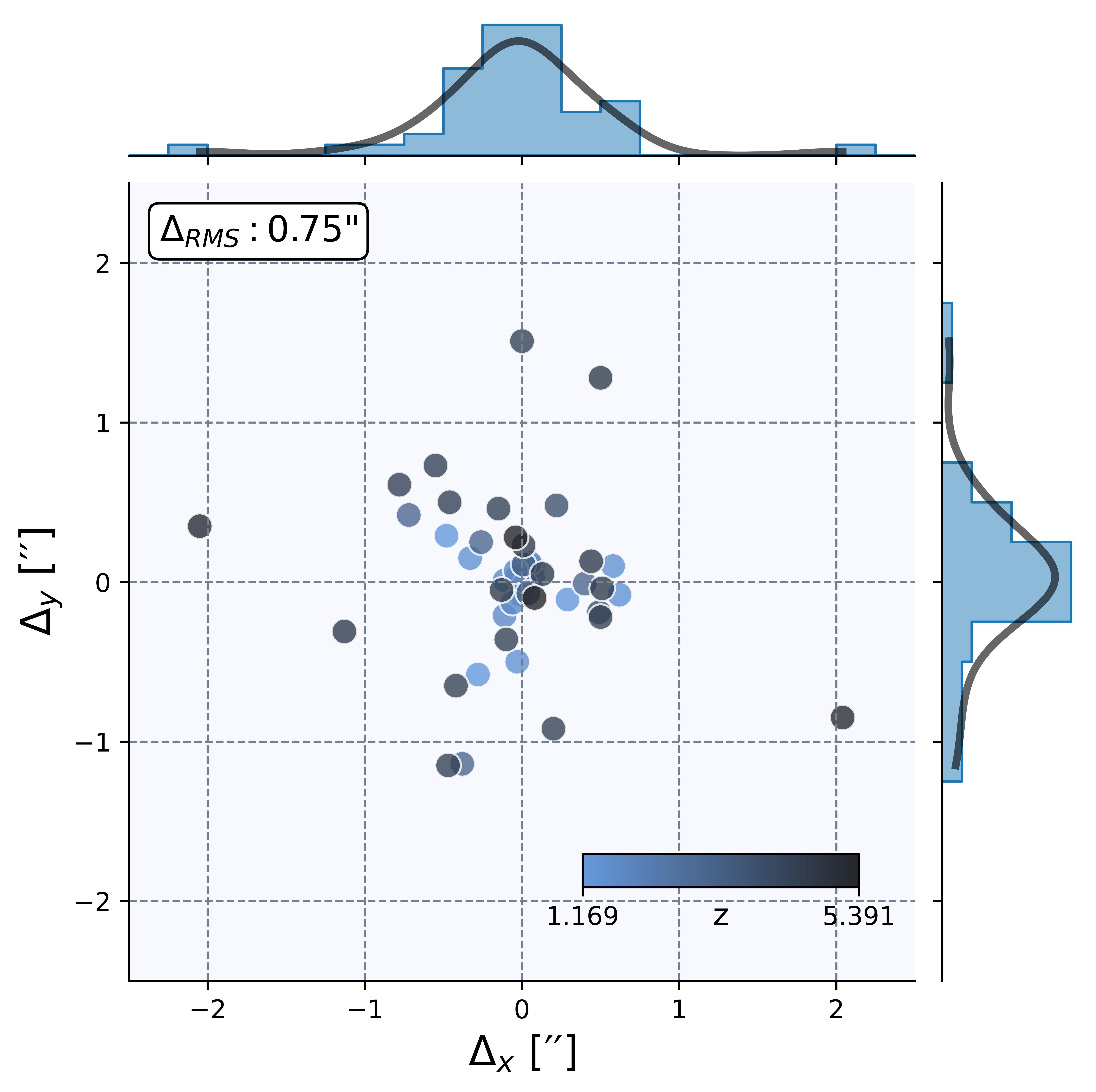}
    \caption{Distribution of the displacement in the image plane, in arcseconds, along the x and y axes between the observed and model-predicted positions of multiple images in the golden sample. The color of the points indicates the redshift of the objects. The histograms are the marginal distributions of the displacements, computed using ten bins of width $\qty{0.25}{\arcsec}$, and the solid black lines are their kernel density estimates. The displacement root mean square (RMS) is $\Delta_{RMS} = \qty{0.75}{\arcsec}$.}
    \label{fig:cimg_displacement}
\end{figure}

\begin{figure}
    \centering
    \includegraphics[width=0.5\textwidth]{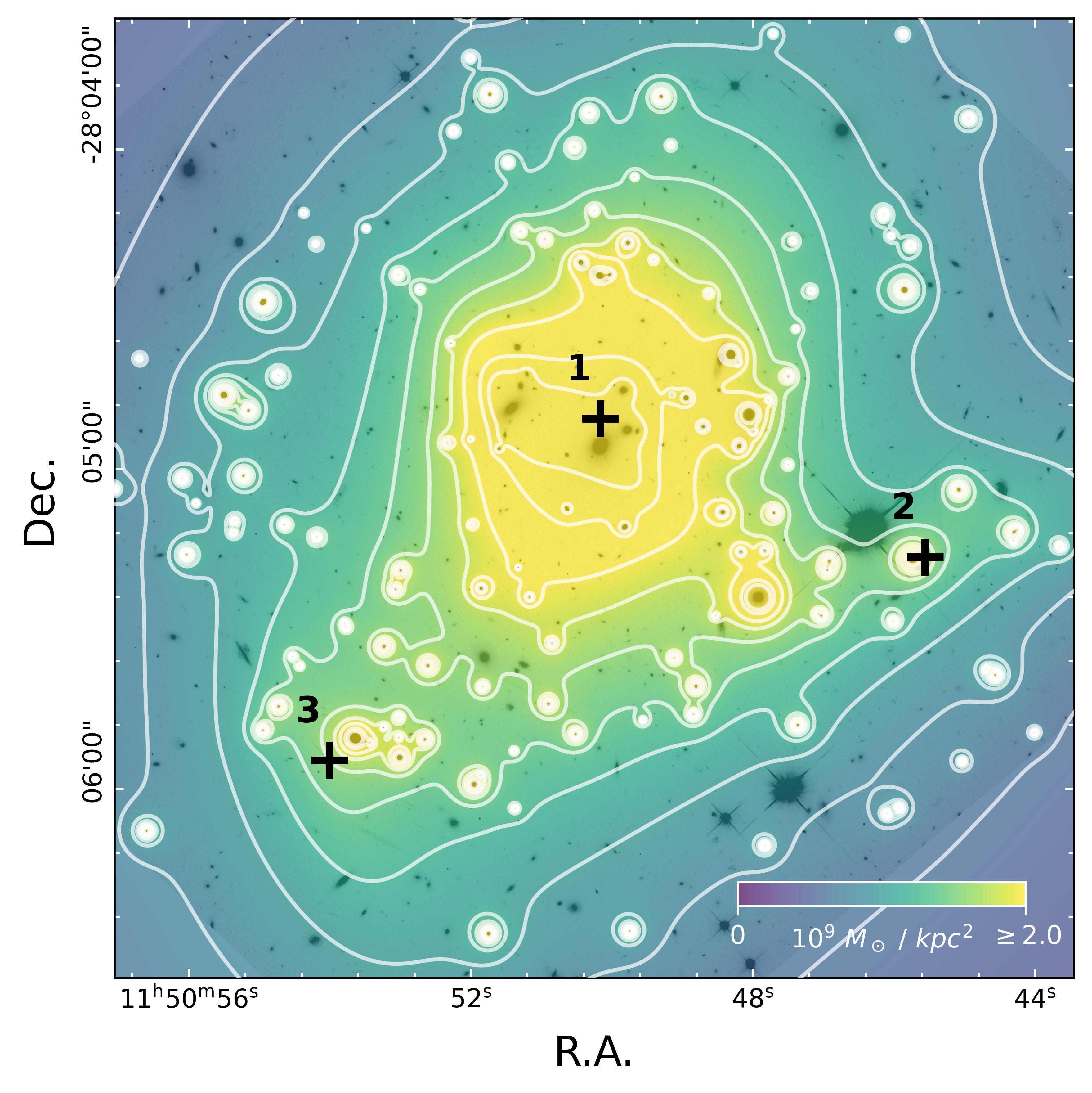}
    \includegraphics[width=0.5\textwidth]{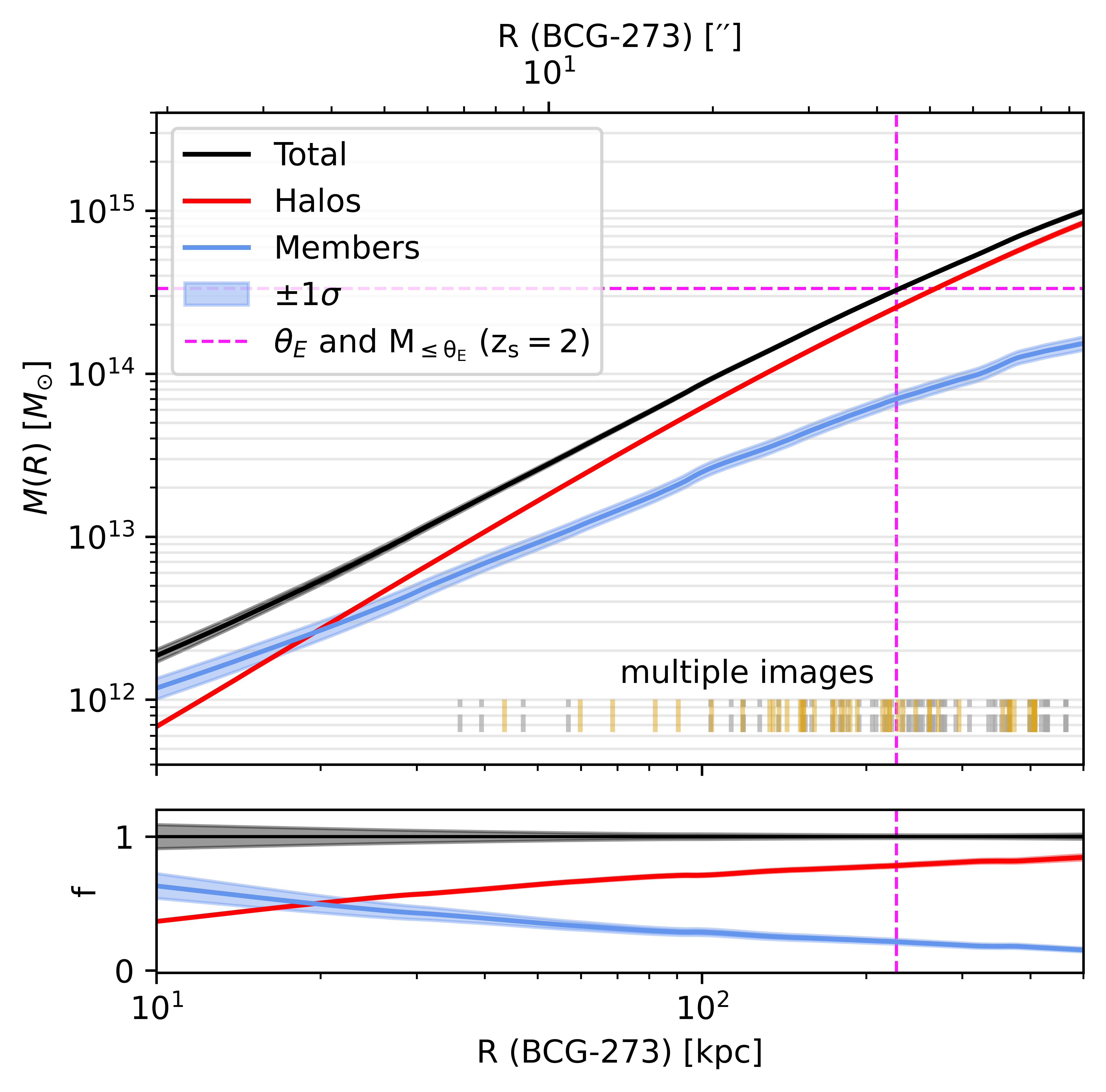}
    \caption{Total projected mass distribution in units of $10^9 ~ \rm{M}_{\odot} ~ / ~ \rm{kpc}^2$ and the central position of the three cluster halos (black + markers), overlaid onto an F814W image (upper panel) and cumulative total median mass profile as a function of the distance $R$ from the BCG 273 (middle panel). Vertical lines mark the radial distance of the multiple images: the golden sample used to optimize the lens model is indicated in gold and other multiple images are indicated by the gray dashed lines. Cumulative mass profiles for the cluster members, including the two BCGs, (in blue), and for the cluster halos (in red) are also shown, along with their ratio with the total mass profile (lower panel). The shaded area indicates, for each mass component, the $\pm1\sigma$ interval. Magenta dashed lines indicate the cluster Einstein radius ($\theta_{\rm{E}}$) and the total mass enclosed by the corresponding critical curve ($\rm{M_{E}}$) for a source at a redshift of $z_s = 2$.}
    \label{fig:model_mass}
\end{figure}

\begin{figure}
    \centering
    \includegraphics[width=0.5\textwidth]{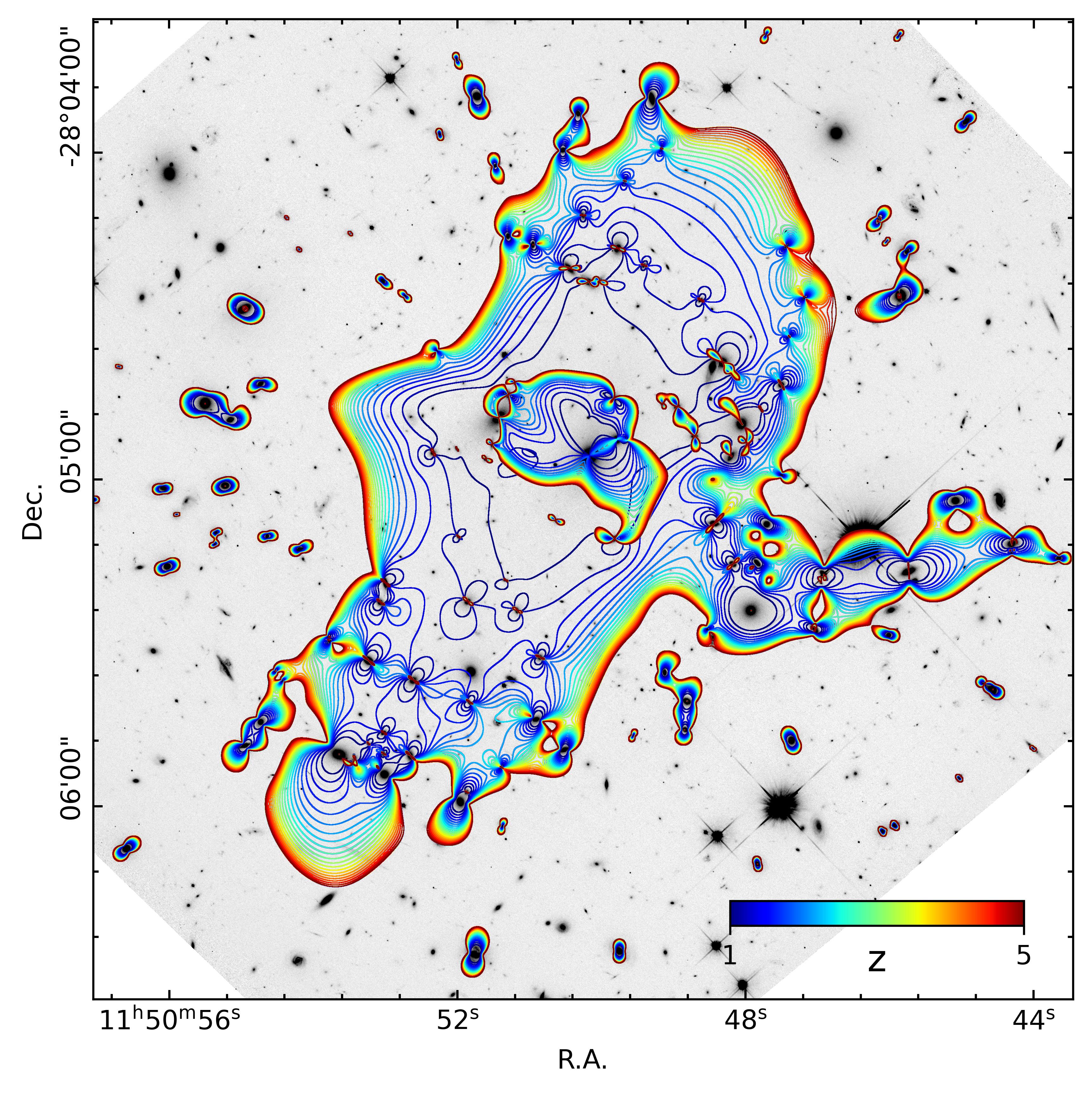}
    \includegraphics[width=0.5\textwidth]{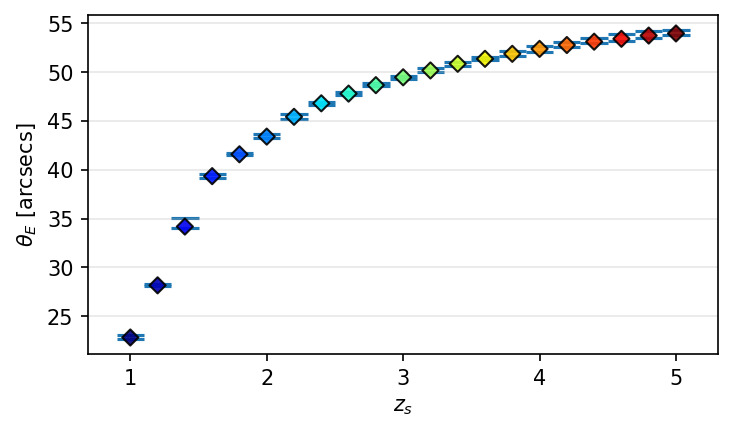}
    \caption{Critical curves, overlaid onto the F814W image, for various redshift values, $z_s$, of the source plane in the range $1 \leq z_s \leq 5$ (upper panel) and the corresponding effective Einstein radii, $\theta_{\rm{E}}$, (lower panel). Error bars in the Einstein radii plot indicate the 16th and 84th percentiles. The values are reported in Table \ref{tab:ein_rad_z}. The color maps the redshift of the source, $z_s$.}
    \label{fig:model_ein_rad}
\end{figure}

In the top panel of Table \ref{tab:model_params}, we report the input parameter values and the assumed uniform priors. In the bottom panel, we report the median values for the parameters of the optimized strong-lensing model, along with the 16th and 84th percentiles from their marginalized posterior distributions. All the percentiles and uncertainties for the quantities reported in this section have been computed using a sample of 150 random realizations of the model extracted from the MCMC chain. For the Einstein radii estimates, we also used a bootstrapping technique \citep{Efron1997} to reduce as much as possible the potential biases caused by a non-normal distribution of the values. The best-fit model is characterized by $\Delta_{RMS} = \ang[angle-symbol-over-decimal]{;;0.75}$, which corresponds to an increase in the accuracy by a factor of $\sim2.5,$ compared to value  of $\ang[angle-symbol-over-decimal]{;;1.9}$ of the model presented by \cite{Zitrin2017}. The goodness of the model in reproducing the observed positions of the multiple images of the golden sample is also illustrated in Fig. \ref{fig:model_absolute_mu}, where both the observed and model-predicted positions of the multiple images are plotted. Moreover, Fig. \ref{fig:cimg_displacement} shows no correlation between the displacements of the positions of the multiple images and the redshift of the sources.\\

Figure \ref{fig:model_mass} shows (in the upper panel) the total projected mass distribution overlaid onto the F814W HST image. In the lower panels, the cumulative total mass profile and its different mass components, within $500~\text{kpc}$ from BCG-273, are illustrated. In Fig. \ref{fig:model_ein_rad}, we show the critical lines, overlaid onto the F814W HST image, for a source at different redshift values in the range $1 \leq z_s \leq 5$ and a scatter plot of the median effective Einstein radius, $\theta_{\rm{E}} = \sqrt{A / \pi}$, where $A$ is the area enclosed within the critical curve. For each value of $z_s$ we considered only the biggest critical curve (i.e., the critical curves generated by isolated cluster members were ignored) and the calculation of the value of $\theta_{\rm{E}}$ was made in the image pixel-space and then the resulting radius is converted into physical angular values. Our lens modeling of this cluster reveals a mass distribution that is slightly elongated along the NW-SE direction, corroborating the evidence of a post-merger scenario suggested by the previous X-ray and radio observations of \cite{Bagchi2011} and \cite{Bonafede2014}, as well as the weak-lensing analysis of \cite{Finner2017} based on Subaru and HST telescopes observations. In particular, we note that the positions of the cluster-scale halos 1, 2, and 3 of our optimized lens model are consistent with the positions of the halos NWc, SEc, and Wc detected by \cite{Finner2017}, respectively\\\\

In Table \ref{tab:ein_rad_z}, we report the median values of $\theta_{\rm{E}}$ and total mass, $\rm{M_{E}}$, enclosed within the critical curve (with the 16th and 84th percentiles) for several values of $z_s$. For a source at $z_s = 2$, we find that this galaxy cluster has an effective Einstein radius of $\theta_{\rm{E}} = 43.4\arcsec \pm 0.1\arcsec$, which corresponds to a total mass enclosed within the critical curve of $3.33_{-0.07}^{+0.02} \times 10^{14} \rm{M}_{\odot}$.  These results are also in agreement with the previous measurements by \cite{Zitrin2017}, who found $\theta_E = (42 \pm 4)"$ and an enclosed mass of $(3.1 \pm 0.5) \times 10^{14} M_{\odot}$. There are currently only four clusters with a confirmed $\theta_{\rm{E}} \geq \qty{40}{\arcsec}$ and our modeling of PLCK-G287 confirms it to be the third largest after MACS J0717.5$+$3745 ($\theta_{\rm{E}} \sim \qty{55}{\arcsec}$, \citealt{Zitrin2009}) and Abell 1689 ($\theta_{\rm{E}} \sim \qty{45}{\arcsec}$, \citealt{Broadhurst2005}), and coming just before RXC J2211.7$-$0349 ($\theta_{\rm{E}} \sim \qty{41}{\arcsec}$, \citealt{Cerny2018}). Clusters like this are of particular interest when comparing Einstein radius distributions from observations to those from theoretical expectations; for example, the semi-analytic analysis of \cite{Oguri2009}, based on the \textLambda-CDM model and cosmological parameters from Wilkinson Microwave Anisotropy Probe five-year data (WMAP5, \citealt{WMAP5}), predicts an all-sky total of $\sim 15 \pm 5$ galaxy clusters with $\theta_{\rm{E}} \geq \qty{40}{\arcsec}$ for a source at $z_s = 3$. Since the Einstein radius increases with the redshift of the lensed source, for a source at $z_s = 2,$ the all-sky total number of galaxy clusters with $\theta_{\rm{E}} \geq \qty{40}{\arcsec}$ should be lower than this quantity. In fact, \cite{Zitrin2012} reported the result of the same analysis, but done using WMAP seven-year data (WMAP7, \citealt{WMAP7}) and for a source at $z_s = 2$, from which we would get a total all-sky count of only $\sim 8 \pm 3$ galaxy clusters.\\

We also computed the probability of galaxy-galaxy strong-lensing (GGSL) events using the same methodologies described in \cite{Meneghetti2023}. We find a median GGSL probability of $P_{\rm{GGLS}} = \left (1.7_{-0.2}^{+0.5} \right) \times 10^{-2}$, for a source at $z_s = 6$ . This value is approximately one order of magnitude larger than the probabilities reported by \cite{Meneghetti2023} for the other galaxy clusters at similar redshift. According to this study, there is a significant discrepancy with the results of cosmological simulations. However, we point out that at the moment, there is no simulated cluster with an Einstein radius as large as that of PLCK-G287. Therefore, the implications of this intriguing result  will be discussed in detail in a future work\\

\begin{figure*}
    \centering
    \includegraphics[width=0.45\textwidth]{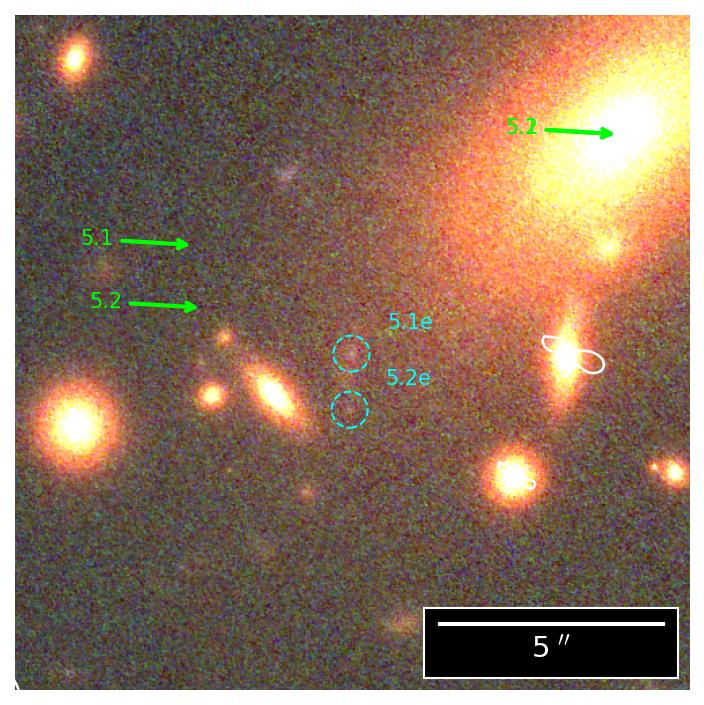}
    \includegraphics[width=0.45\textwidth]{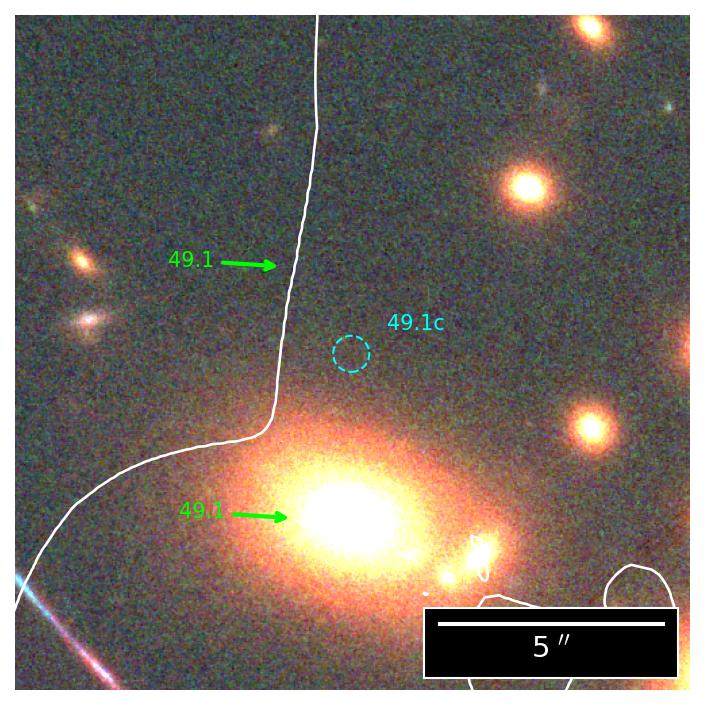}
    \caption{Predicted (green arrows) and observed (cyan dashed circles) positions for multiple lensed images 5.1e, 5.2e (left panel), and 49.1c (right panel). These multiple images were not detected originally and have been found by visually inspecting the data-cube around the positions predicted by the strong-lensing model. Note: System 49 is a Lyman-\textalpha{} emitter that is too faint to be visible in the RGB cutout, but it is clearly visible in the spectral data-cube (see the spectrum plots for system 49 in Fig. \ref{fig:cimg_sys_49} in Appendix \ref{appendix:multiple_images}).}
    \label{fig:validation_fams}
\end{figure*}

\begin{figure*}
    \centering
    \includegraphics[width=0.45\textwidth]{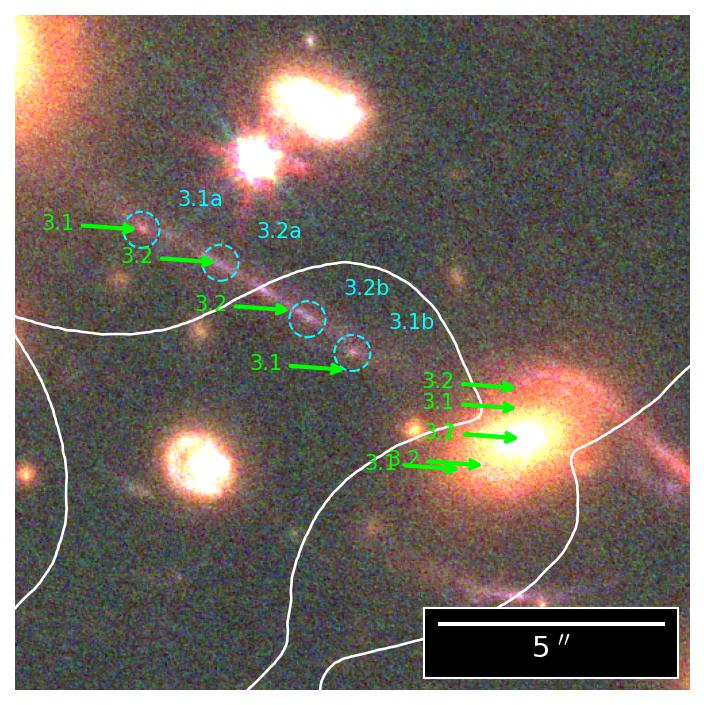}
    \includegraphics[width=0.45\textwidth]{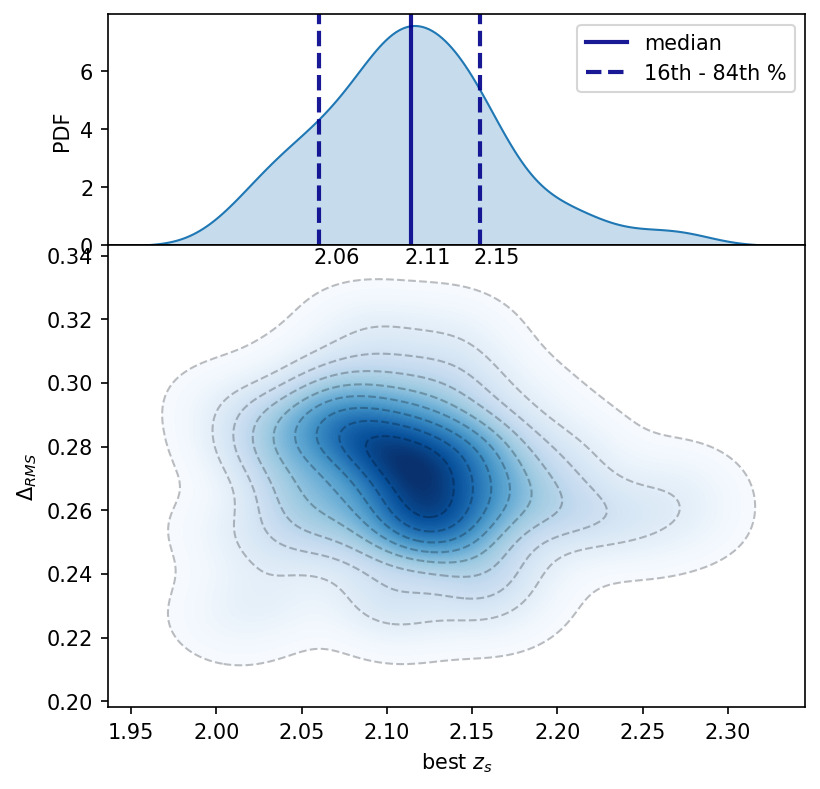}
    \caption{Predicted (green arrows) and observed (cyan dashed circles) positions for multiple lensed images of system 3 (left panel) and the PDF of the redshift obtained from the strong-lensing model (right panel). From the redshift PDF we get a median value of $z_{s3} = 2.11 \pm 0.05$ with a $\Delta_{RMS} = \ang[angle-symbol-over-decimal]{;;0.27} \pm \ang[angle-symbol-over-decimal]{;;0.02}$. The critical line for a source at this redshift is also shown in the right panel.}
    \label{fig:validation_zbests_sys_3}
\end{figure*}

\begin{figure*}
    \centering
    \includegraphics[width=0.45\textwidth]{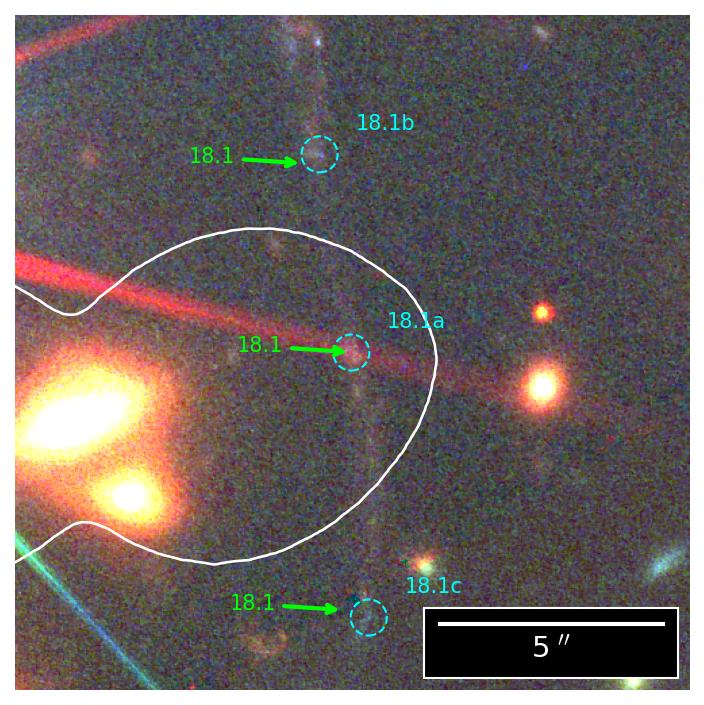}
    \includegraphics[width=0.45\textwidth]{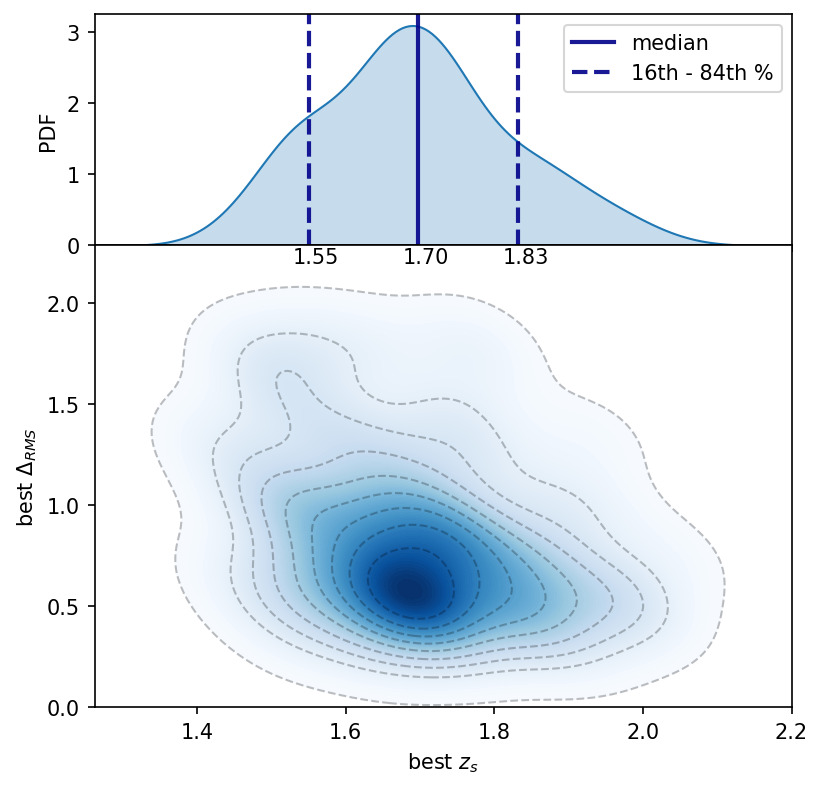}
    \caption{Predicted (green arrows) and observed (cyan dashed circles) positions for multiple lensed images of system 18 (left panel) and the PDF of the redshift obtained from the strong-lensing model (right panel). From the redshift PDF we get a median value of $z_{s18} = 1.70 \pm 0.15$ with a $\Delta_{RMS} = \ang[angle-symbol-over-decimal]{;;0.7} \pm \ang[angle-symbol-over-decimal]{;;0.5}$. The critical line for a source at this redshift is also shown in the right panel.}
    \label{fig:validation_zbests_sys_18}
\end{figure*}

\begin{figure*}
    \centering
    \includegraphics[width=0.45\textwidth]{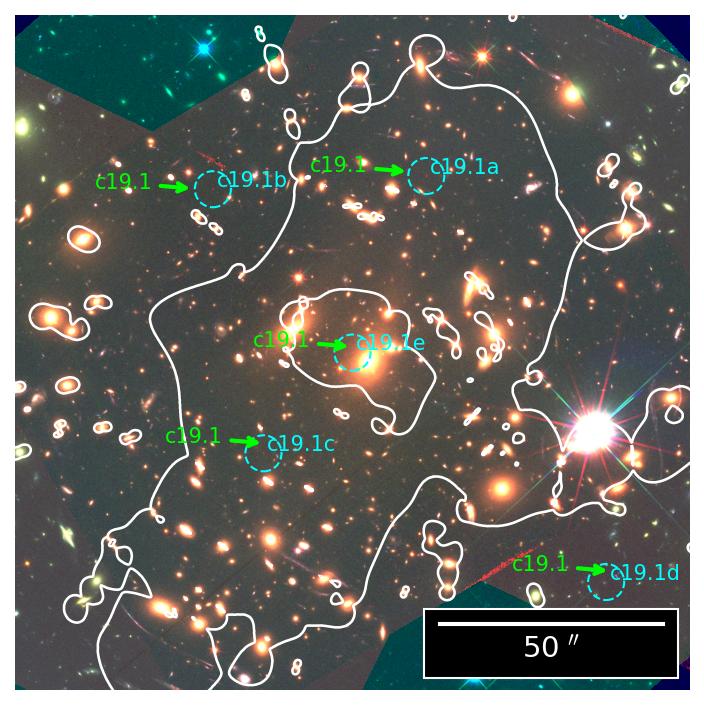}
    \includegraphics[width=0.45\textwidth]{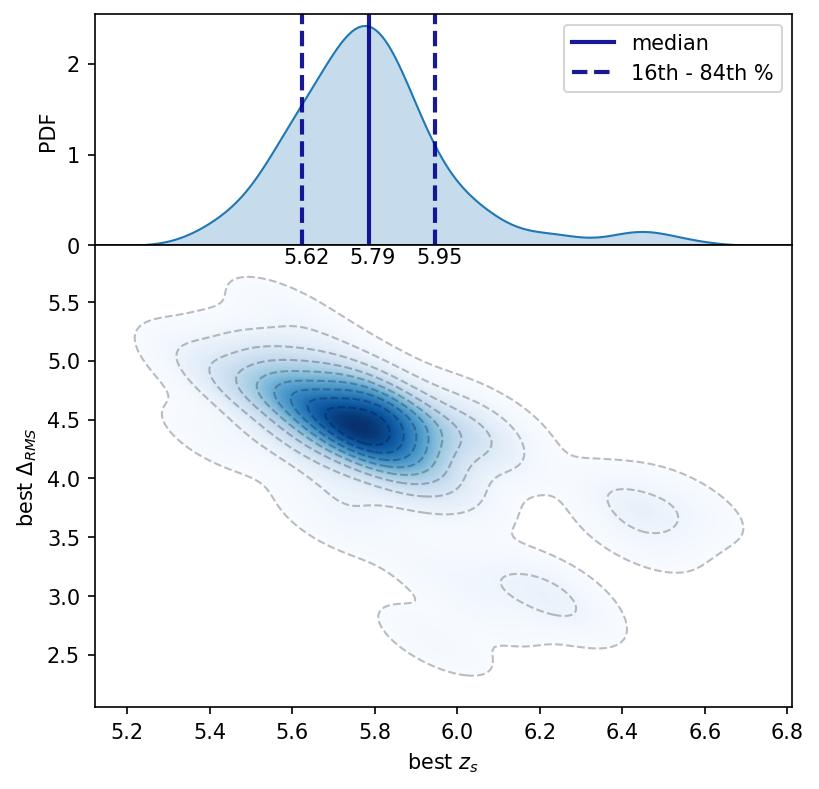}\\
    \caption{Predicted (green arrows) and observed (cyan dashed circles) positions for multiple lensed images of system 19 (left panel) and the PDF of the redshift obtained from the strong-lensing model (right panel). From the redshift PDF we get a median value of $z_{s19} = 5.79 \pm 0.17$ with a $\Delta_{RMS} = \ang[angle-symbol-over-decimal]{;;4.5} \pm \ang[angle-symbol-over-decimal]{;;0.3}$. The critical line for a source at this redshift is also shown in the images.}
    \label{fig:validation_zbests_sys_19}
\end{figure*}

Using the positions of the multiple images predicted by the best-fitting model, we also discovered the new multiple images 49.1c, 5.1e, and 5.2e that were not previously identified (see Fig. \ref{fig:validation_fams}). These observed new multiple images are a few arcseconds off the predicted position, but this is expected since they are not considered in the model optimization. We also tried to predict the position of the counter-images for systems 3, 18, and 19 (see Fig. \ref{fig:validation_zbests_sys_3}, Fig. \ref{fig:validation_zbests_sys_18}, and Fig. \ref{fig:validation_zbests_sys_19}). These systems are not part of the golden sample due to the uncertainty on the quality of the redshift estimation caused by the low signal-to-noise ratio (S/N) of their extracted spectra. Therefore, we used the strong-lensing model to predict the positions of the multiple images of 3.1a, 3.2a, 18.1a, and 19.1e for various redshift values and then selecting (for each system) the one that produced the lowest value for the $\Delta_{RMS}$. We repeated the process also using 150 random realizations of the model built by randomly extracting the configurations from the MCMC chain. In this way, we were able to compute the probability density functions (PDF) of the redshifts for these two systems. This gives us a median value of the redshift of $z_{s3} = 2.11 \pm 0.05,$ with  $\Delta_{RMS} = \ang[angle-symbol-over-decimal]{;;0.27} \pm \ang[angle-symbol-over-decimal]{;;0.02}$ for system 3, $z_{s50} = 1.70_{-0.15}^{+0.13}$ with  $\Delta_{RMS} = \ang[angle-symbol-over-decimal]{;;0.7} \pm \ang[angle-symbol-over-decimal]{;;0.5}$ for system 18, and  $z_{s19} = 5.79 \pm 0.17$ with $\Delta_{RMS} = \ang[angle-symbol-over-decimal]{;;4.5} \pm \ang[angle-symbol-over-decimal]{;;0.3}$ for system 19. It is worth noting that the redshifts predicted for systems 3 and 18 are compatible with the spectroscopic ones within $2\sigma$ and $1\sigma$, respectively, while the redshift predicted for system 19 is compatible within $2\sigma$ with the photometric redshift from \cite{Zitrin2017}. System 19 was also identified in by \cite{Salmon2020} during their search for high-redshift objects in the RELICS survey. The prediction of our model is compatible within $3\sigma$ with their photometric redshift estimation of $6.8_{-0.2}^{+0.4}$.

\begingroup
\renewcommand{\arraystretch}{1.75} 
\begin{table*}[h]
    \centering
    \caption{\label{tab:ein_rad_z} Effective Einstein radii and masses enclosed in the critical curves.}
    \begin{tabular}{ccc|ccc|ccc}
    $z_s$ & ~$\theta_{\rm{E}}~[\arcsec]$ & ~$\rm{M_{E}}~[10^{14}~\rm{M}_{\odot}]$ & $z_s$ & ~$\theta_{\rm{E}}~[\arcsec]$ & ~$\rm{M_{E}}~[10^{14}~\rm{M}_{\odot}]$ & $z_s$ & ~$\theta_{\rm{E}}~[\arcsec]$ & ~$\rm{M_{E}}~[10^{14}~\rm{M}_{\odot}]$\\
    \hline
    \hline
    1.0 & ${22.82}~_{-0.04}^{+0.09}~$ & ${1.19}~_{-0.02}^{+0.01}~$  & 2.4 & ${46.82}~_{-0.04}^{+0.05}~$ & ${3.76}~_{-0.08}^{+0.03}~$  & 3.8 & ${51.89}~_{-0.11}^{+0.06}~$ & ${4.38}~_{-0.09}^{+0.04}~$  \\
    1.2 & ${28.16}~_{-0.03}^{+0.03}~$ & ${1.67}~_{-0.03}^{+0.01}~$  & 2.6 & ${47.81}~_{-0.04}^{+0.02}~$ & ${3.88}~_{-0.08}^{+0.03}~$  & 4.0 & ${52.40}~_{-0.11}^{+0.09}~$ & ${4.46}~_{-0.09}^{+0.04}~$  \\
    1.4 & ${34.19}~_{-0.04}^{+0.10}~$ & ${2.26}~_{-0.05}^{+0.01}~$  & 2.8 & ${48.68}~_{-0.03}^{+0.06}~$ & ${4.00}~_{-0.08}^{+0.03}~$  & 4.2 & ${52.79}~_{-0.06}^{+0.08}~$ & ${4.51}~_{-0.09}^{+0.04}~$  \\
    1.6 & ${39.35}~_{-0.04}^{+0.04}~$ & ${2.85}~_{-0.06}^{+0.02}~$  & 3.0 & ${49.48}~_{-0.06}^{+0.05}~$ & ${4.08}~_{-0.08}^{+0.03}~$  & 4.4 & ${53.13}~_{-0.06}^{+0.06}~$ & ${4.55}~_{-0.09}^{+0.04}~$  \\
    1.8 & ${41.59}~_{-0.02}^{+0.04}~$ & ${3.12}~_{-0.06}^{+0.02}~$  & 3.2 & ${50.19}~_{-0.03}^{+0.06}~$ & ${4.16}~_{-0.08}^{+0.04}~$  & 4.6 & ${53.43}~_{-0.06}^{+0.10}~$ & ${4.58}~_{-0.09}^{+0.04}~$  \\
    2.0 & ${43.41}~_{-0.06}^{+0.05}~$ & ${3.33}~_{-0.07}^{+0.02}~$  & 3.4 & ${50.87}~_{-0.05}^{+0.05}~$ & ${4.24}~_{-0.08}^{+0.04}~$  & 4.8 & ${53.76}~_{-0.08}^{+0.12}~$ & ${4.62}~_{-0.09}^{+0.04}~$  \\
    2.2 & ${45.43}~_{-0.07}^{+0.05}~$ & ${3.59}~_{-0.07}^{+0.03}~$  & 3.6 & ${51.38}~_{-0.03}^{+0.06}~$ & ${4.33}~_{-0.08}^{+0.04}~$  & 5.0 & ${53.98}~_{-0.07}^{+0.08}~$ & ${4.65}~_{-0.09}^{+0.05}~$  \\
    \hline
    \end{tabular}
    \tablefoot{Median values of effective Einstein radius ($\theta_{\rm{E}}$) for several redshifts of the source ($z_s$) and corresponding median total mass enclosed in the critical curve ($\rm{M_{E}}$) from the best-fit model. We also quote also the 16th - 84th percentiles.}
\end{table*}
\endgroup

\section{Summary}
\label{sec:summary}

We present a new high-precision strong-lensing model for the galaxy cluster PLCK G287.0+32.9, including a catalog of spectroscopic redshift containing 490 objects with a quality flag $\rm{QF} \geq 3$. The combined use of HST ACS and WFC3 images, as well as spectroscopic data from VLT MUSE and VIMOS and from KECK DEIMOS allow us to confirm 30 multiple images of 12 background sources previously identified by \cite{Zitrin2017} and to identify 16 new multiply lensed background sources, resulting in a total sample of 114 multiple images of 28 multiply lensed sources. Of these, a golden sample of 47 multiple images of 12 multiply lensed sources is used to optimize our best-fitting strong-lensing model. To date, this is the largest spectroscopic sample of multiple images available for this cluster.\\

The final $\Delta_{RMS}$ is equal to $\ang[angle-symbol-over-decimal]{;;0.75}$, which corresponds to an improvement of a factor of $\sim 2.5$ compared to the previous lens model by \cite{Zitrin2017} in terms of the accuracy in reconstructing the position of the multiple images. The derived total mass distribution and Einstein radius confirm this cluster to be a very prominent gravitational lens with an effective $\theta_{\rm{E}} = 43.4\arcsec \pm 0.1\arcsec$, for a source at redshift of $z_s = 2$, and a corresponding total mass enclosed in the critical curve $\rm{M_{E}} = {3.33}_{-0.07}^{+0.02} \times{10^{14} \rm{M}_{\odot}}$.\\

We also validated the lens model by searching for possible new multiple images that had not previously been identified and, subsequently, we also checked its ability to reproduce the multiple images of three systems not used in the optimization process. The predictions of the model lead us to the identification of three new multiple images (5.1e, 5.2e, and 49.1c) and allow us to reproduce the complex configurations of the systems 3 and 18, with a low value of their $\Delta_{RMS}$, while also producing  redshift estimates that are compatible with the spectroscopic ones. The model is also able to reproduce the geometry of system 19 and provides a redshift that is compatible with the photometric estimates of previous studies, thus confirming that this is a high-redshift object at $z \sim 6$.

\begin{acknowledgements}

This work is based on observations taken by the RELICS Treasury Program (GO 14096) with the NASA/ESA HST, which is operated by the Association of Universities for Research in Astronomy, Inc., under NASA contract NAS5-26555.
Based on observations collected at the European Southern Observatory under ESO programme(s) 0102.A-0640(A) and/or data obtained from the ESO Science Archive Facility with DOI(s) under https://doi.org/10.18727/archive/41.
We acknowledge financial support through grants PRIN-MIUR 2017WSCC32 and 2020SKSTHZ. AM acknowledges financial support through grant NextGenerationEU" RFF M4C2 1.1 PRIN 2022 project 2022ZSL4BL INSIGHT.
AA has received funding from the European Union’s Horizon 2020 research and innovation programme under the Marie Skłodowska-Curie grant agreement No 101024195 — ROSEAU.
The data published in this paper have been obtained using the pandora.ez software developed by INAF IASF-Milano.
This research made use of Photutils, an Astropy package for the detection and photometry of astronomical sources \citep{Photutils180}.
\end{acknowledgements}

\begin{appendix}

\section{Masking star diffraction spikes}
\label{appendix:start_spike}
\begin{figure}
    \centering
    \includegraphics[width=0.5\textwidth]{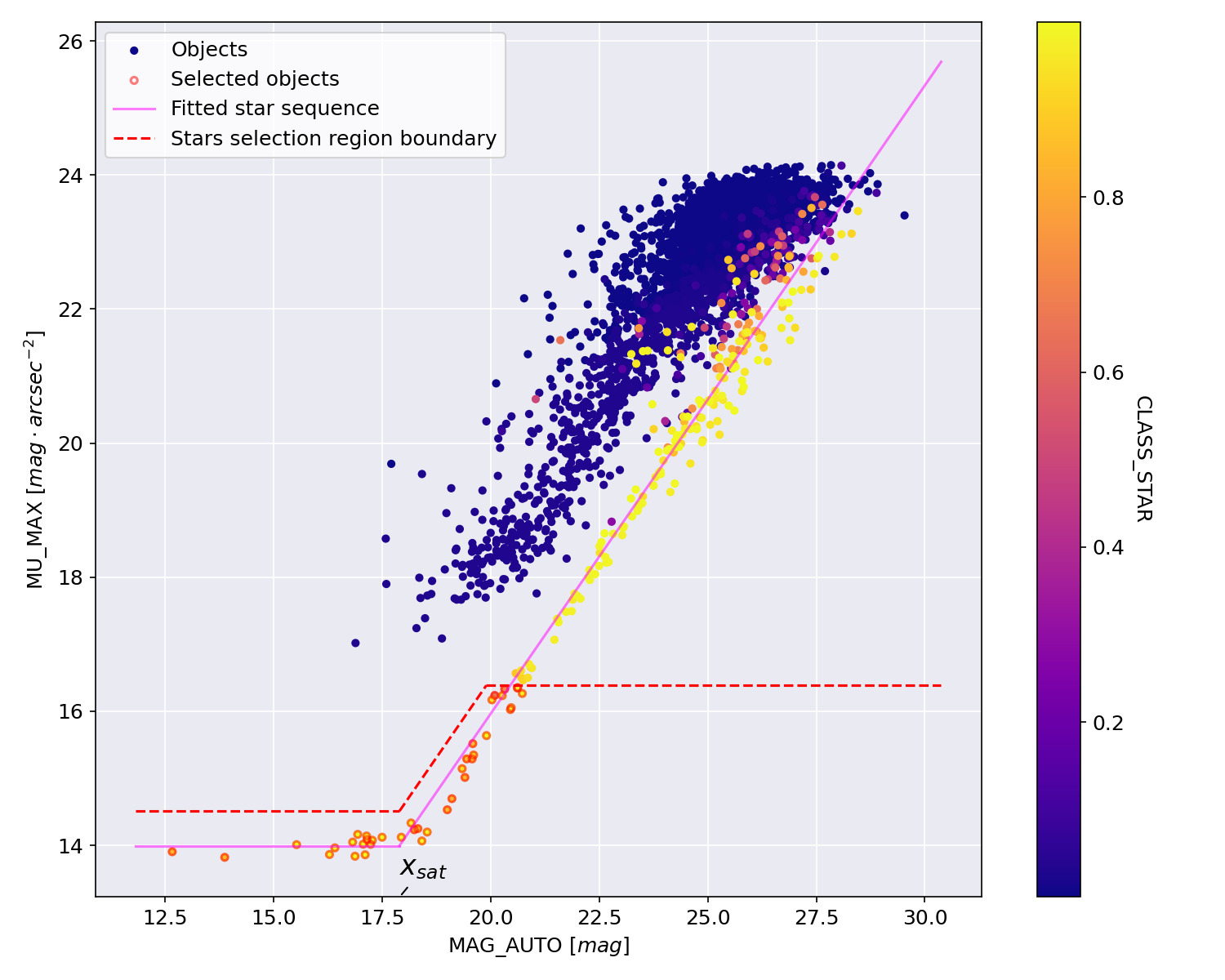}
    \caption{Objects distribution in the plane  MU\_MAX vs. MAG\_AUTO for the F814W band. Color maps to Sextractor stellarity index CLASS\_STAR, where yellow indicates a star and blue indicates not a star. The magenta solid line is the fitted model of the star sequence, while the red dashed line indicates the thresholds used to select the brightest stars that may contain diffraction spikes. Objects circled in red are the actually selected stars.}
    \label{fig:bright_stars_selection}
\end{figure}

Even if we have a single detection image, we test a semi-automatic procedure that requires only a minimum of human intervention and that can be easily reused in other cases. The procedure starts by identifying the brightest stars in the plane MU\_MAX vs. MAG\_AUTO for the detection image. In this plane, bright stars form a sequence made by two intersecting lines (Fig. \ref{fig:bright_stars_selection}): saturated stars form a horizontal line at low values of MU\_MAX, this indicates that their peak surface brightness does not change even if the Kron magnitude does and this is, in fact, due to saturated pixels. At some point, when the brightness is low enough so that there are no more saturated pixels, MU\_MAX increases almost linearly with MAG\_AUTO. Using the subsample of objects with the stellarity index $\rm{CLASS\_STAR} \geq 0.9$, we fit a simple two-segment linear model defined by (\ref{eq:saturated_linear_model}) with a Levenberg-Marquardt minimizer \citep{Astropy2022}. Because the quality of the fit is sensitive to the initial values, we used the Hough transform \citep{Hough1962} to get an approximate estimate of the slope of the linear part of the star sequence.

\begin{equation}
    f(x) = 
    \begin{cases}
    A + B \cdot x &\text{if{}}\, x > x_{sat} \\
    A + B \cdot x_{sat} &\text{if{}}\, x \leq x_{sat}
    \end{cases}.\\
    \label{eq:saturated_linear_model}
\end{equation}

We then fixed an upper threshold for Kron magnitude equal to $x_{sat} + 2$. This value has been chosen to include most of the brightest stars that may show spikes, while minimizing the total number of stars that would require further checking. For all objects in the star subsample with $x_{sat} < \rm{MAG\_AUTO} < x_{sat} + 2$, we computed the distance from the fitted model and the standard deviation, $\sigma_d$, of these distances is used to compute the offset $\Delta_{mu} = 5\sigma_d$. Finally, we identified a selection region in the MU\_MAX - MAG\_AUTO plane defined by Eq. (\ref{eq:selection_region}):

\begin{equation}
    y \leq
    \begin{cases}
    f(x) + \Delta_{mu} &\text{if{}} x < x_{sat} + 2 \\
    f(x_{sat} + 2) + \Delta_{mu} & \text{if{}} x >= x_{sat} + 2 \\
    \end{cases}.\\
    \label{eq:selection_region}
\end{equation}

\begin{figure*}
    \centering
    \includegraphics[width=\textwidth]{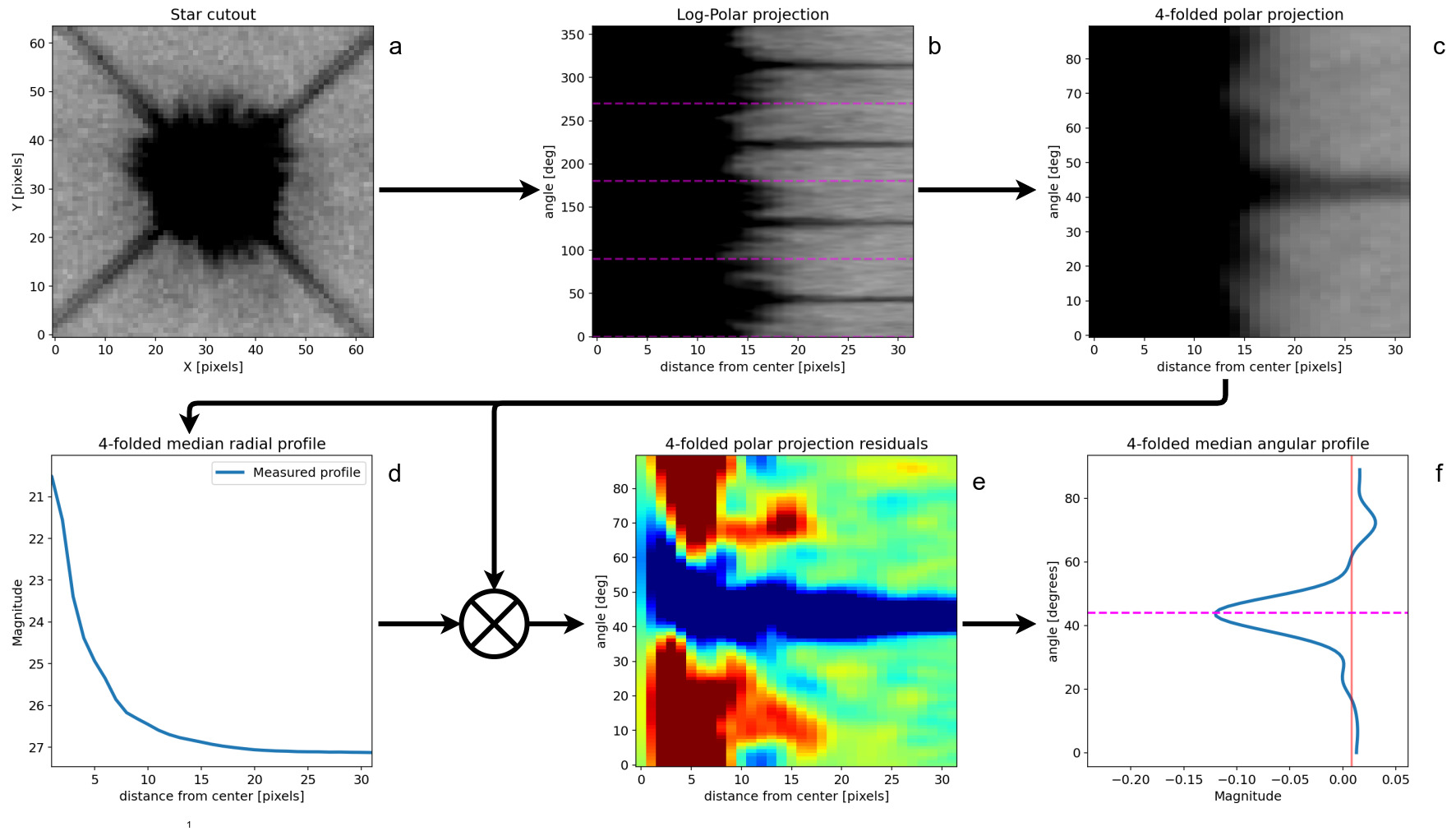}
    \caption{Flowchart \textbf{of the} spike detection algorithm: a) original cutout of the star; b) log-Polar transformed cutout. The dashed magenta lines indicate the division into four stripes of same size; c) median stacking of the stripes; d)  radial profile of the median stripe; e) residual image obtained by first subtracting the radial profile from the median stripe and then dividing the result by the same radial profile;  f)  angular profile of the residual image, obtained by computing the median value along the radial axis. The magenta dashed line indicates the angular position of the spike.}
    \label{fig:spike_detection_flowchart}
\end{figure*}

All objects within this region, independently of their value of the stellarity index, were checked for the presence of diffraction spikes as follows. For each object, a cutout $C(x, y)$ of size $W=60$ pixels is extracted from the detection image, converted from fluxes to magnitudes, so that the object is in the center (Fig. \ref{fig:spike_detection_flowchart}a). The cutout is transformed to a log-polar representation $C(r, \theta),$ where $\theta \in [0, 2\pi]$ with an angular resolution of $\ang{1;;}$ and $r \in [0, W/2]$. It is then divided into $N$ stripes of equal height $C_k(r, \theta')$ with $\theta' \in \left[\frac{2\pi}{N}k, \frac{2\pi}{N}(k + 1)\right]$ and $k \in \{0, ..., N-1\}$  (Fig. \ref{fig:spike_detection_flowchart}b). The number of stripes $N$ is equal to the number of spikes that are expected to be produced (four in this case) and it is known beforehand since it depends only on the optics of the telescope. The stripes are then median-stacked to reduce the noise and the influence of any nearby object, obtaining a median stripe $\overline{C}(r, \theta'')$ with $\theta'' \in \left[0, \frac{2\pi}{N}\right]$ (Fig. \ref{fig:spike_detection_flowchart}c). This median stripe is further stacked along the polar axis to get a median radial profile of the star $\overline{I}(r) = {med} _\theta(\overline{C}(r, \theta''))$ (Fig. \ref{fig:spike_detection_flowchart}d). The radial profile is used to make a 2D model of the star $S(r, \theta'') = \overline{I}(r)\forall\theta''$ that is then subtracted from the median stripe and the result is weighted by the inverse of the radial profile, leaving $R(r, \theta'') = [\overline{C}(r, \theta'') - S(r, \theta'')]/S(r, \theta'')$ only the noise and the diffraction spike in the residual image, if it is bright enough to be detected (Fig. \ref{fig:spike_detection_flowchart}e). The angular profile $I(\theta'') = med_r(\overline{C}(r, \theta''))$, obtained by computing median along the radial axis of this residual image (Fig. \ref{fig:spike_detection_flowchart}f), can be used to detect the presence of a diffraction spike, as well as its angular position, by applying a simple peak finding algorithm like find\_peaks from the scipy python library \citep{Scipy2001}. If other than one prominent peak is detected, it means that there are no bright diffraction spikes; otherwise, the position of the peak indicates the angular position of the diffraction spikes. It is worth pointing out that since a spike has usually a small angular extension and therefore should have a low, but still non-vanishing influence on the radial profile, making it slightly biased. However, this bias will: a) make the generated circular masks slightly bigger with respect to the ones computed with unbiased radial profiles; b) affect the spike detection only for faint stars. Both these issues should not negatively affect the masking of bright stars, which are the ones that generate most of the spurious detections.\\

\begin{figure}
    \centering
    \includegraphics[width=0.5\textwidth]{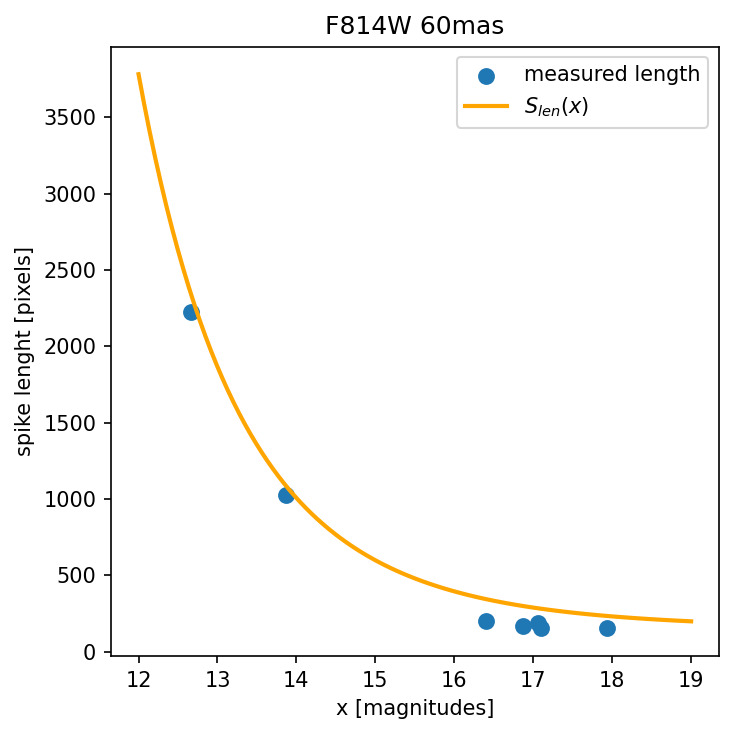}
    \caption{Comparison between the length of the diffraction spikes of a random sample of stars in the 60 mas detection image (band F814W) and the estimated length of the mask, $S_{len}$, as a function of the magnitude, MAG\_AUTO}
    \label{fig:spike_lenght_plot}
\end{figure}

For each star that presents a diffraction spike, a masking region is constructed by generating a series of N rectangular masks centered on the object itself and rotated using the angular position of the spike that has been found previously. The length of the rectangular regions is computed using the function in Eq. (\ref{eq:spike_lenght}):
\begin{equation}
    S_{len} (m) = a + (b \cdot m)^{\gamma} 
    \label{eq:spike_lenght}
,\end{equation}
where $S_{len}$ is the length in pixels, $m$ is the value of MAG\_AUTO and $a$, $b,$ and $\gamma$ are parameters that depend on the instrumentation and the photometric setup used. In our case, for the 60 mas F814W band, we used $a = 100$, $b=4.7 \cdot 10^{13}$, and $\gamma = -9.37$. Our goal is not to have a precise estimation of the length of the spikes but to make a masking region long enough to contain at least the bright visible part of the spikes. To check that the values we have chosen satisfy this requirement, from our detection images, we selected  a sample of seven stars of different magnitudes and manually measured the length of the diffraction spikes. From Fig. \ref{fig:spike_lenght_plot}, we can see that in every case, the length $S_{len}$ is always slightly bigger than the true length. We also construct a circular mask for the central bright part of the star and its radius, $r_c$, is derived from the radial profile, $\overline{I}(r),$ so that the difference between the maximum of $\overline{I}$ and $\overline{I}(r_c)$ is the 97\% of the range of $\overline{I}$  (\ref{eq:core_size}):\\

\begin{equation}
    r_c {} : {} \frac{max(\overline{I}(r)) - \overline{I}(r_c)}{max(\overline{I}(r)) - min(\overline{I(r)}} = 0.97
    \label{eq:core_size}
.\end{equation}

\begin{figure*}
    \begin{subfigure}[b]{0.5\textwidth}
        \centering
        \includegraphics[width=1.0\textwidth]{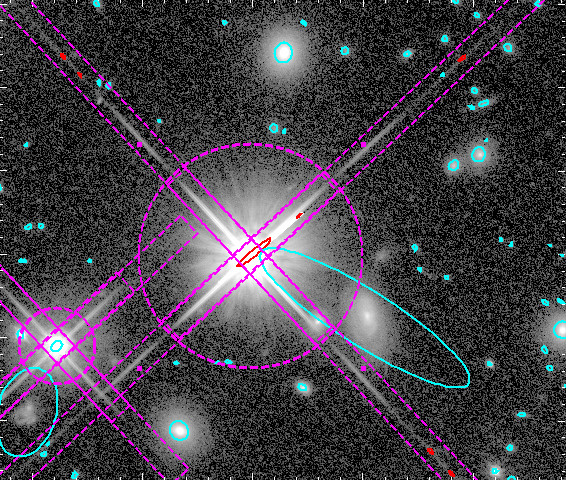}
        \caption{\label{fig:spike_maks_examples_1}}
    \end{subfigure}
    \begin{subfigure}[b]{0.5\textwidth}
        \centering
        \includegraphics[width=1.0\textwidth]{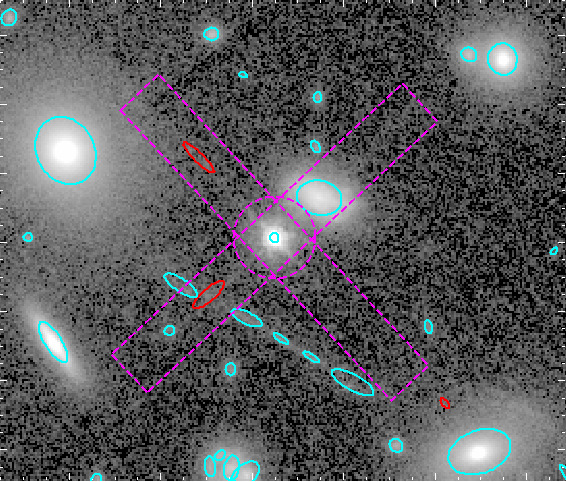}
        \caption{\label{fig:spike_maks_examples_2}}
    \end{subfigure}
    \caption{Two examples of masks used to remove spurious detections near bright stars in the band F814W: removed objects are highlighted in red. Note: in (b) a bright galaxy is not marked as a spurious detection, despite being in the masked region, because its ellipticity is lower than the threshold value of $0.5$.}
    \label{fig:spike_maks_examples}
\end{figure*}

In each masked region, we mark any object with an ellipticity of $e = 1 - \rm{A\_IMAGE} / \rm{B\_IMAGE} \geq 0.5$ as a spurious detection. We also marked as spurious detection objects with an unrealistic value of $\rm{MAG\_AUTO} > 40$  (Fig. \ref{fig:spike_maks_examples}), where this threshold is chosen to be as conservative as possible.\\

\clearpage

\section{Extrapolation of the missing magnitudes in band F160W}
\label{appendix:imputation}
Cluster members that have missing data for the magnitude in band F160W are inferred using a set of linear regressors: using all objects in the photometric catalog, we first compute the Pearson correlation coefficient (PCC) between the Kron magnitudes, $m_{\rm{F160W}}$, in band F160W and the ones in any other photometric band, $m_{x}$. We then used the photometric band with the greatest value of the PCC to train a linear regressor that is used to impute the missing data in the cluster member sample. We repeat the process using the remaining photometric bands sorted according to their value of the PCC with the condition that $PCC \geq 0.5$. With this procedure, 55 objects were imputed using the band F110W and 14 using the band F814W, for a total of 69 objects\\

\begin{figure}[th]
    \centering
    \includegraphics[width=0.5\textwidth]{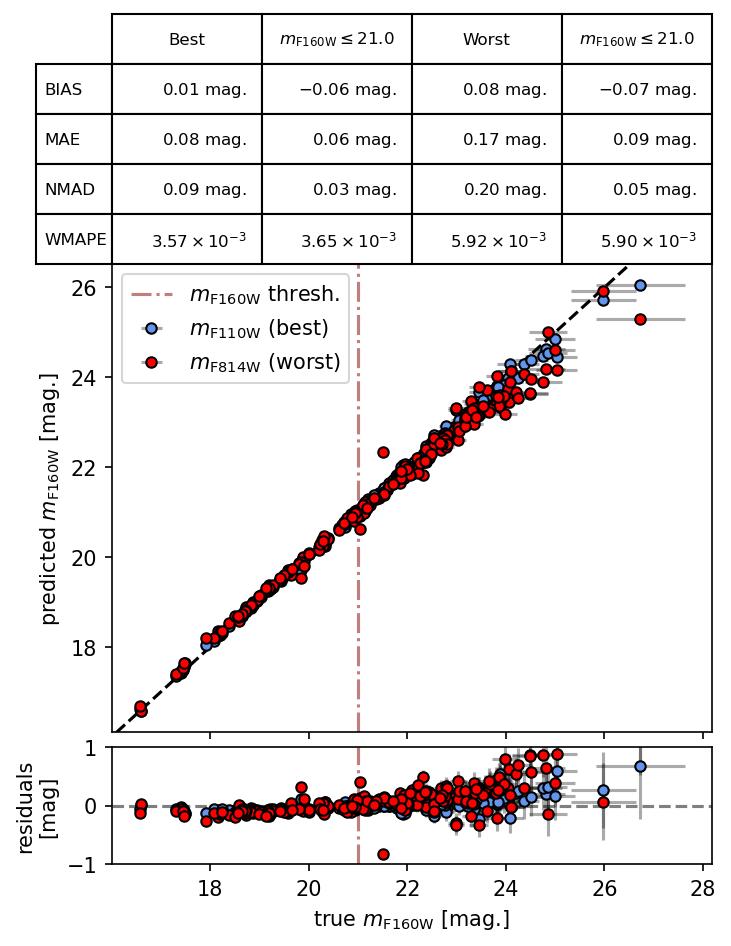}
    \caption{True versus predicted Kron magnitudes $m_{\rm{F160W}}$ plots (upper panel) and residuals (lower panel) for 222 spectroscopic cluster members. The predicted values were obtained using two linear regressors trained using Kron magnitudes, $m_{\rm{F110W}}$ and $m_{\rm{F814W}}$, (red and blue points, respectively). The brown dash-dotted line indicates the magnitude threshold used to select the cluster members sample used for the strong-lensing model. The table shows various statistical indicators for the best and worst regressor, considering all the values (first and third column) or only the objects for which $m_{\rm{F160W}} \leq 21$ (second and last column).}
    \label{fig:members_f160w_imputation}
\end{figure}

To assess the goodness of the imputation process, we use the two regressors that are trained on $m_{\rm{F110W}}$ and $m_{\rm{F814W}}$ to predict the $m_{\rm{F160W}}$ of 222 spectroscopic members that have no missing value in any of the previous three features and the statistical indicators computed on the residuals. Fig. \ref{fig:members_f160w_imputation} seems to confirm the reliability of the regressors, especially for magnitudes lower than the threshold used to select the cluster members sample.\\

\clearpage

{%
    \section{Multiple Images}
    \label{appendix:multiple_images}
    In Table C.1, only available in electronic form at the CDS, we summarize the properties of the identified multiple images, reporting also the eventual photometric redshift estimations from \cite{Zitrin2017}. The first column (name) reports the name of the images in the format X.Yk, where X indicates the background source, Y indicates the clump and substructure, and k differentiates among the various counter-images. A leading \textit{c} in the name indicates a candidate counter-image that is difficult to confirm owing to the low quality of the redshift estimation and/or the complex geometry of the system; the second column (ID) reports the corresponding ID in the spectroscopic and photometric catalogs; the fourth and fifth columns show the J2000 ICRS right ascension and declination  (R.A and Dec.); the sixth and seventh columns ($z_{\rm{spec}}$ and QF) report the measured spectroscopic redshift and its estimation quality flag (1=insecure, 2=likely, 3=secure, and 9=based on one emission line); the eighth column (Gold) indicates whether an image belongs to the golden sample used to optimize the lens model; The ninth column ($z_{\rm{model}}$) reports the median redshift estimated using the best fitting lens model along with the 16th and 84th percentiles; the last two columns (Arc ID and $z_{\rm{phot}}$) are the original ID from \cite{Zitrin2017} and the photometric redshift estimation from \cite{Zitrin2017}, along with the associated $95\%$ confidence interval.\\
    
    The figures from Fig. \ref{fig:cimg_sys_1} to Fig. \ref{fig:cimg_sys_49} show, for each system of multiple images, a portion of the spectrum centered on the most prominent emission line (if any) used to determine the redshift, along with two $\qty{5}{\arcsec}$ cut-outs, centered on the position of each multiple source. These cut-outs are taken from the HST RGB image (in the upper-left panel), and from the white-light image obtained by stacking a slice of the spectral data cube corresponding to the wavelengths range shown in the plot of the spectrum (in the upper right corner) \\

    \onecolumn
    \begin{figure}[H]
        \raggedright{}
        \begin{subfigure}{.25\textwidth}
            \centering
            \includegraphics[width=1.0\textwidth]{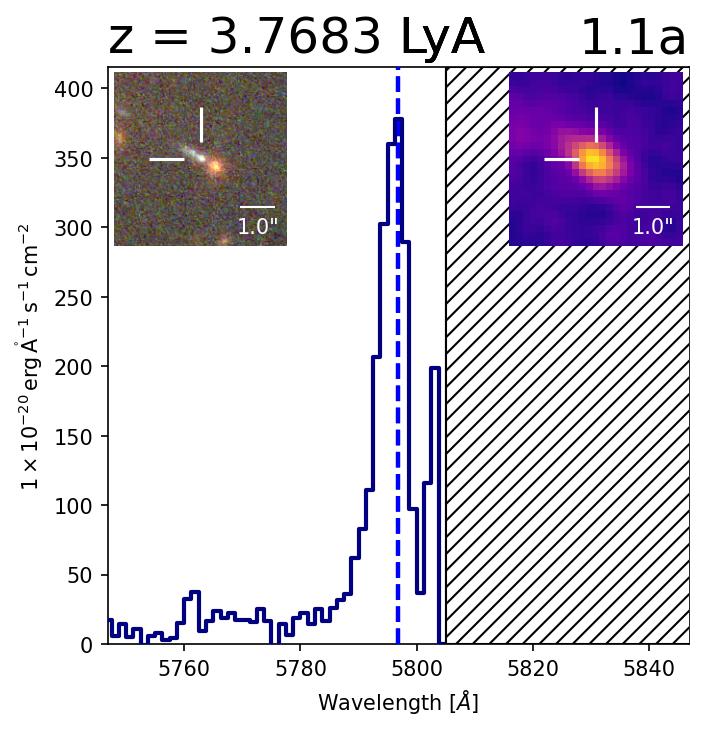}
        \end{subfigure}
        \begin{subfigure}{.25\textwidth}
            \centering
            \includegraphics[width=1.0\textwidth]{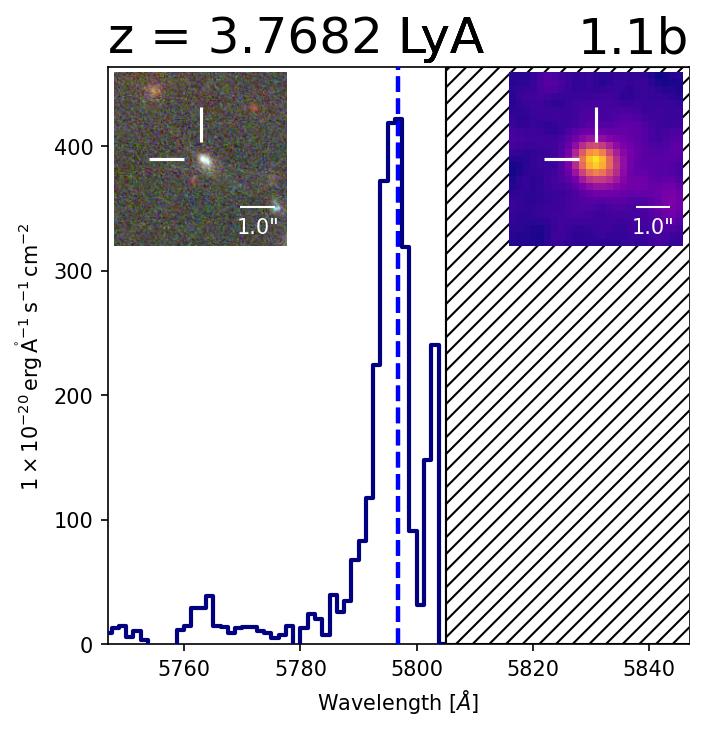}
        \end{subfigure}%
        \begin{subfigure}{.25\textwidth}
            \centering
            \includegraphics[width=1.0\textwidth]{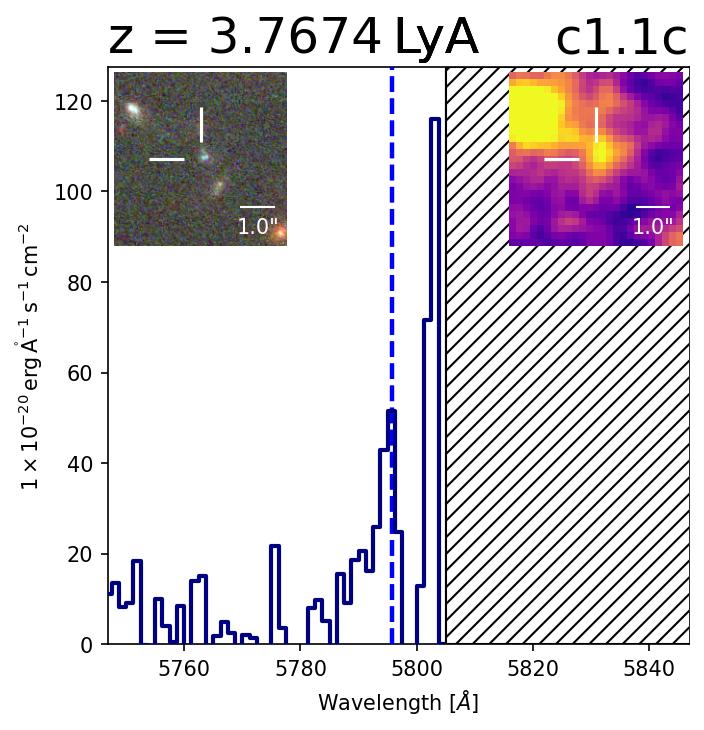}
        \end{subfigure}%
        \caption{\label{fig:cimg_sys_1}System 1}
    \end{figure}
    
    \begin{figure}[H]
        \raggedright{}
        \begin{subfigure}{.25\textwidth}
            \centering
            \includegraphics[width=1.0\textwidth]{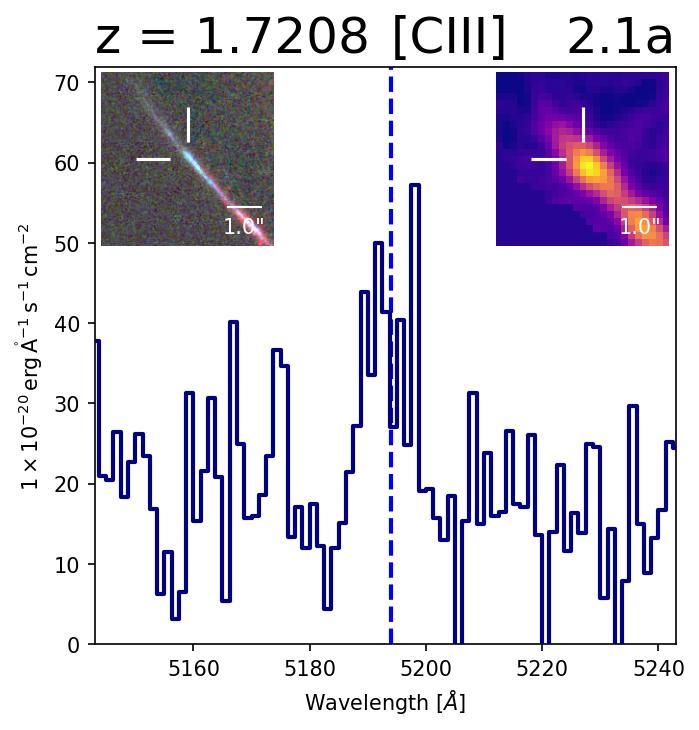}
        \end{subfigure}
        \begin{subfigure}{.25\textwidth}
            \centering
            \includegraphics[width=1.0\textwidth]{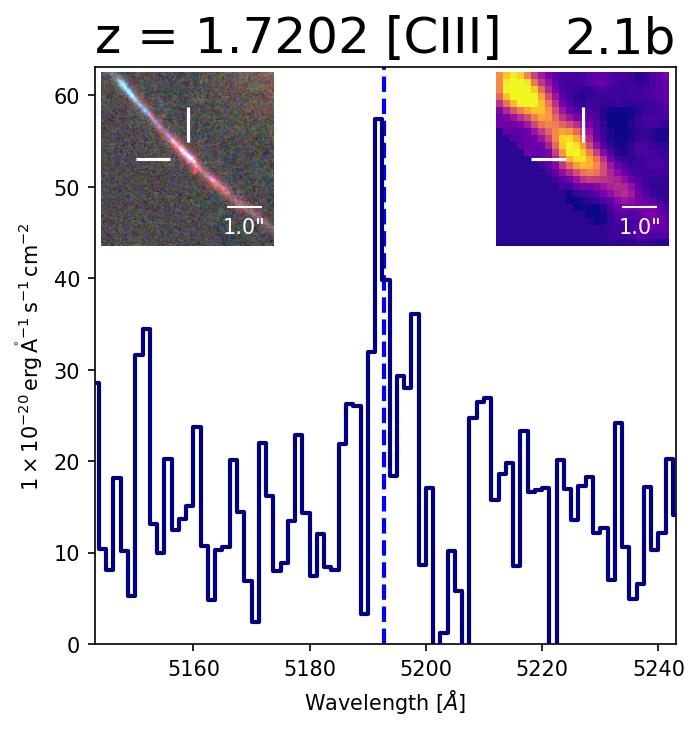}
        \end{subfigure}%
        \begin{subfigure}{.25\textwidth}
            \centering
            \includegraphics[width=1.0\textwidth]{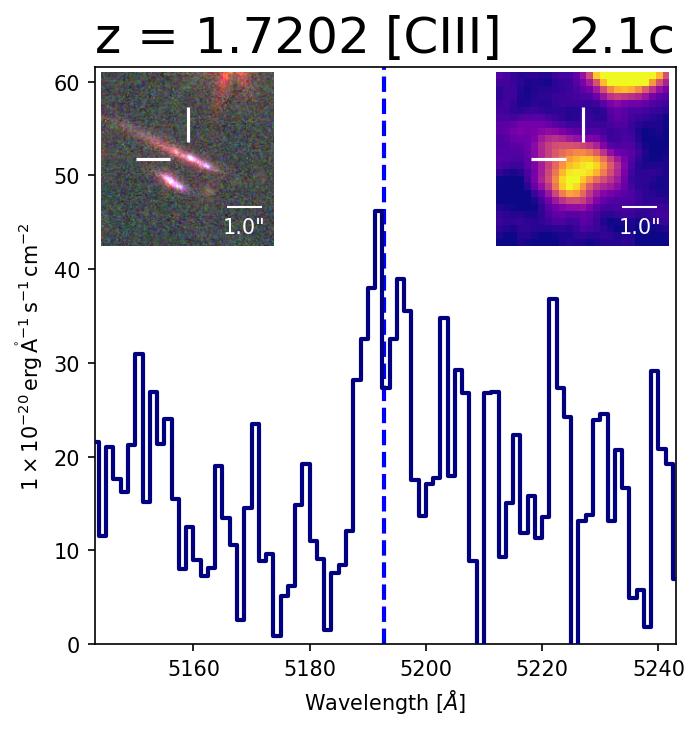}
        \end{subfigure}%
        \begin{subfigure}{.25\textwidth}
            \centering
            \includegraphics[width=1.0\textwidth]{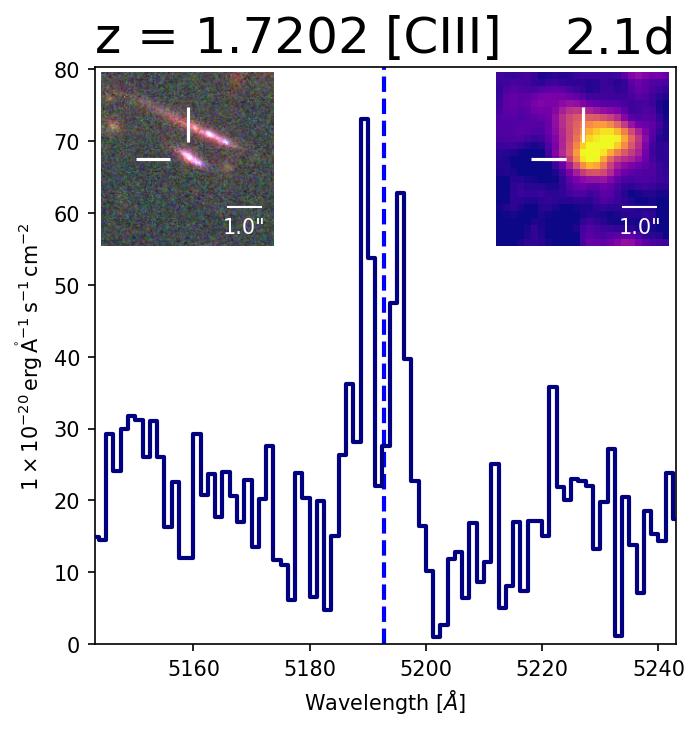}
        \end{subfigure}
    
        \begin{subfigure}{.25\textwidth}
            \centering
            \includegraphics[width=1.0\textwidth]{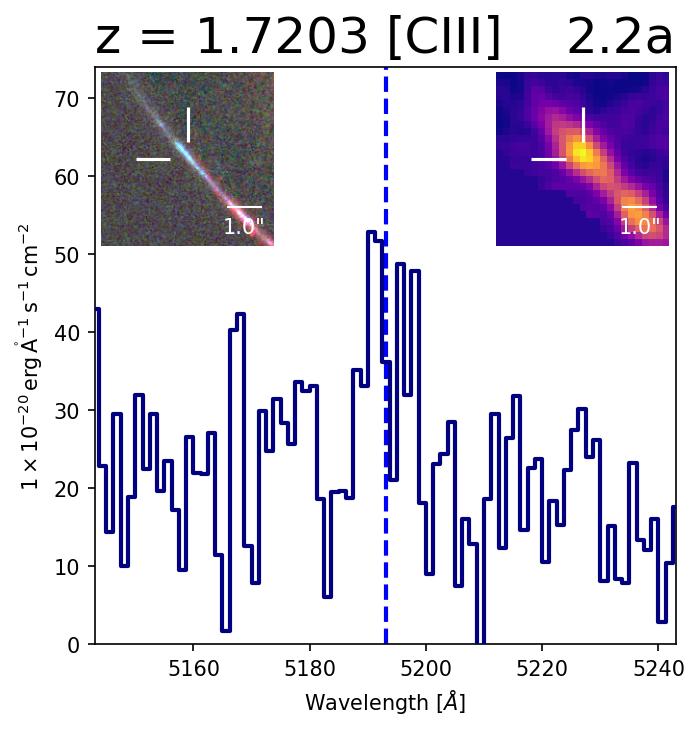}
        \end{subfigure}
        \begin{subfigure}{.25\textwidth}
            \centering
            \includegraphics[width=1.0\textwidth]{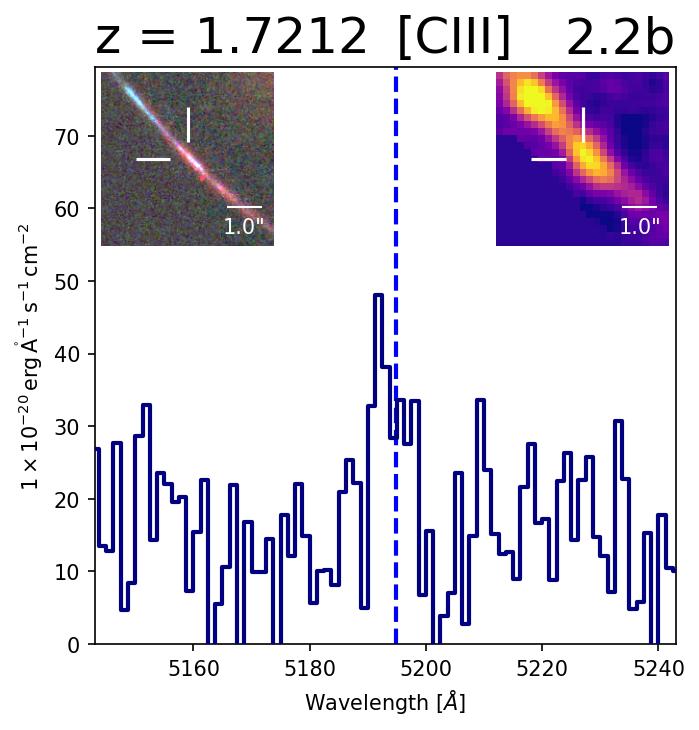}
        \end{subfigure}%
        \begin{subfigure}{.25\textwidth}
            \centering
            \includegraphics[width=1.0\textwidth]{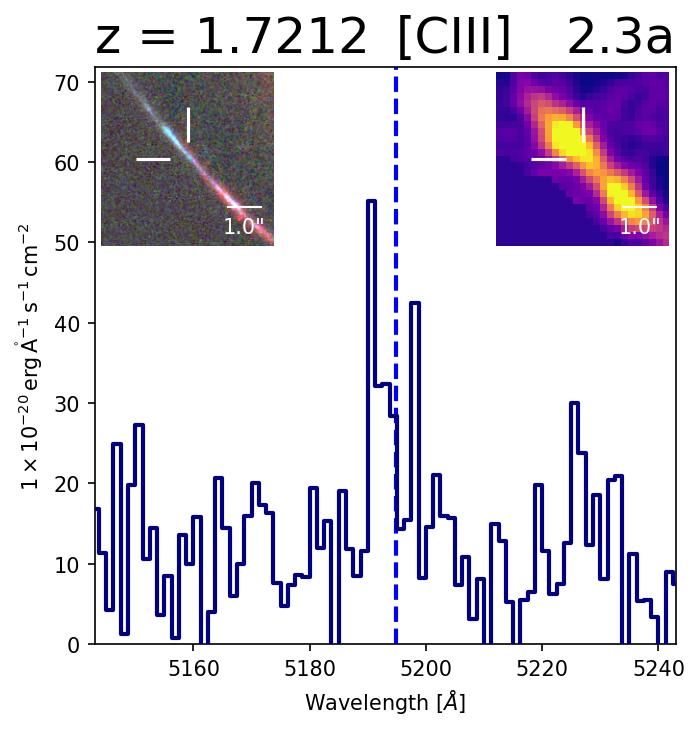}
        \end{subfigure}%
        \begin{subfigure}{.25\textwidth}
            \centering
            \includegraphics[width=1.0\textwidth]{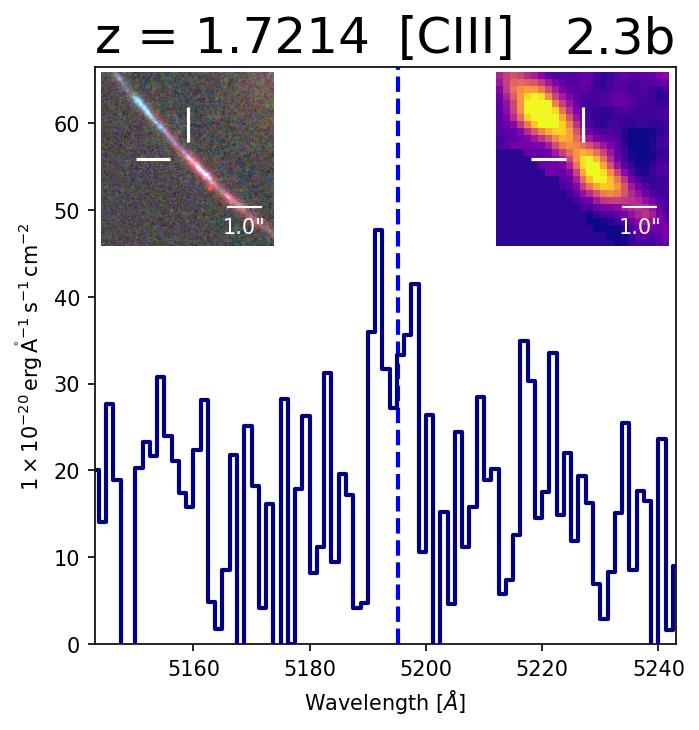}
        \end{subfigure}
    
        \begin{subfigure}{.25\textwidth}
            \centering
            \includegraphics[width=1.0\textwidth]{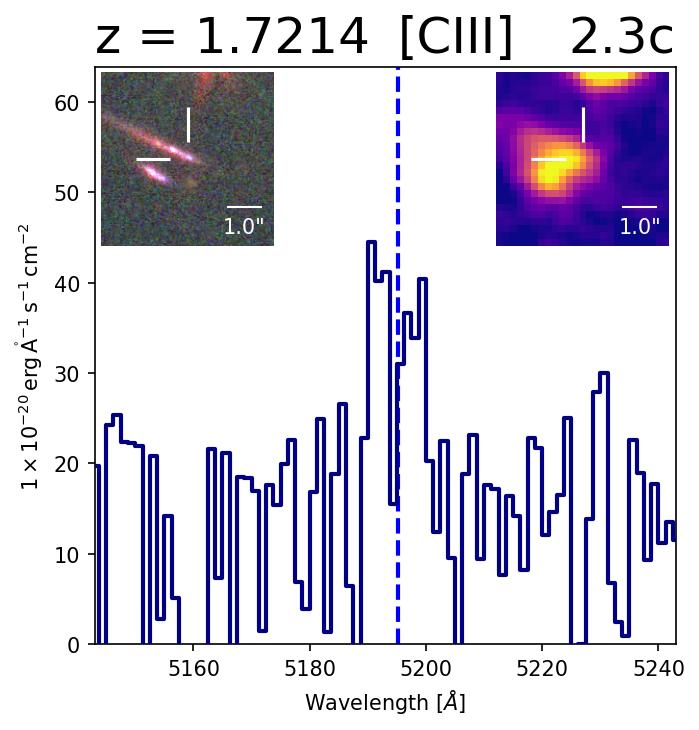}
        \end{subfigure}
        \begin{subfigure}{.25\textwidth}
            \centering
            \includegraphics[width=1.0\textwidth]{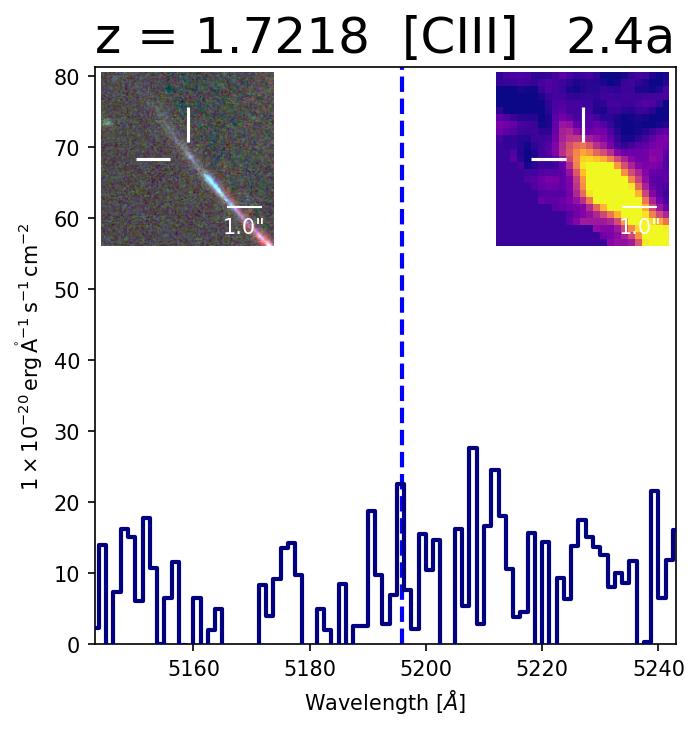}
        \end{subfigure}%
        \begin{subfigure}{.25\textwidth}
            \centering
            \includegraphics[width=1.0\textwidth]{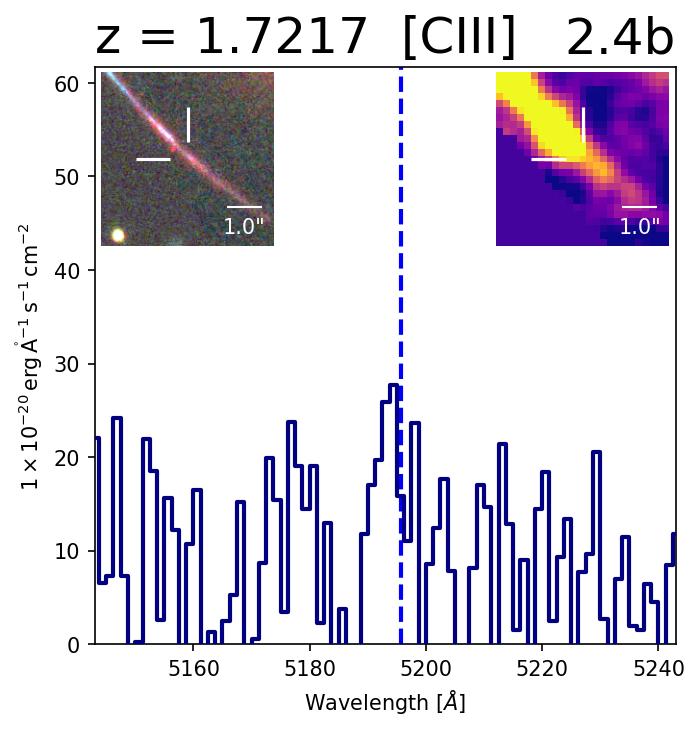}
        \end{subfigure}%
        \begin{subfigure}{.25\textwidth}
            \centering
            \includegraphics[width=1.0\textwidth]{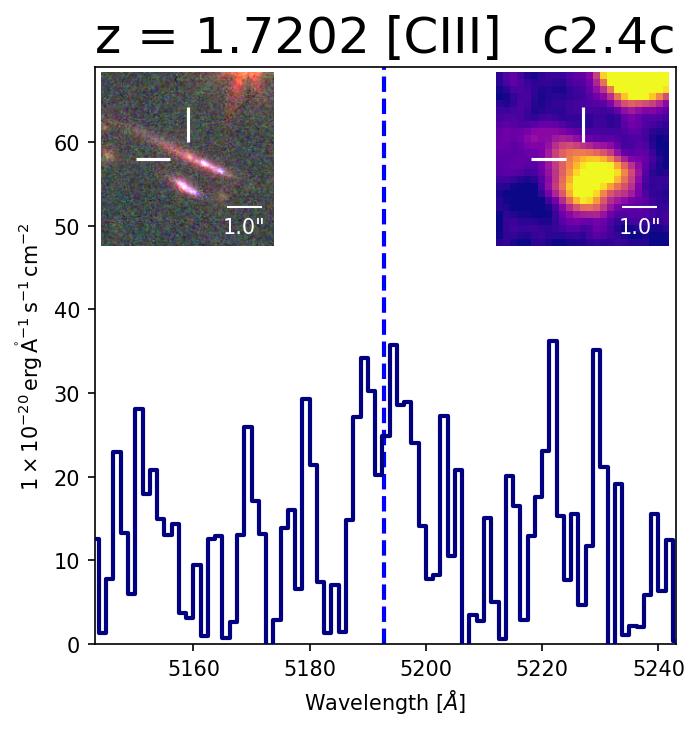}
        \end{subfigure}%
        \caption{\label{fig:cimg_sys_2} System 2}
    \end{figure}

    \begin{figure}[H]
        \raggedright{}
        \begin{subfigure}{.25\textwidth}
            \centering
            \includegraphics[width=1.0\textwidth]{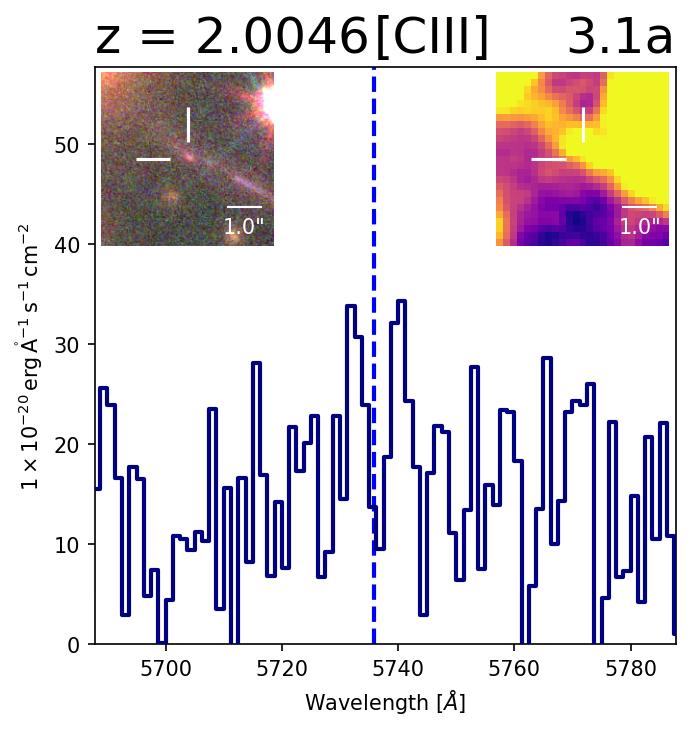}
        \end{subfigure}
        \begin{subfigure}{.25\textwidth}
            \centering
            \includegraphics[width=1.0\textwidth]{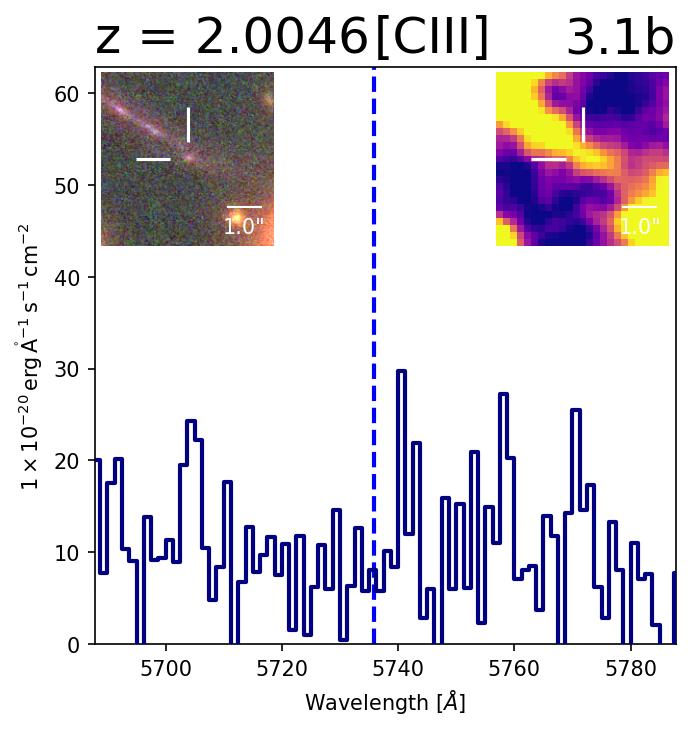}
        \end{subfigure}%
        \begin{subfigure}{.25\textwidth}
            \centering
            \includegraphics[width=1.0\textwidth]{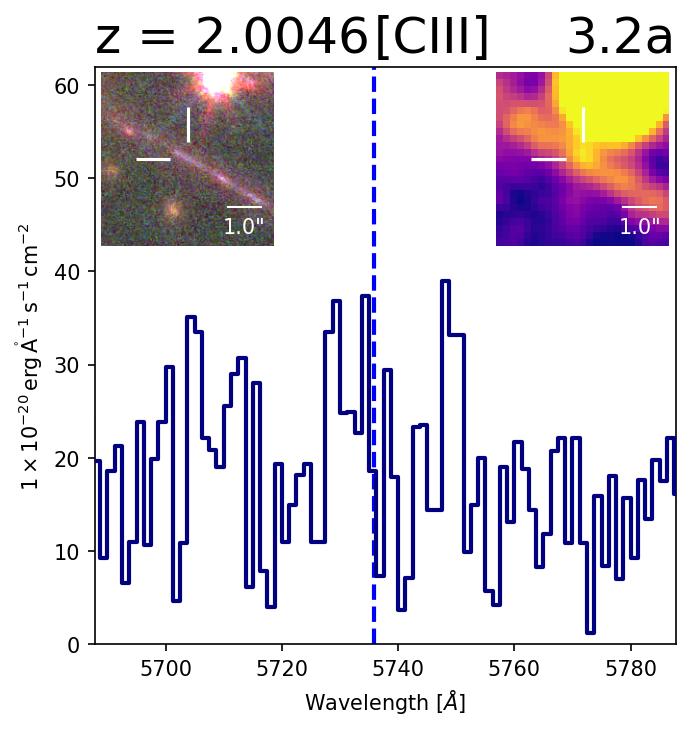}
        \end{subfigure}%
        \begin{subfigure}{.25\textwidth}
            \centering
            \includegraphics[width=1.0\textwidth]{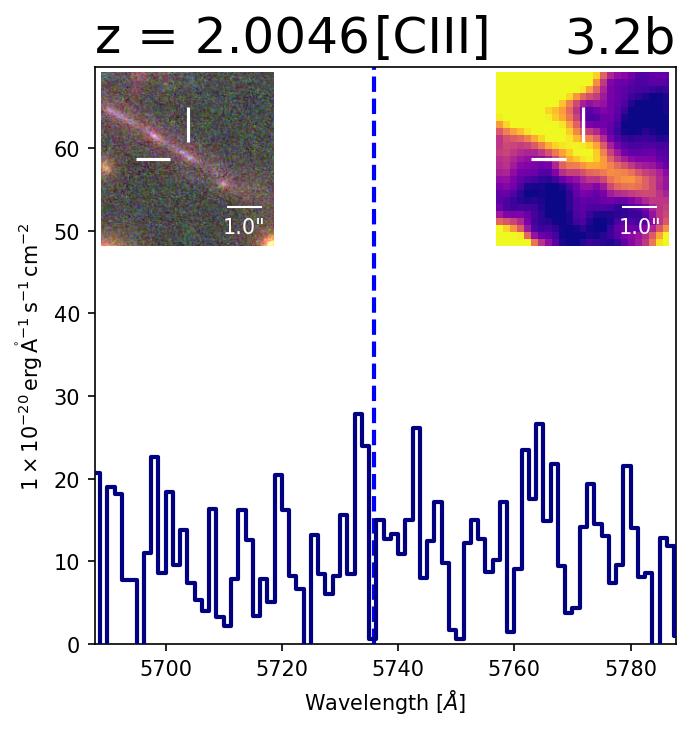}
        \end{subfigure}%
        \caption{\label{fig:cimg_sys_3}System 3}
    \end{figure}
    
    \begin{figure}[H]
        \raggedright{}
        \begin{subfigure}{.25\textwidth}
            \centering
            \includegraphics[width=1.0\textwidth]{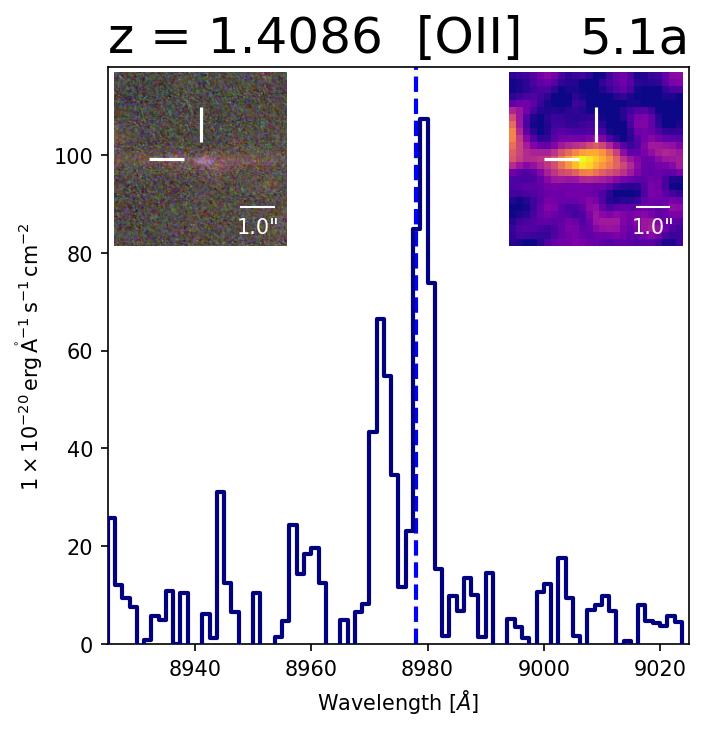}
        \end{subfigure}
        \begin{subfigure}{.25\textwidth}
            \centering
            \includegraphics[width=1.0\textwidth]{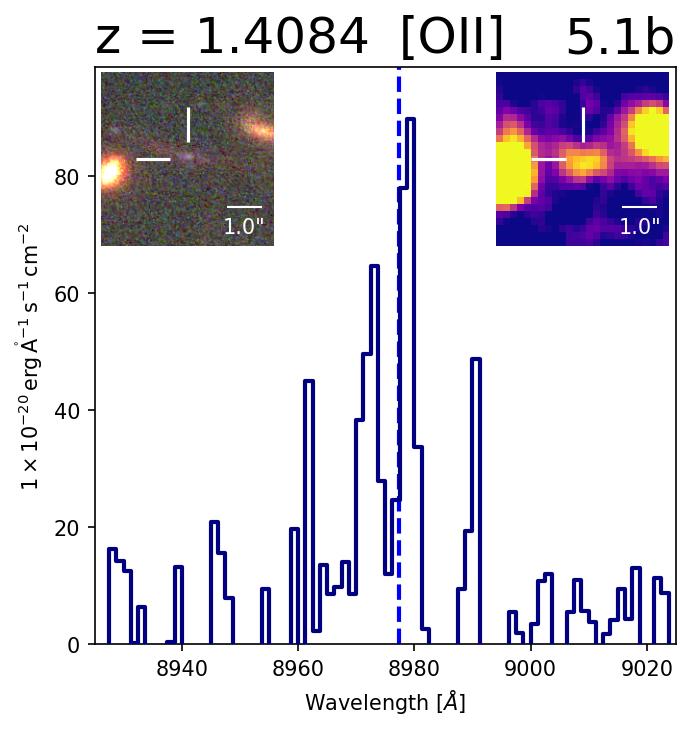}
        \end{subfigure}%
        \begin{subfigure}{.25\textwidth}
            \centering
            \includegraphics[width=1.0\textwidth]{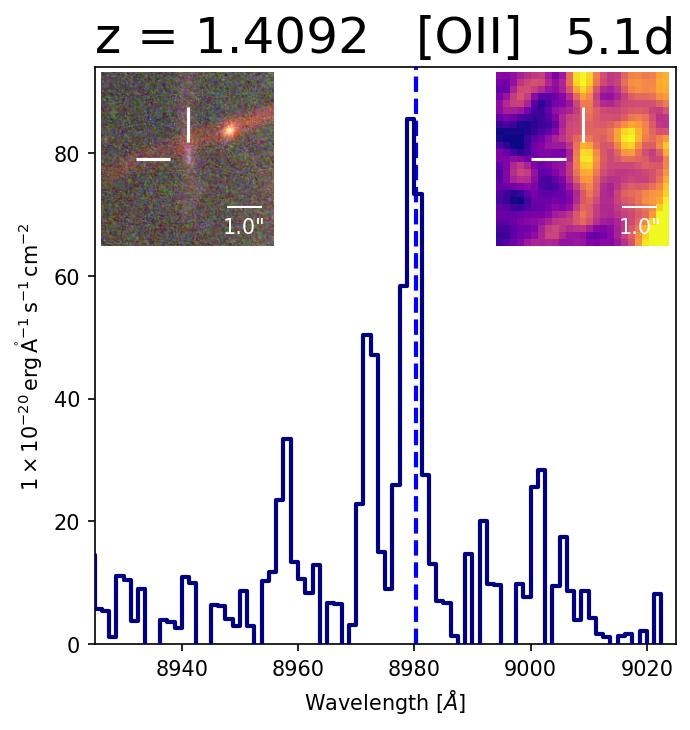}
        \end{subfigure}%
        \begin{subfigure}{.25\textwidth}
            \centering
            \includegraphics[width=1.0\textwidth]{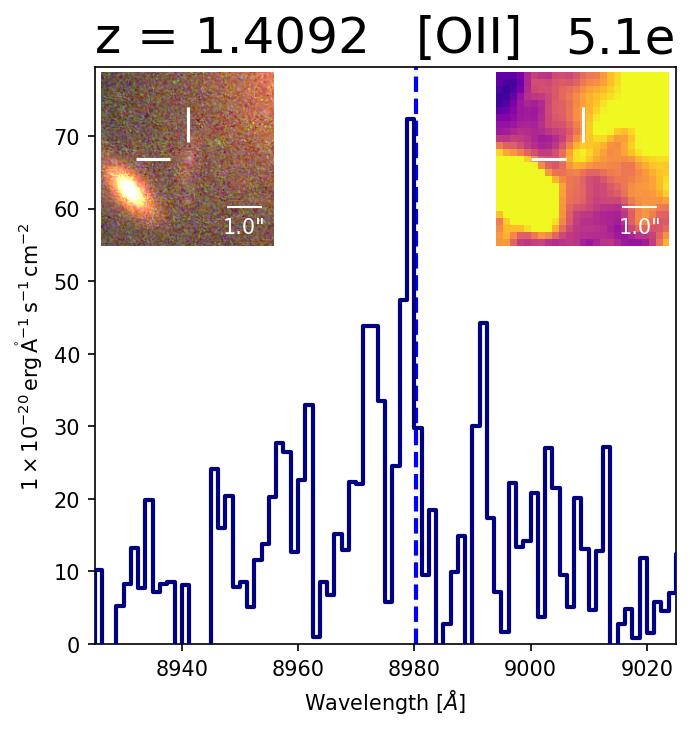}
        \end{subfigure}
    
        \begin{subfigure}{.25\textwidth}
            \centering
            \includegraphics[width=1.0\textwidth]{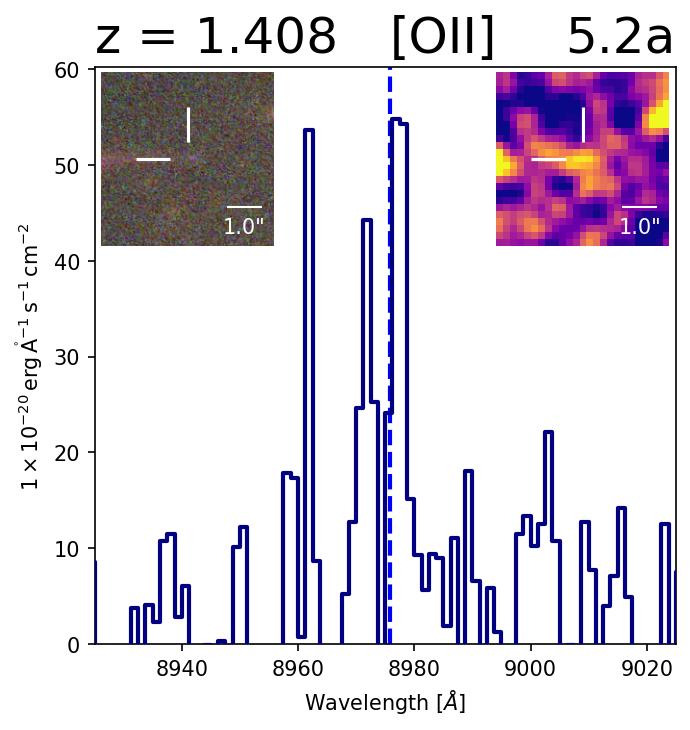}
        \end{subfigure}
        \begin{subfigure}{.25\textwidth}
            \centering
            \includegraphics[width=1.0\textwidth]{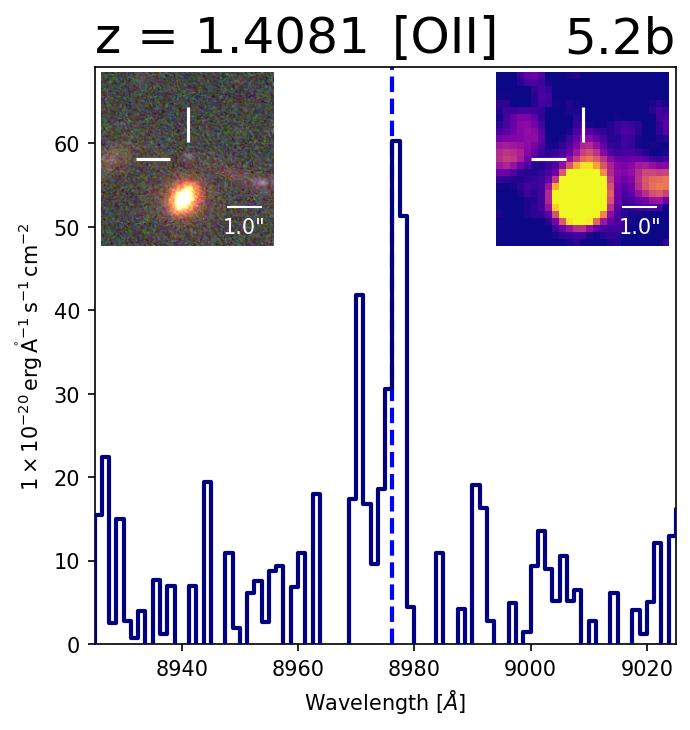}
        \end{subfigure}%
        \begin{subfigure}{.25\textwidth}
            \centering
            \includegraphics[width=1.0\textwidth]{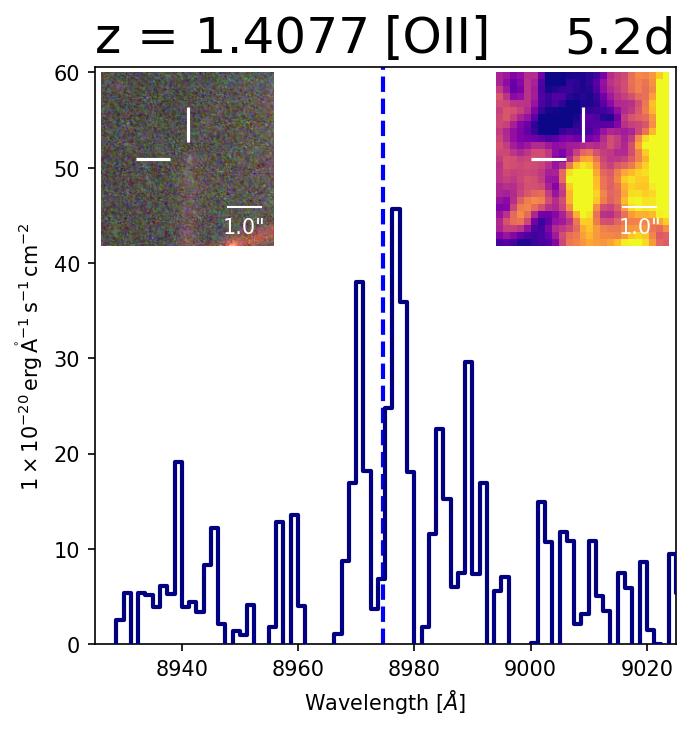}
        \end{subfigure}%
        \begin{subfigure}{.25\textwidth}
            \centering
            \includegraphics[width=1.0\textwidth]{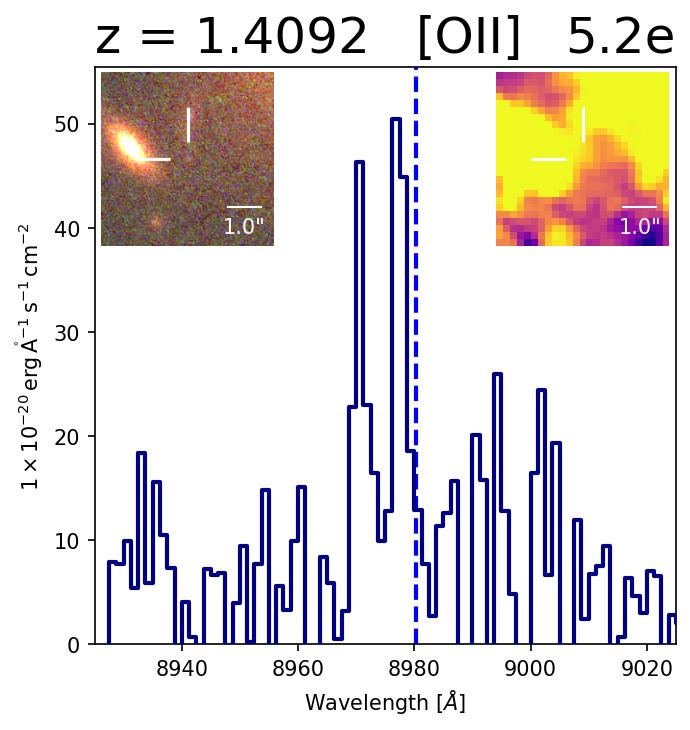}
        \end{subfigure}%
        \caption{\label{fig:cimg_sys_5}System 5}
    \end{figure}

    \begin{figure}[H]
        \raggedright{}
        \begin{subfigure}{.25\textwidth}
            \centering
            \includegraphics[width=1.0\textwidth]{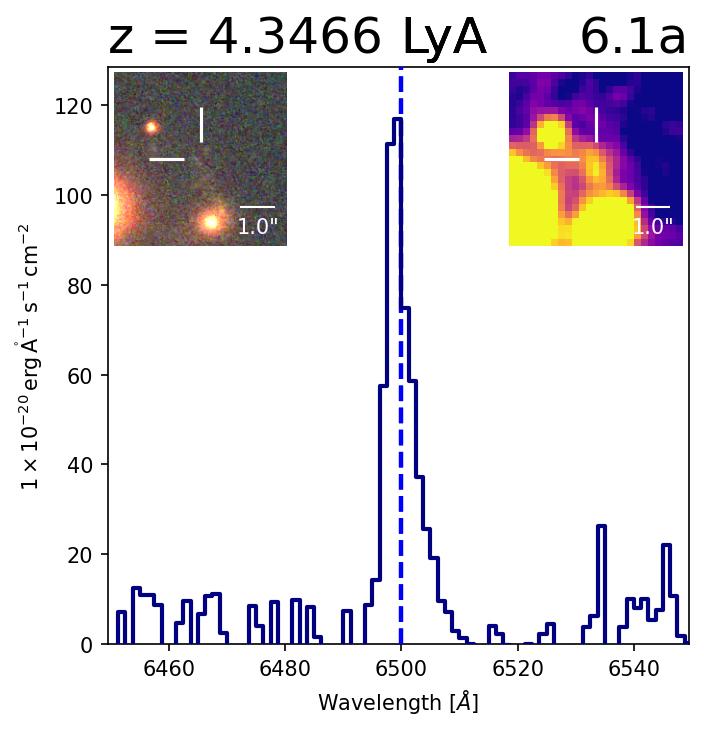}
        \end{subfigure}
        \begin{subfigure}{.25\textwidth}
            \centering
            \includegraphics[width=1.0\textwidth]{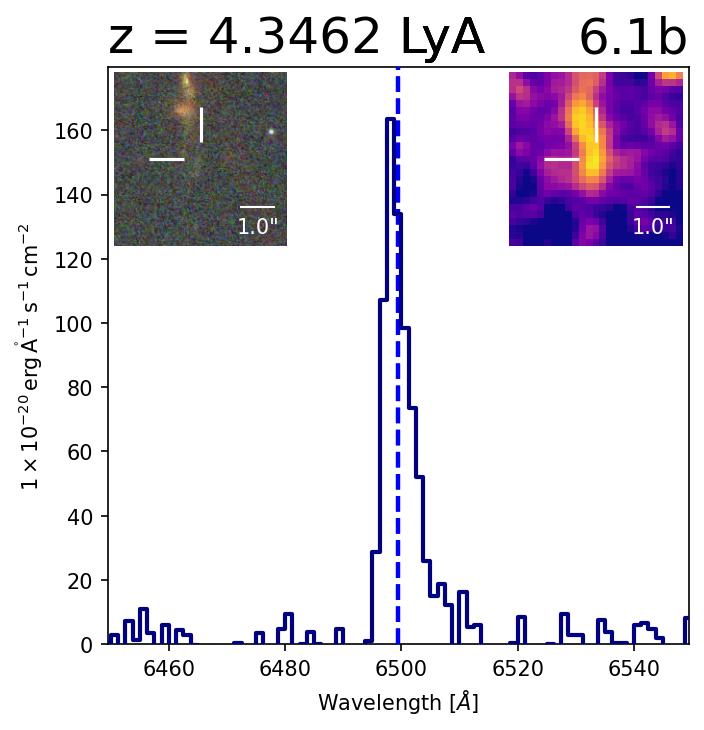}
        \end{subfigure}%
        \begin{subfigure}{.25\textwidth}
            \centering
            \includegraphics[width=1.0\textwidth]{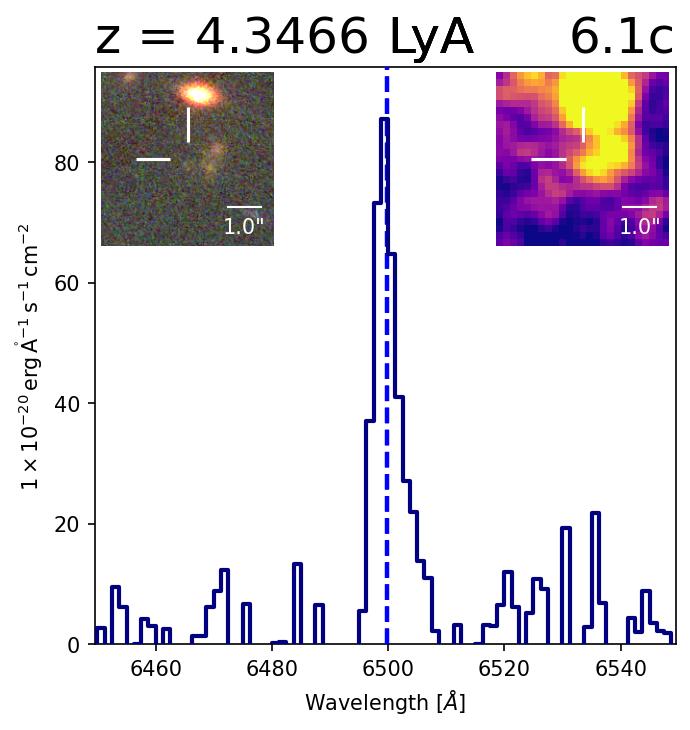}
        \end{subfigure}%
        \begin{subfigure}{.25\textwidth}
            \centering
            \includegraphics[width=1.0\textwidth]{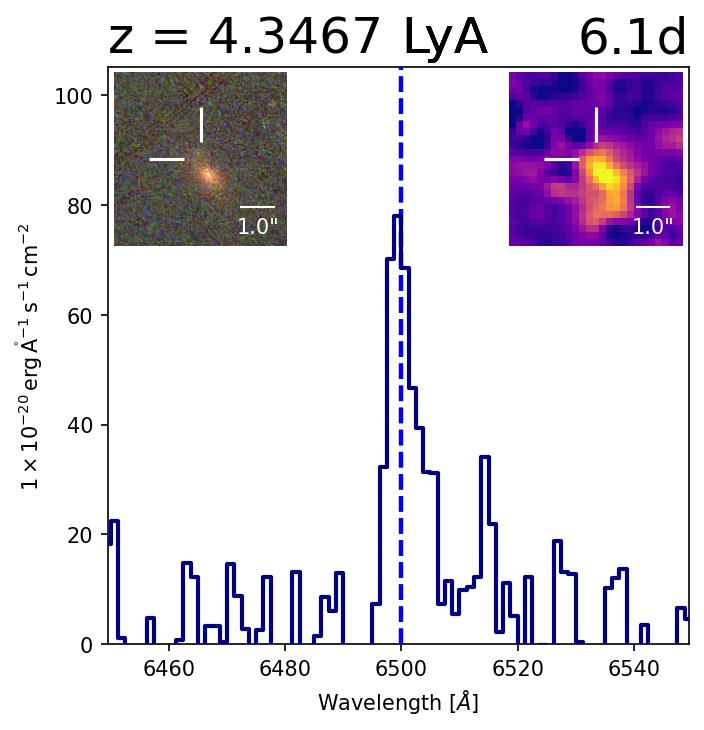}
        \end{subfigure}
    
        \begin{subfigure}{.25\textwidth}
            \centering
            \includegraphics[width=1.0\textwidth]{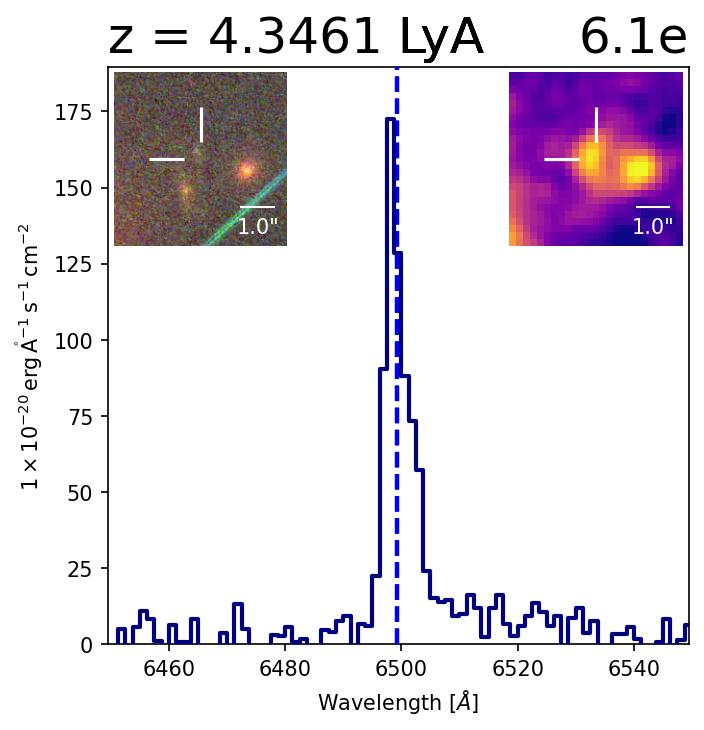}
        \end{subfigure}
        \begin{subfigure}{.25\textwidth}
            \centering
            \includegraphics[width=1.0\textwidth]{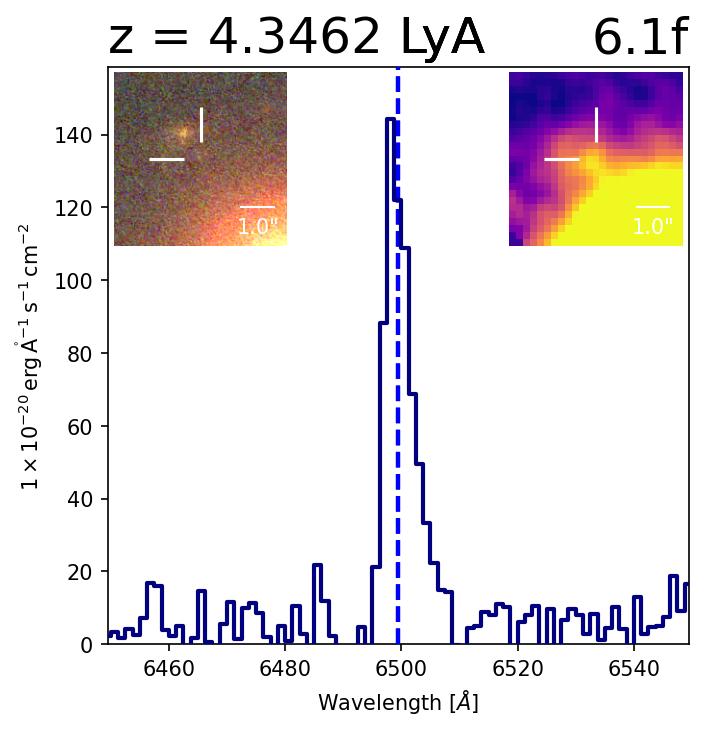}
        \end{subfigure}%
        \caption{\label{fig:cimg_sys_6}System 6}
    \end{figure}
    
    \begin{figure}[H]
        \raggedright{}
        \begin{subfigure}{.25\textwidth}
            \centering
            \includegraphics[width=1.0\textwidth]{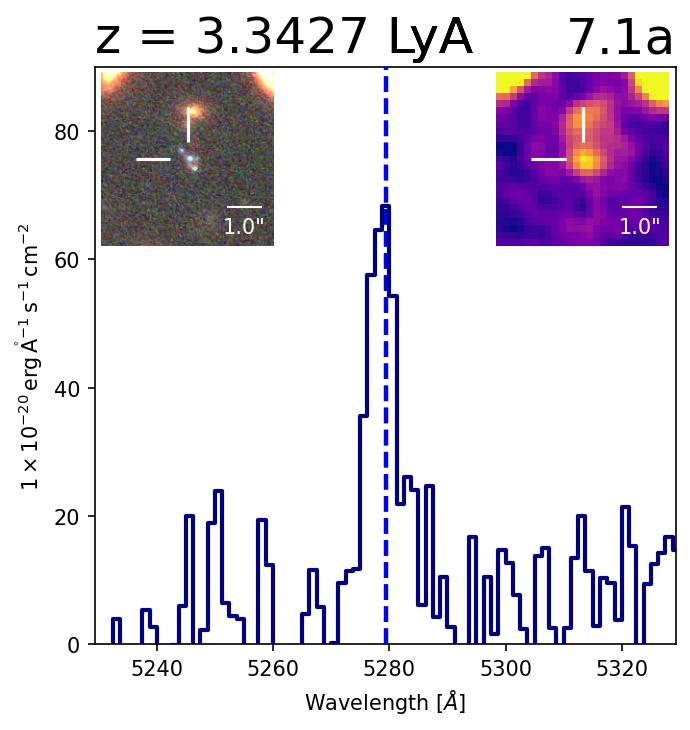}
        \end{subfigure}
        \begin{subfigure}{.25\textwidth}
            \centering
            \includegraphics[width=1.0\textwidth]{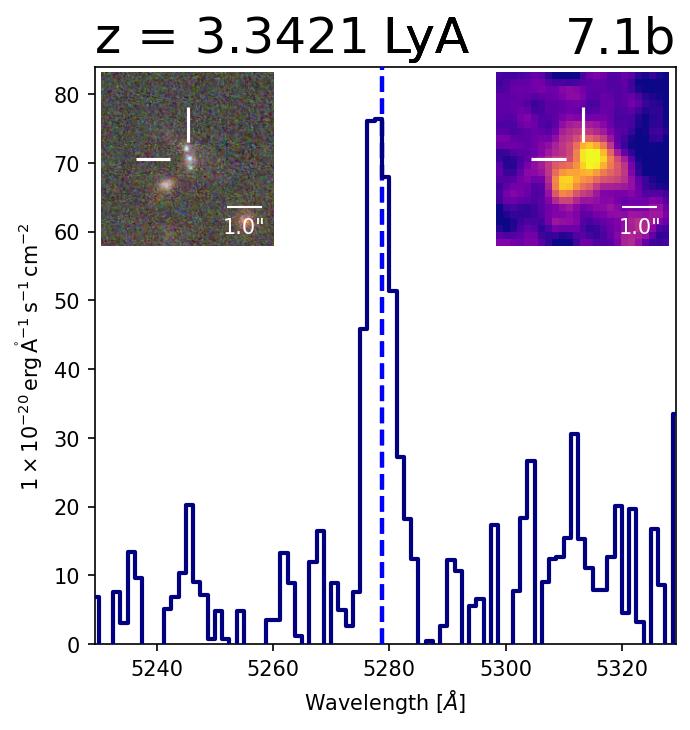}
        \end{subfigure}%
        \begin{subfigure}{.25\textwidth}
            \centering
            \includegraphics[width=1.0\textwidth]{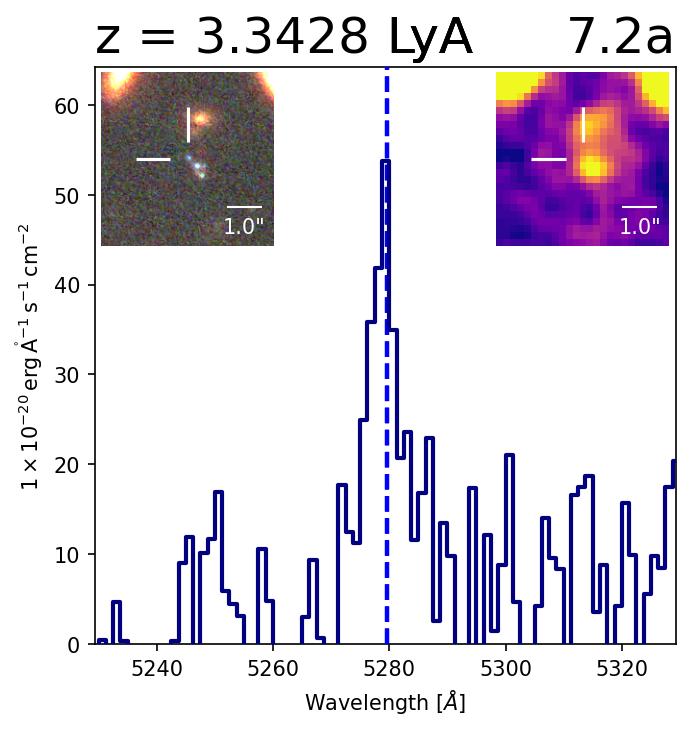}
        \end{subfigure}%
        \begin{subfigure}{.25\textwidth}
            \centering
            \includegraphics[width=1.0\textwidth]{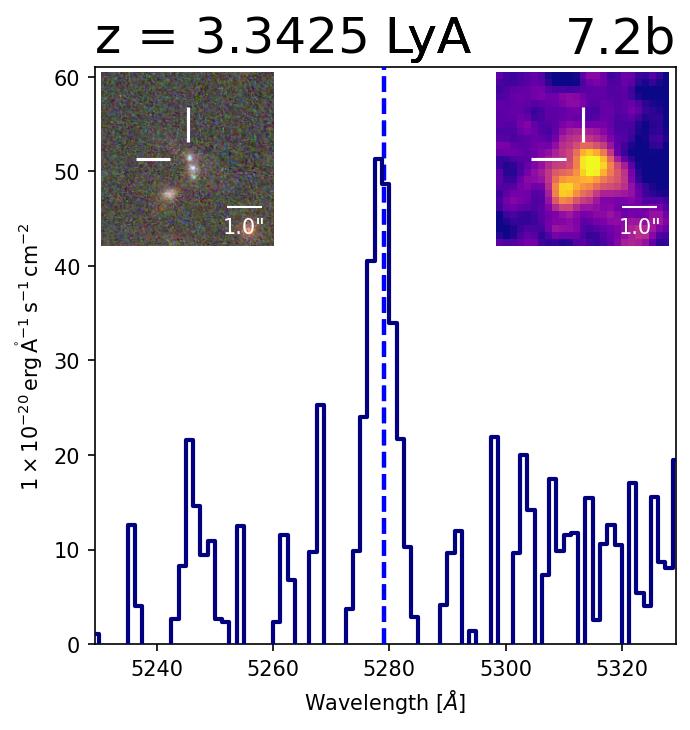}
        \end{subfigure}
    
        \begin{subfigure}{.25\textwidth}
            \centering
            \includegraphics[width=1.0\textwidth]{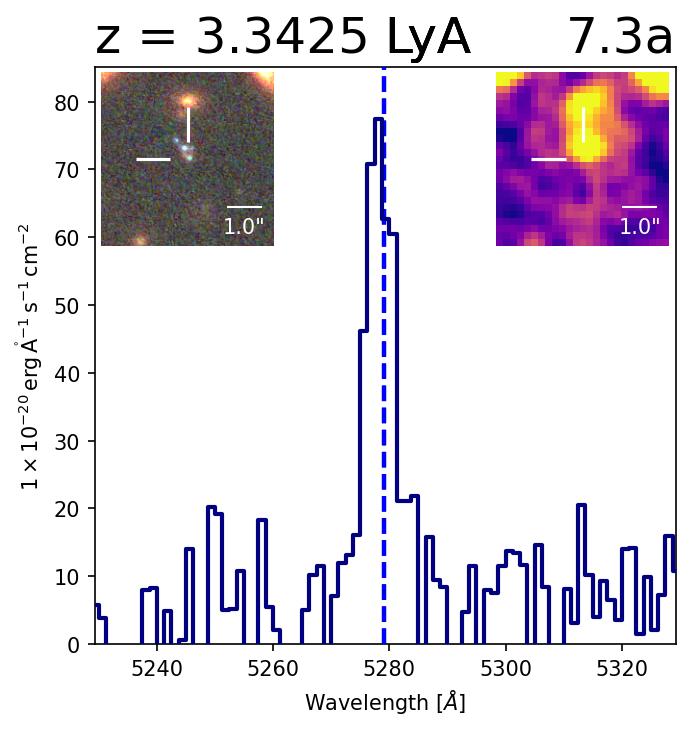}
        \end{subfigure}
        \begin{subfigure}{.25\textwidth}
            \centering
            \includegraphics[width=1.0\textwidth]{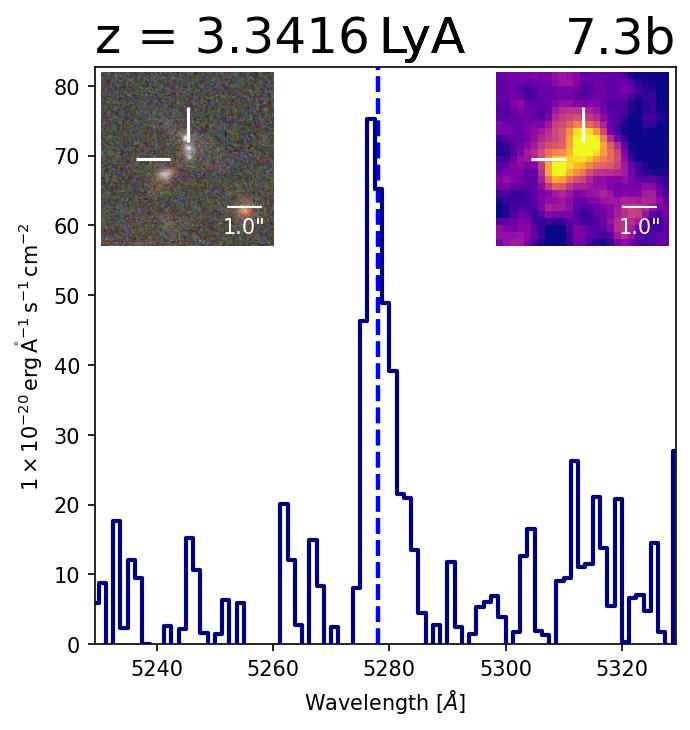}
        \end{subfigure}%
        \caption{\label{fig:cimg_sys_7}System 7}
    \end{figure}

    \begin{figure}[H]
        \raggedright{}
        \begin{subfigure}{.25\textwidth}
            \centering
            \includegraphics[width=1.0\textwidth]{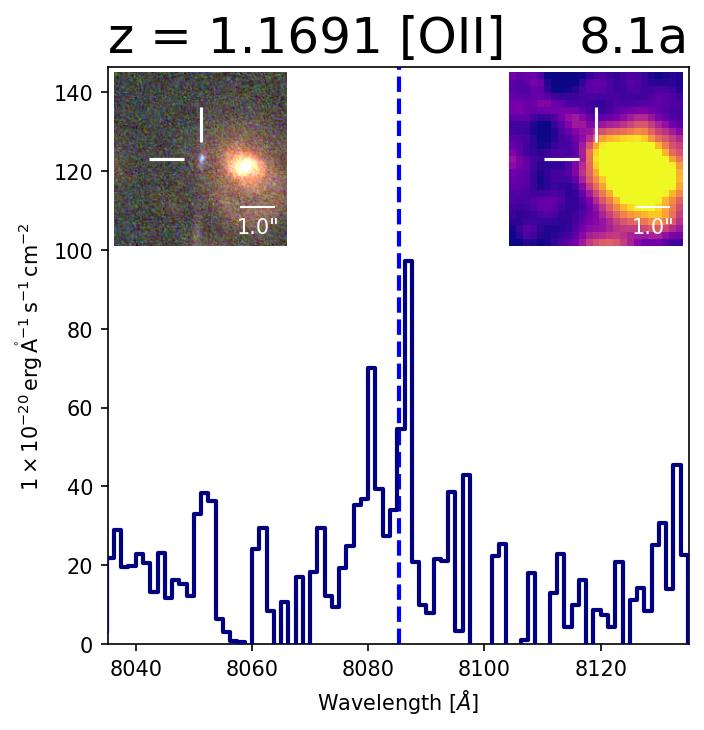}
        \end{subfigure}
        \begin{subfigure}{.25\textwidth}
            \centering
            \includegraphics[width=1.0\textwidth]{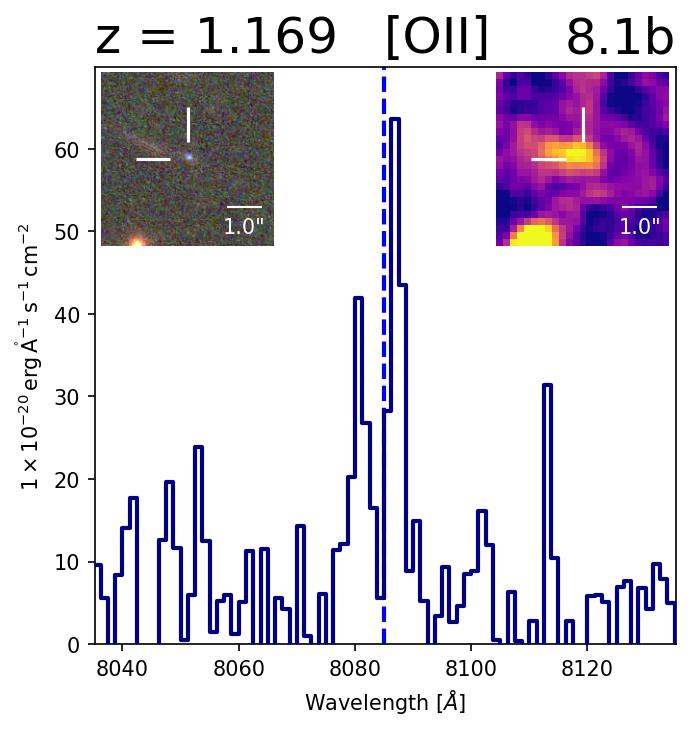}
        \end{subfigure}%
        \begin{subfigure}{.25\textwidth}
            \centering
            \includegraphics[width=1.0\textwidth]{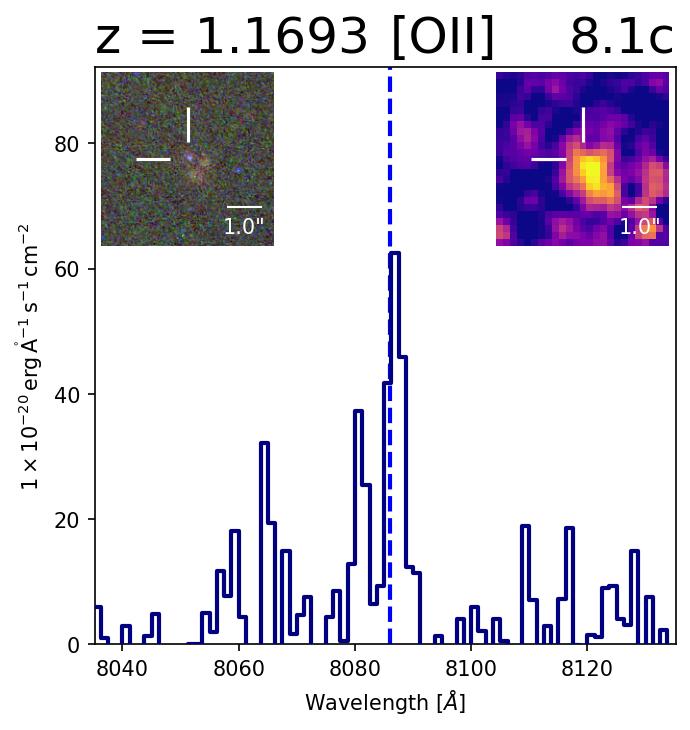}
        \end{subfigure}%
        \caption{\label{fig:cimg_sys_8}System 8}
    \end{figure}
    
    \begin{figure}[H]
        \raggedright{}
        \begin{subfigure}{.25\textwidth}
            \centering
            \includegraphics[width=1.0\textwidth]{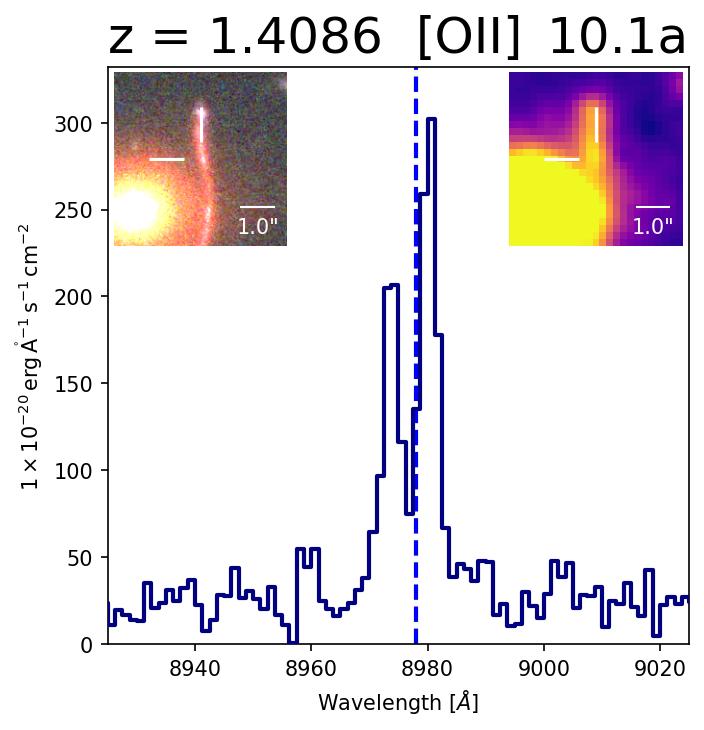}
        \end{subfigure}
        \begin{subfigure}{.25\textwidth}
            \centering
            \includegraphics[width=1.0\textwidth]{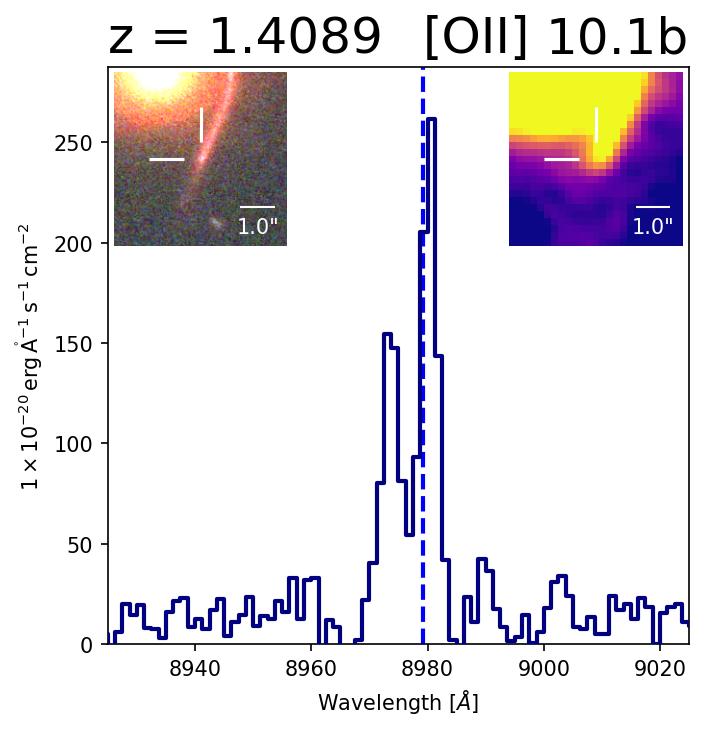}
        \end{subfigure}%
        \begin{subfigure}{.25\textwidth}
            \centering
            \includegraphics[width=1.0\textwidth]{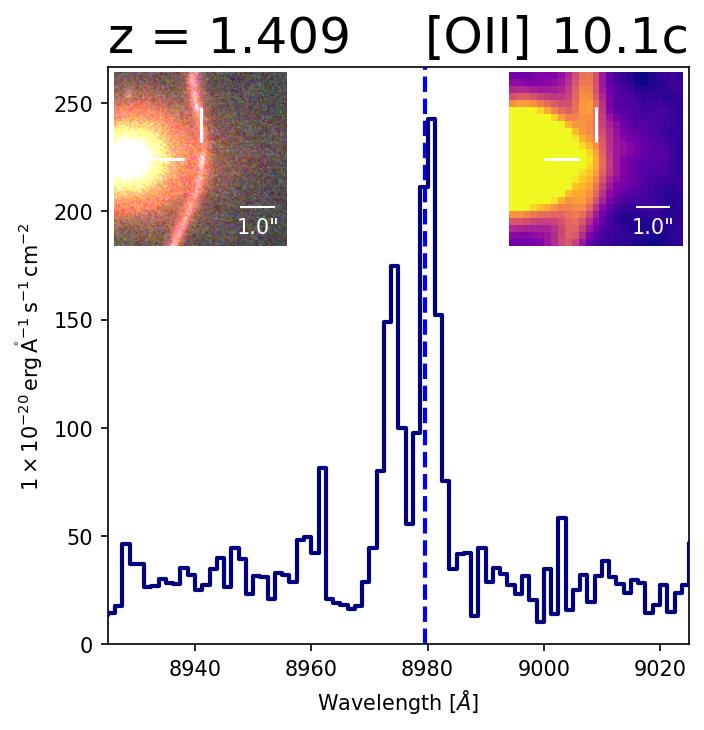}
        \end{subfigure}%
        \begin{subfigure}{.25\textwidth}
            \centering
            \includegraphics[width=1.0\textwidth]{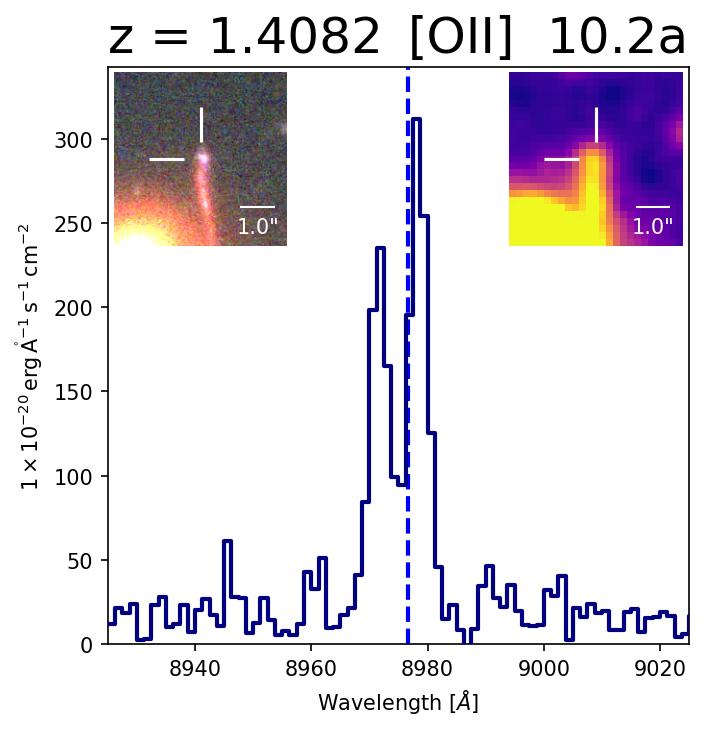}
        \end{subfigure}
        \caption{\label{fig:cimg_sys_10}System 10}
    \end{figure}
    
    \begin{figure}[H]
        \raggedright{}
        \begin{subfigure}{.25\textwidth}
            \centering
            \includegraphics[width=1.0\textwidth]{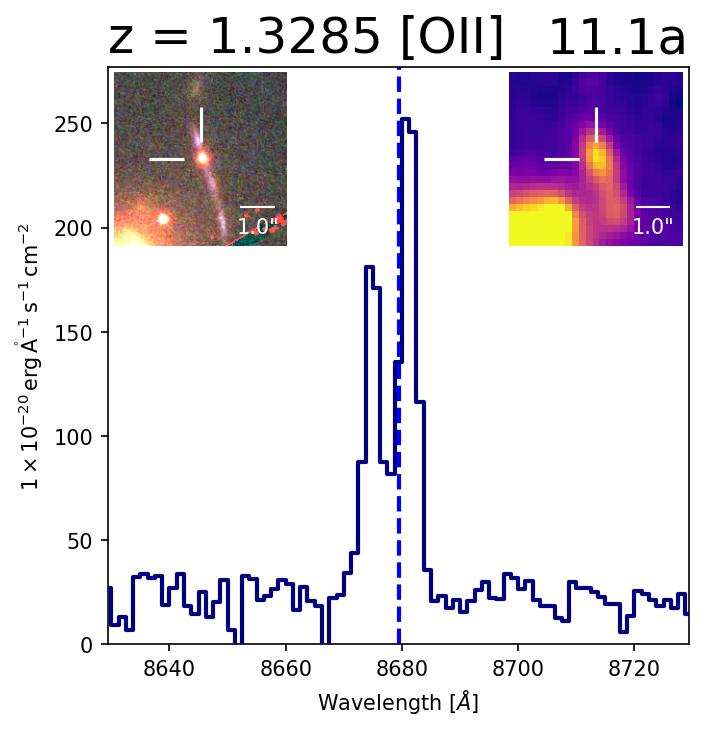}
        \end{subfigure}
        \begin{subfigure}{.25\textwidth}
            \centering
            \includegraphics[width=1.0\textwidth]{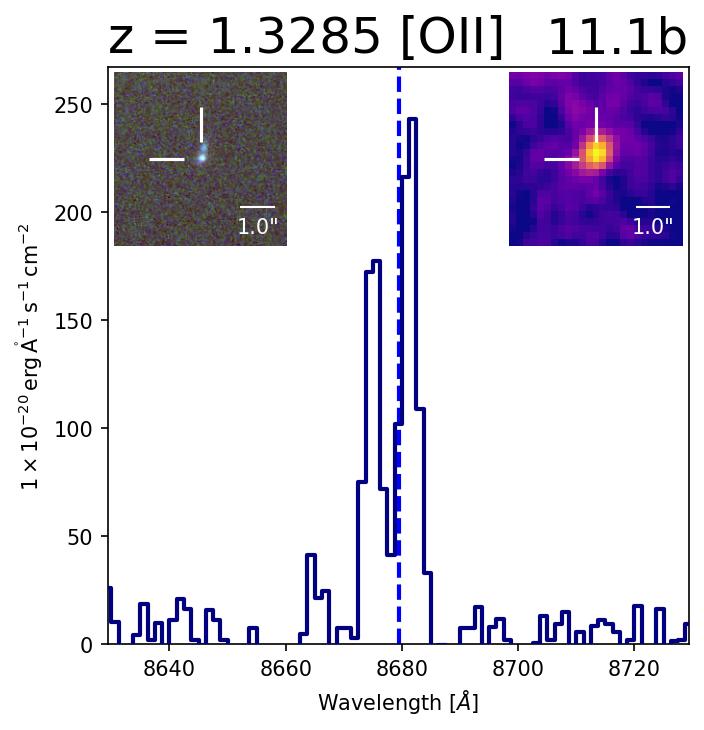}
        \end{subfigure}%
            \begin{subfigure}{.25\textwidth}
            \centering
            \includegraphics[width=1.0\textwidth]{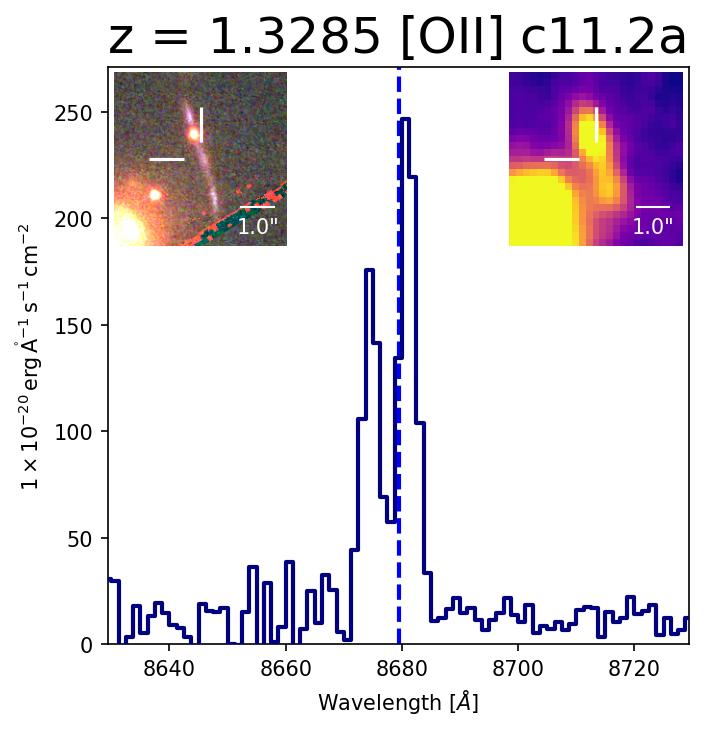}
        \end{subfigure}
        \begin{subfigure}{.25\textwidth}
            \centering
            \includegraphics[width=1.0\textwidth]{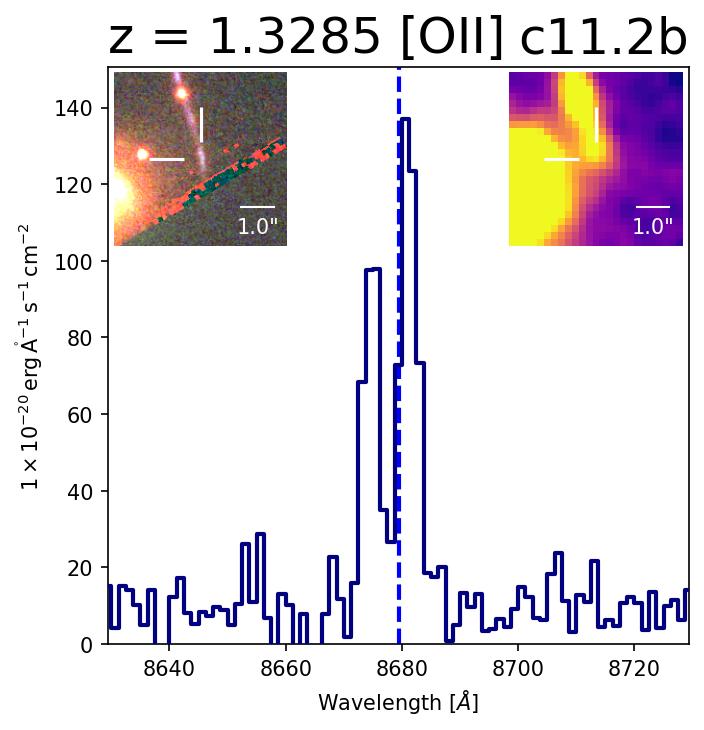}
        \end{subfigure}
        \begin{subfigure}{.25\textwidth}
            \centering
            \includegraphics[width=1.0\textwidth]{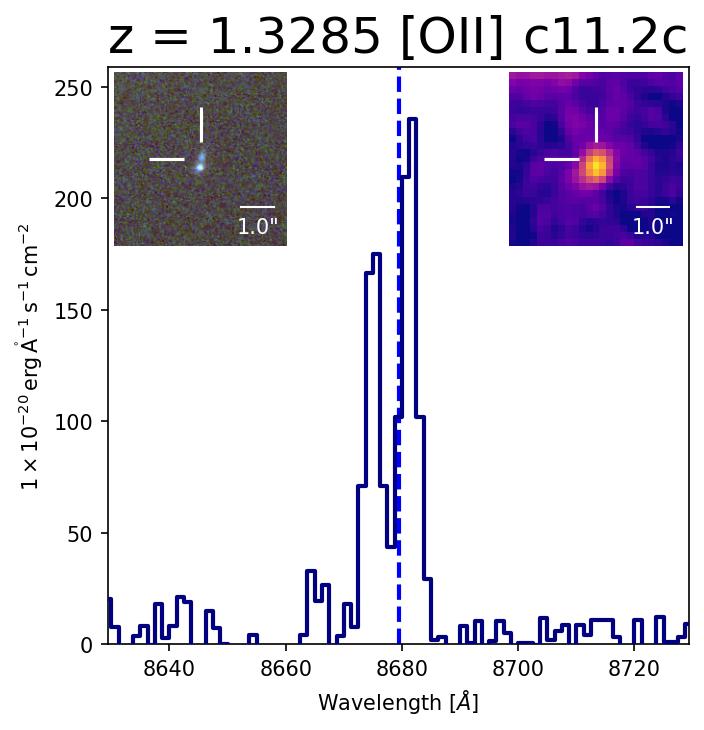}
        \end{subfigure}%
        \caption{\label{fig:cimg_sys_11}System 11}
    \end{figure}
    
    \begin{figure}[H]
        \raggedright{}
        \begin{subfigure}{.25\textwidth}
            \centering
            \includegraphics[width=1.0\textwidth]{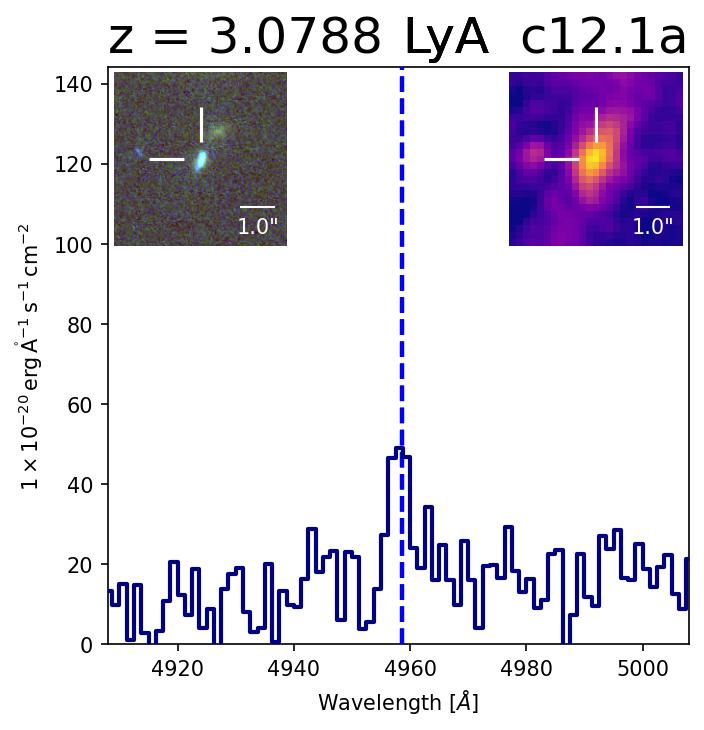}
        \end{subfigure}
        \begin{subfigure}{.25\textwidth}
            \centering
            \includegraphics[width=1.0\textwidth]{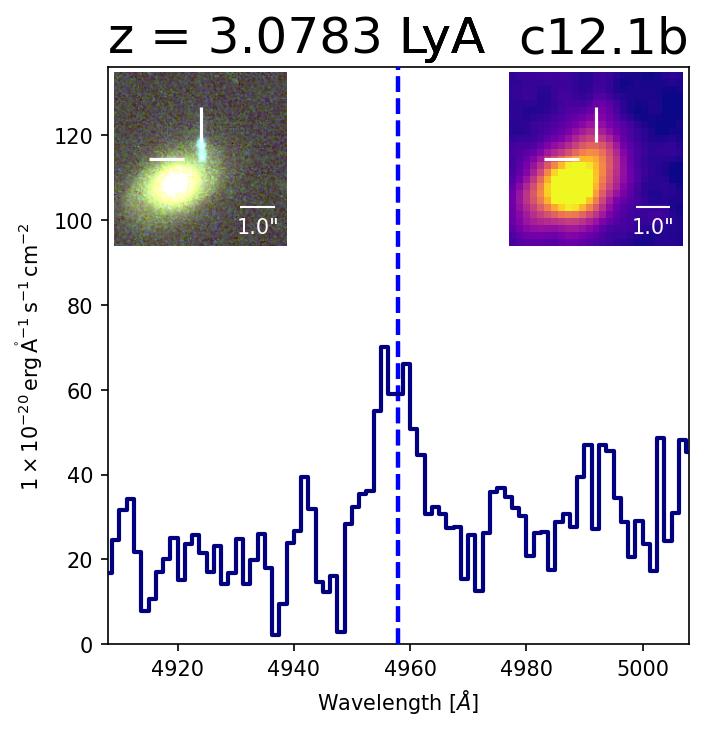}
        \end{subfigure}%
        \begin{subfigure}{.25\textwidth}
            \centering
            \includegraphics[width=1.0\textwidth]{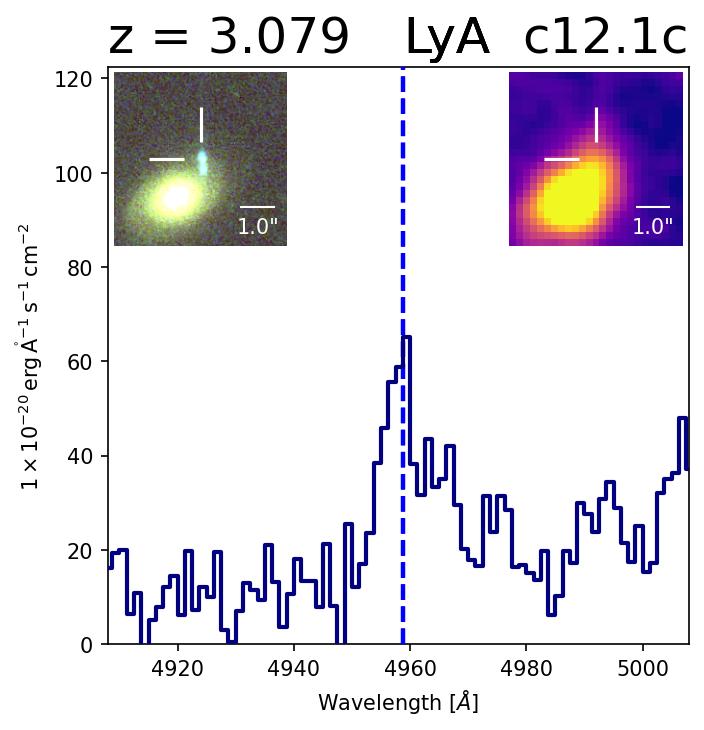}
        \end{subfigure}%
        \caption{\label{fig:cimg_sys_12}System 12}
    \end{figure}
    
    \begin{figure}[H]
        \raggedright{}
        \begin{subfigure}{.25\textwidth}
            \centering
            \includegraphics[width=1.0\textwidth]{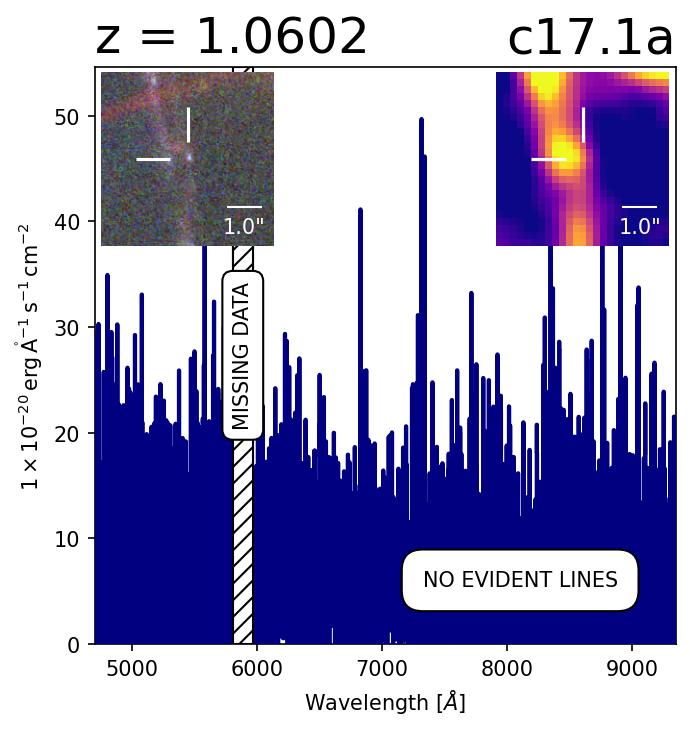}
        \end{subfigure}
        \begin{subfigure}{.25\textwidth}
            \centering
            \includegraphics[width=1.0\textwidth]{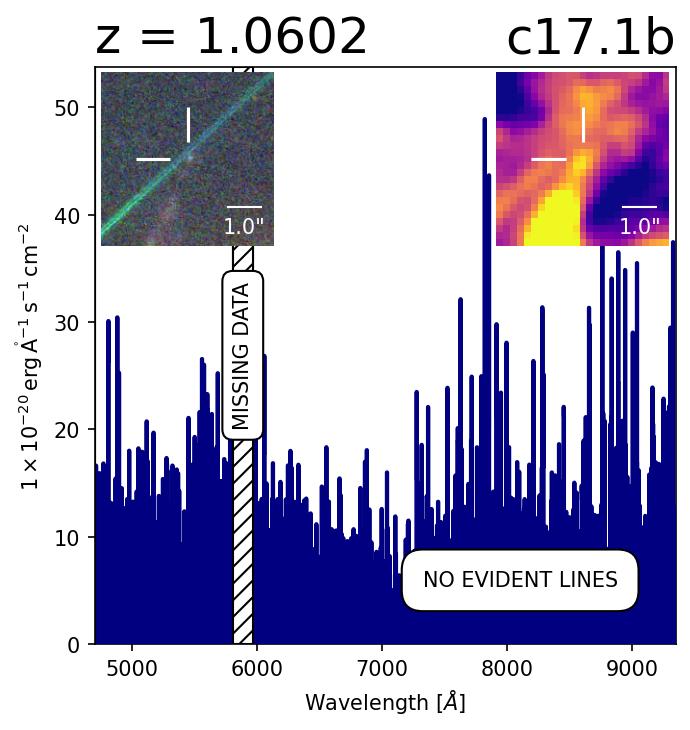}
        \end{subfigure}%
        \caption{\label{fig:cimg_sys_17}System 17}
    \end{figure}
    
    \begin{figure}[H]
        \raggedright{}
        \begin{subfigure}{.25\textwidth}
            \centering
            \includegraphics[width=1.0\textwidth]{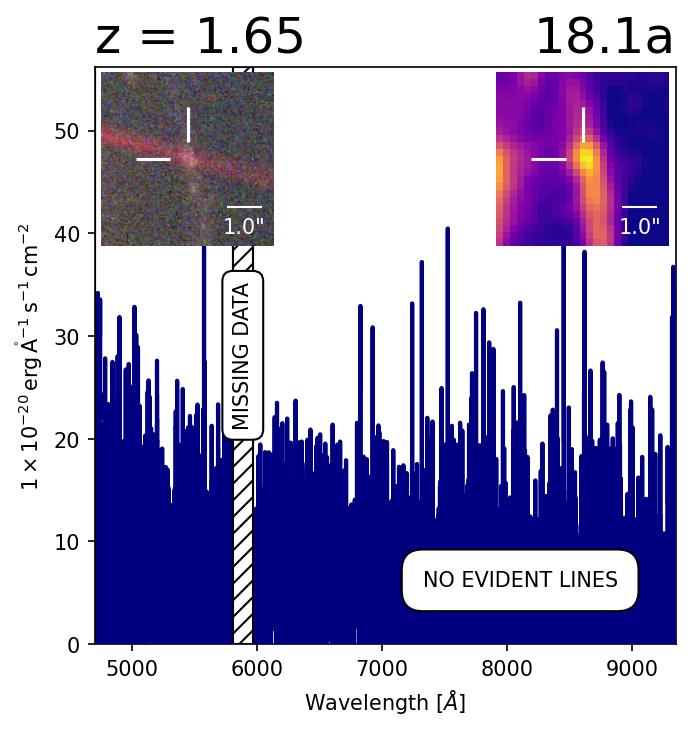}
        \end{subfigure}
        \begin{subfigure}{.25\textwidth}
            \centering
            \includegraphics[width=1.0\textwidth]{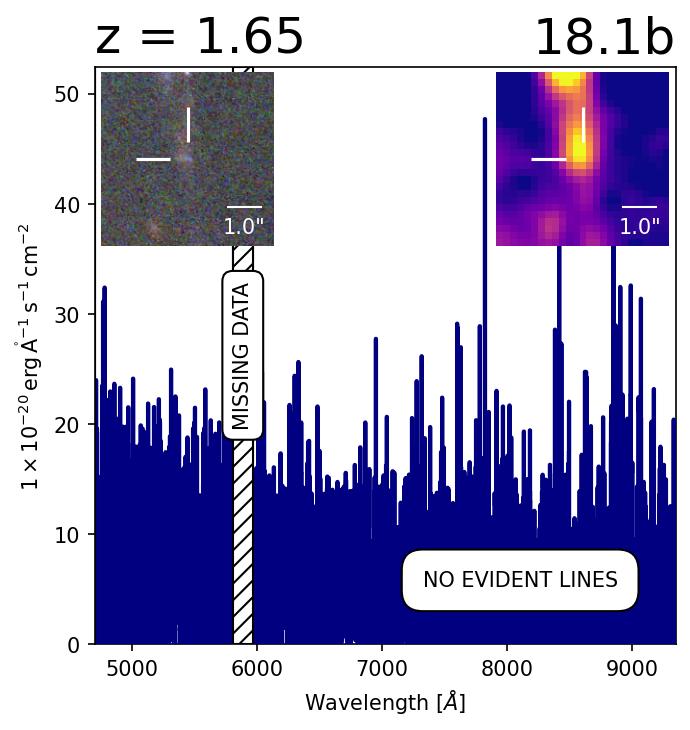}
        \end{subfigure}%
        \begin{subfigure}{.25\textwidth}
            \centering
            \includegraphics[width=1.0\textwidth]{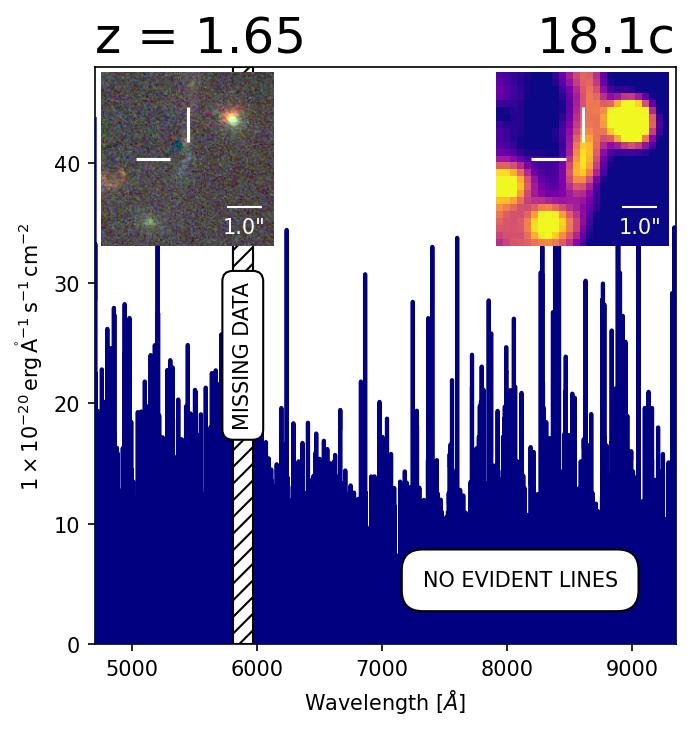}
        \end{subfigure}
        \caption{\label{fig:cimg_sys_50}System 18}
    \end{figure}
    
    \begin{figure}[H]
        \raggedright{}
        \begin{subfigure}{.25\textwidth}
            \centering
            \includegraphics[width=1.0\textwidth]{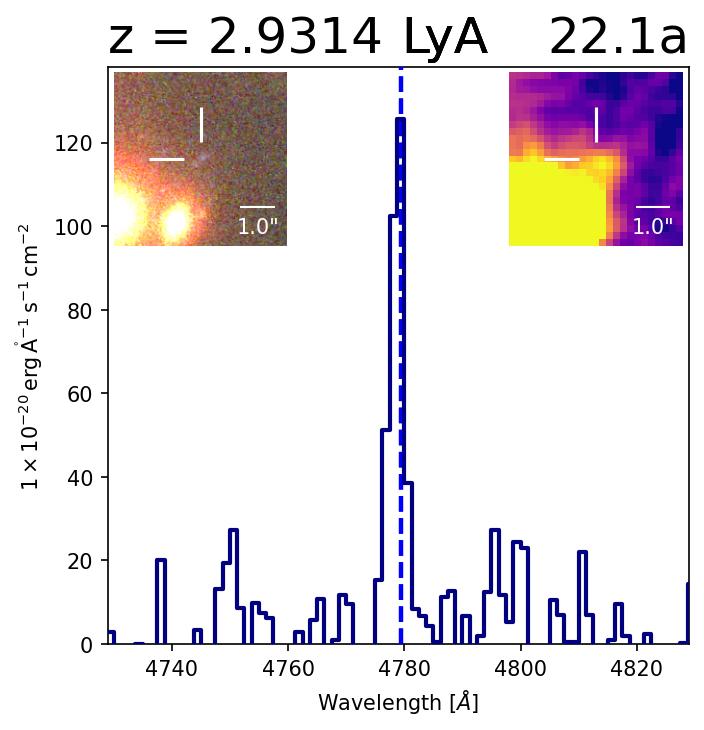}
        \end{subfigure}
        \begin{subfigure}{.25\textwidth}
            \centering
            \includegraphics[width=1.0\textwidth]{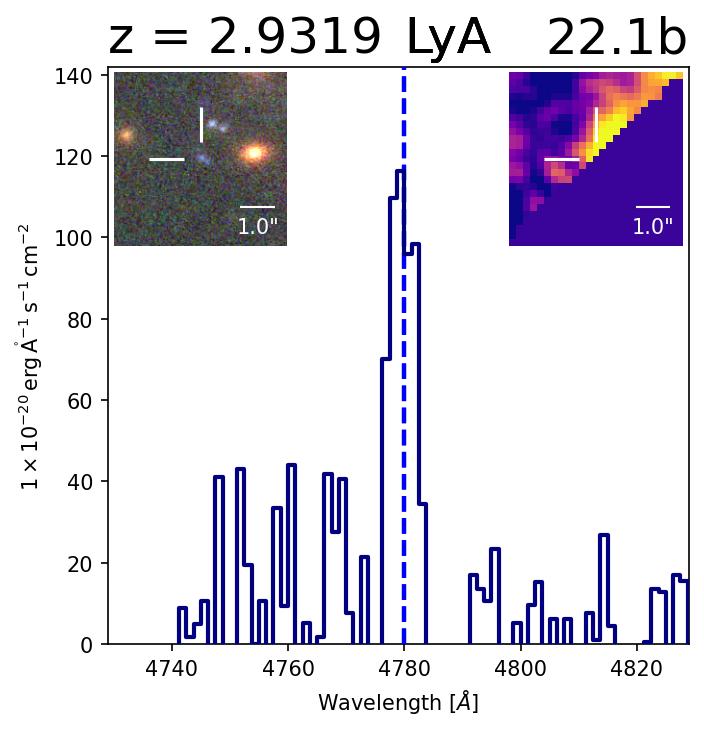}
        \end{subfigure}%
        \begin{subfigure}{.25\textwidth}
            \centering
            \includegraphics[width=1.0\textwidth]{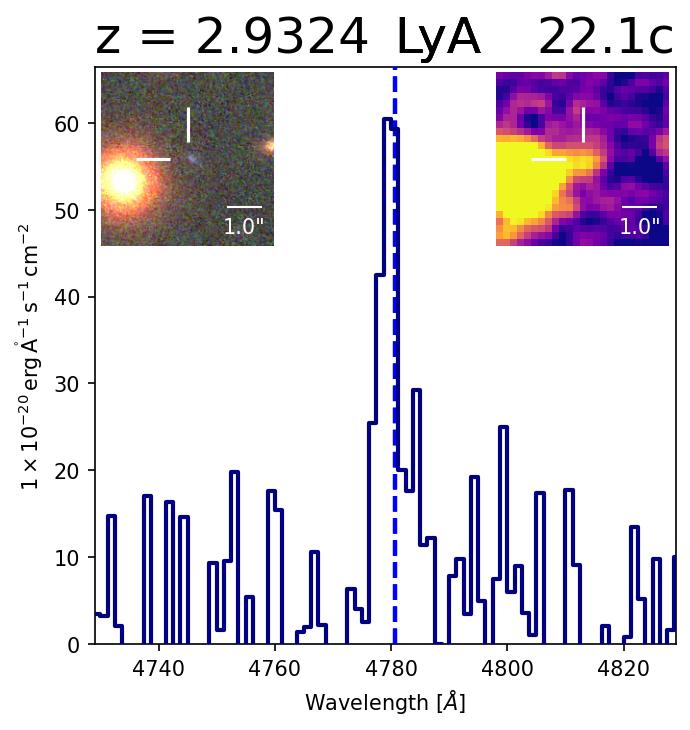}
        \end{subfigure}
        \begin{subfigure}{.25\textwidth}
            \centering
            \includegraphics[width=1.0\textwidth]{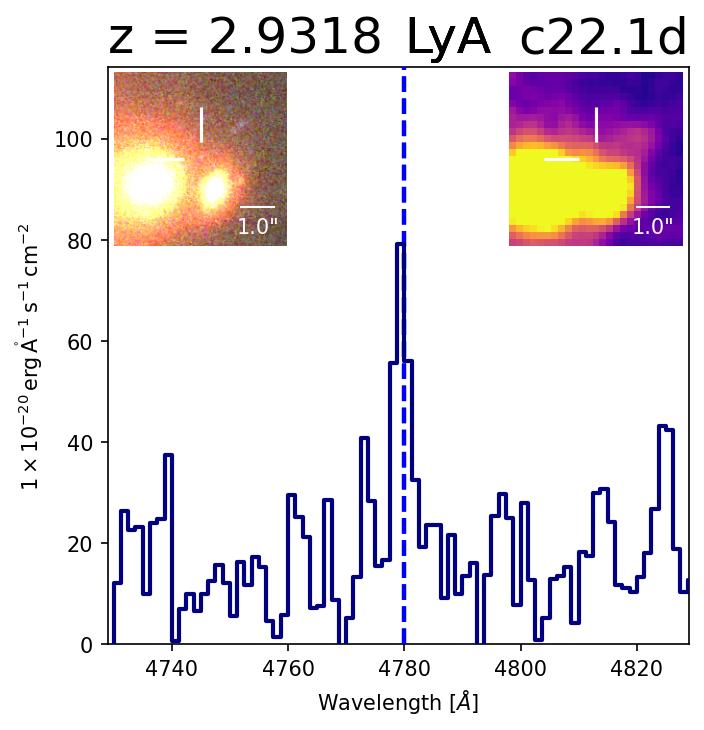}
        \end{subfigure}%
        \begin{subfigure}{.25\textwidth}
            \centering
            \includegraphics[width=1.0\textwidth]{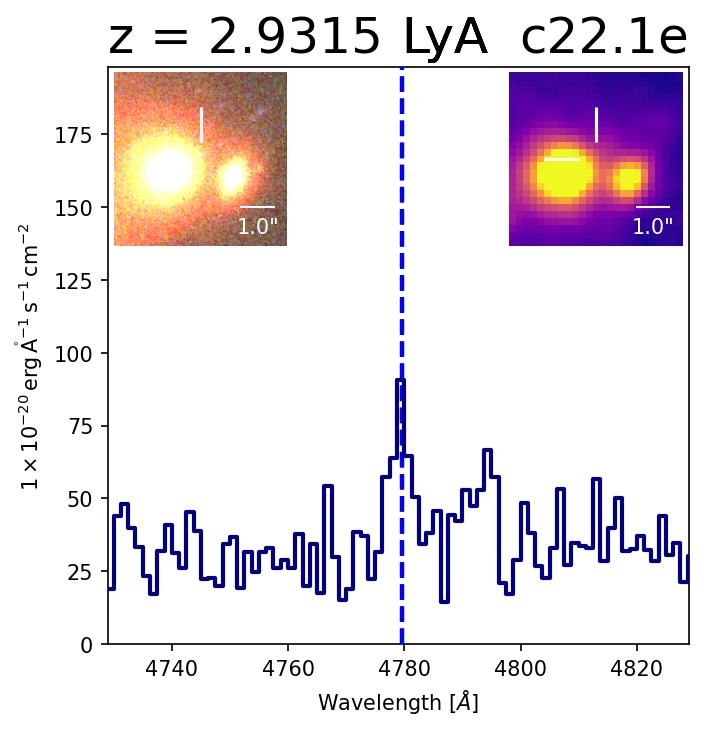}
        \end{subfigure}%
        \caption{\label{fig:cimg_sys_22}System 22}
    \end{figure}
    
    \begin{figure}[H]
        \raggedright{}
        \begin{subfigure}{.25\textwidth}
            \centering
            \includegraphics[width=1.0\textwidth]{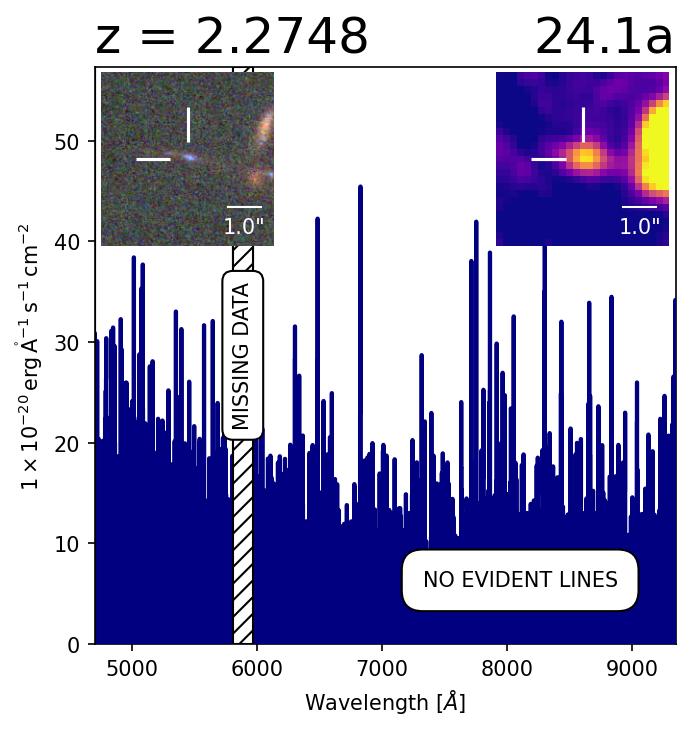}
        \end{subfigure}
        \begin{subfigure}{.25\textwidth}
            \centering
            \includegraphics[width=1.0\textwidth]{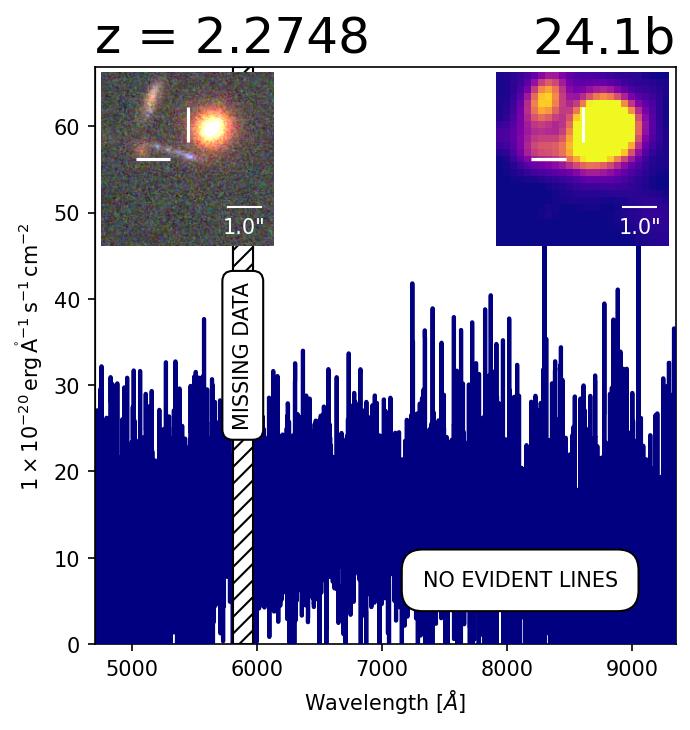}
        \end{subfigure}%
        \caption{\label{fig:cimg_sys_24}System 24}
    \end{figure}
    
    \begin{figure}[H]
        \raggedright{}
        \begin{subfigure}{.25\textwidth}
            \centering
            \includegraphics[width=1.0\textwidth]{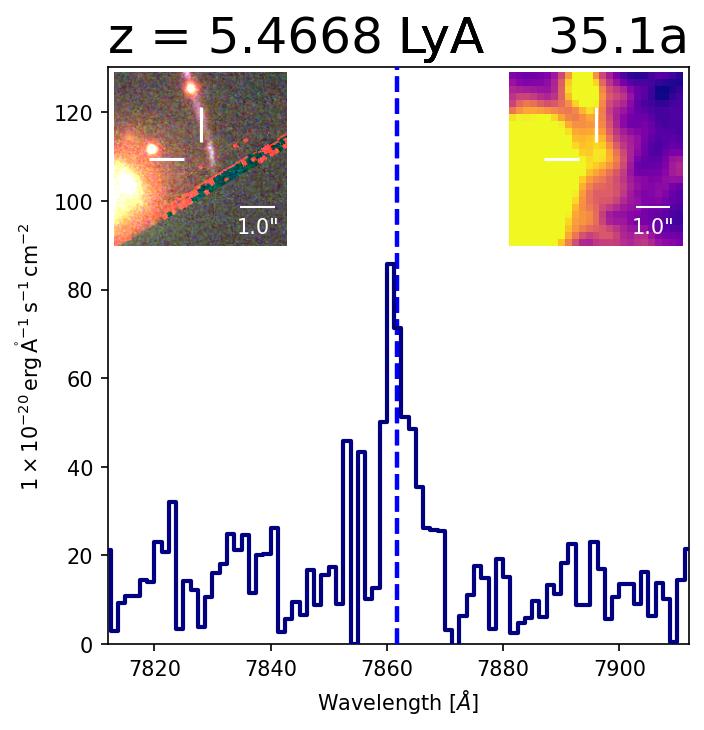}
        \end{subfigure}
        \begin{subfigure}{.25\textwidth}
            \centering
            \includegraphics[width=1.0\textwidth]{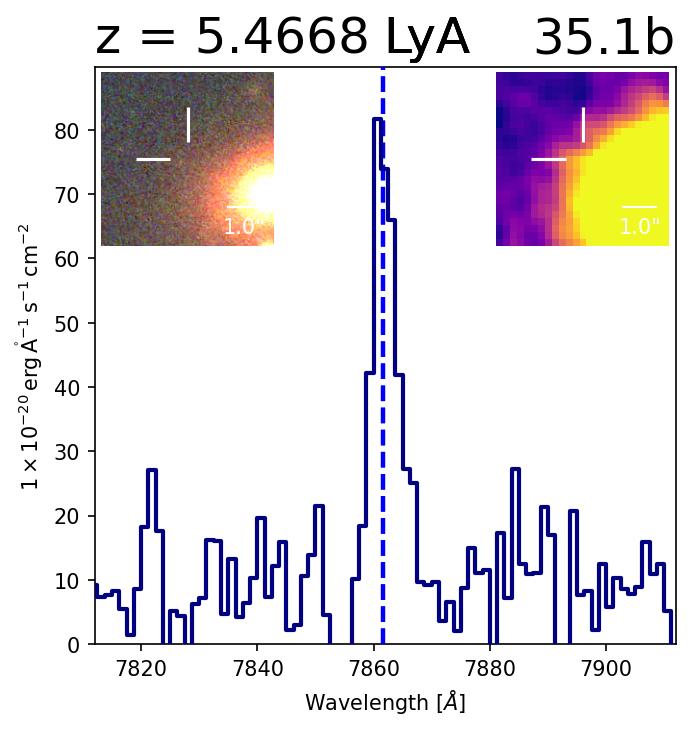}
        \end{subfigure}%
        \begin{subfigure}{.25\textwidth}
            \centering
            \includegraphics[width=1.0\textwidth]{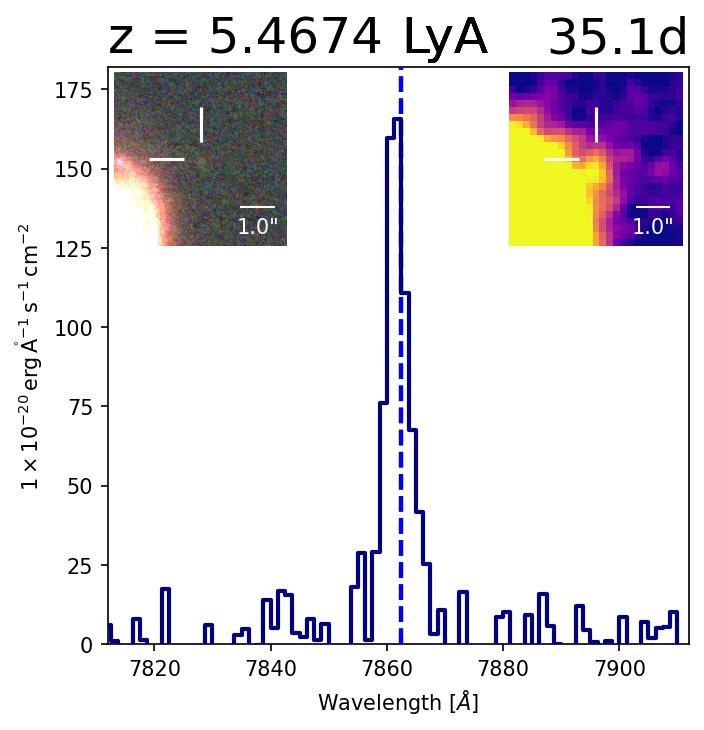}
        \end{subfigure}%
        \begin{subfigure}{.25\textwidth}
            \centering
            \includegraphics[width=1.0\textwidth]{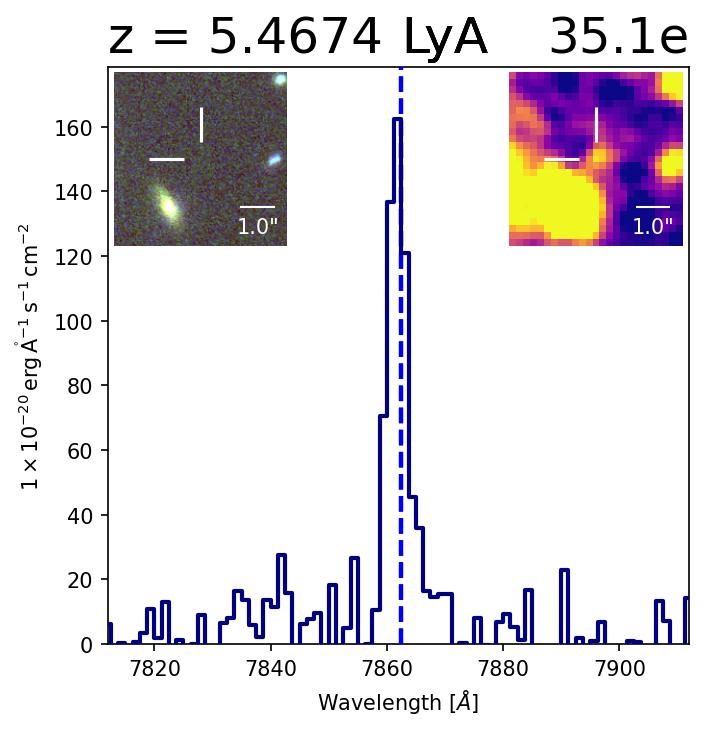}
        \end{subfigure}%
        \caption{\label{fig:cimg_sys_35}System 35}
    \end{figure}
    
    \begin{figure}[H]
        \raggedright{}
        \begin{subfigure}{.25\textwidth}
            \centering
            \includegraphics[width=1.0\textwidth]{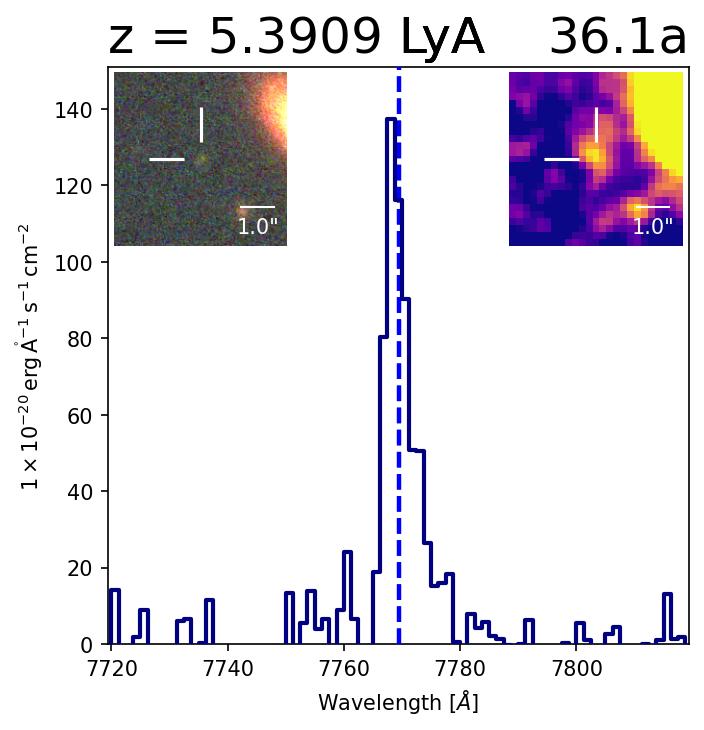}
        \end{subfigure}
        \begin{subfigure}{.25\textwidth}
            \centering
            \includegraphics[width=1.0\textwidth]{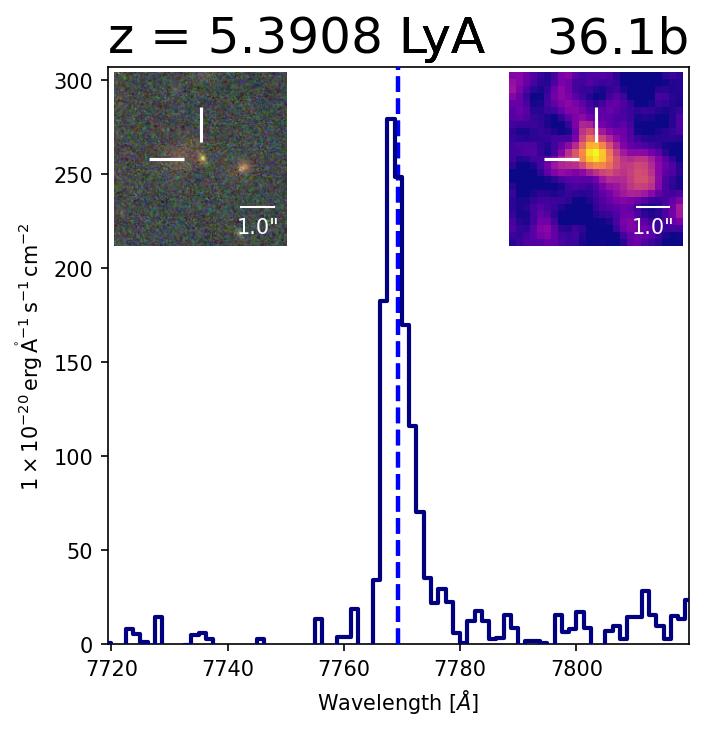}
        \end{subfigure}%
        \caption{\label{fig:cimg_sys_36}System 36}
    \end{figure}

    \begin{figure}[H]
        \raggedright{}
        \begin{subfigure}{.25\textwidth}
            \centering
            \includegraphics[width=1.0\textwidth]{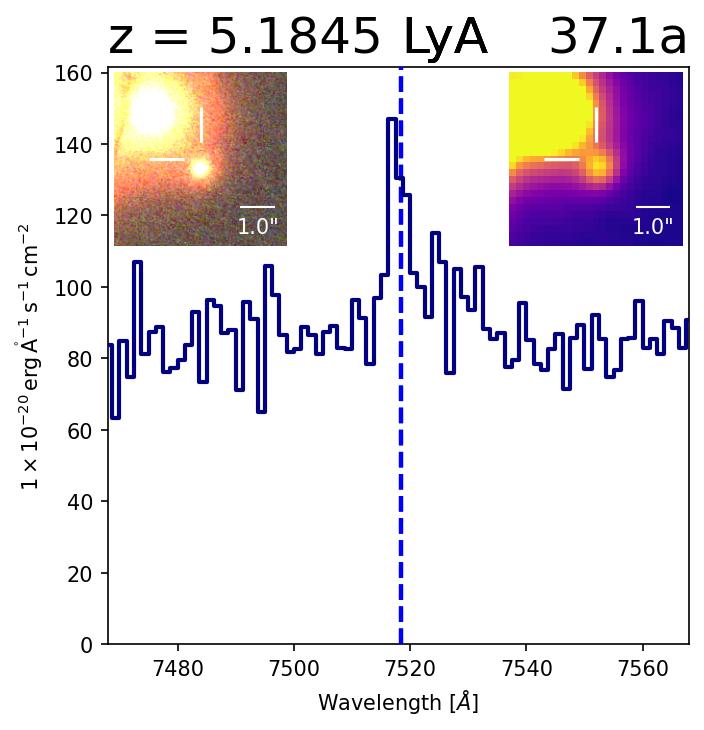}
        \end{subfigure}
        \begin{subfigure}{.25\textwidth}
            \centering
            \includegraphics[width=1.0\textwidth]{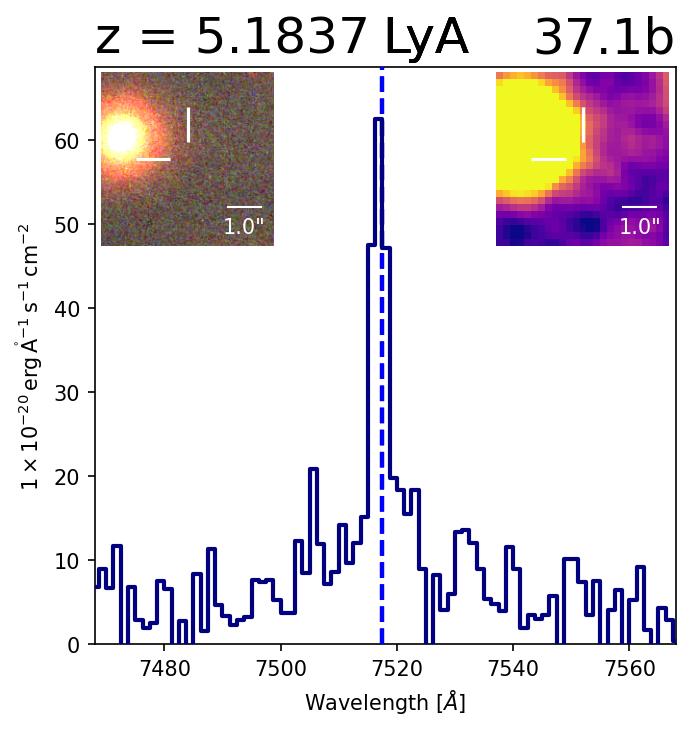}
        \end{subfigure}%
        \begin{subfigure}{.25\textwidth}
            \centering
            \includegraphics[width=1.0\textwidth]{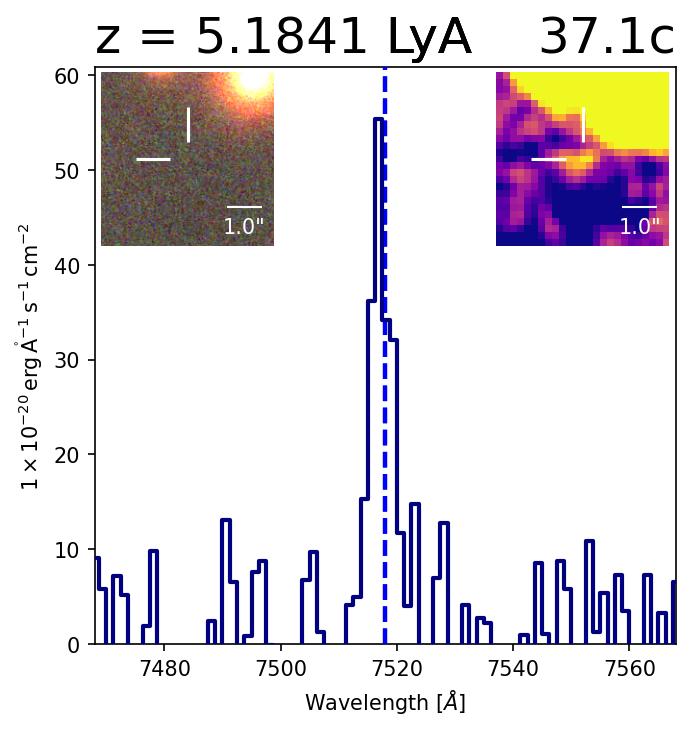}
        \end{subfigure}%
        \begin{subfigure}{.25\textwidth}
            \centering
            \includegraphics[width=1.0\textwidth]{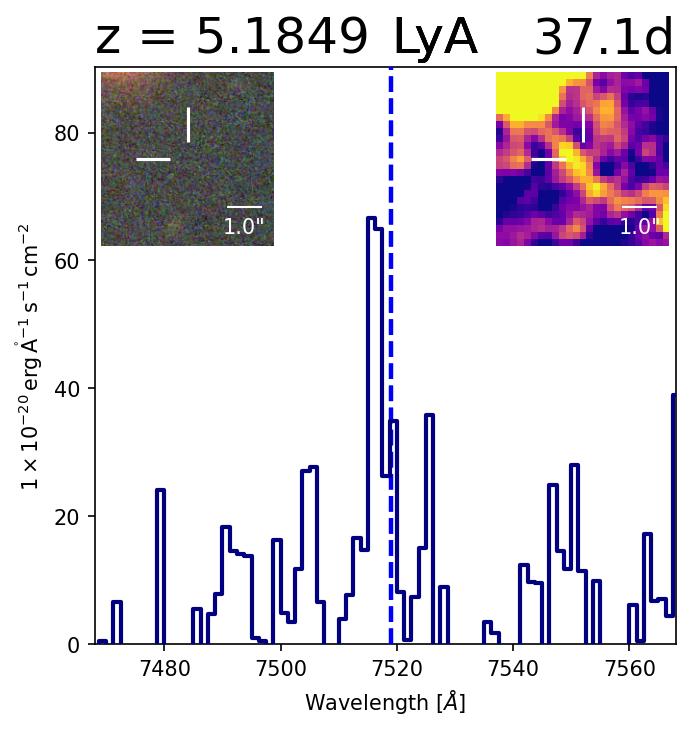}
        \end{subfigure}%
        \caption{\label{fig:cimg_sys_37}System 37}
    \end{figure}
    
    \begin{figure}[H]
        \raggedright{}
        \begin{subfigure}{.25\textwidth}
            \centering
            \includegraphics[width=1.0\textwidth]{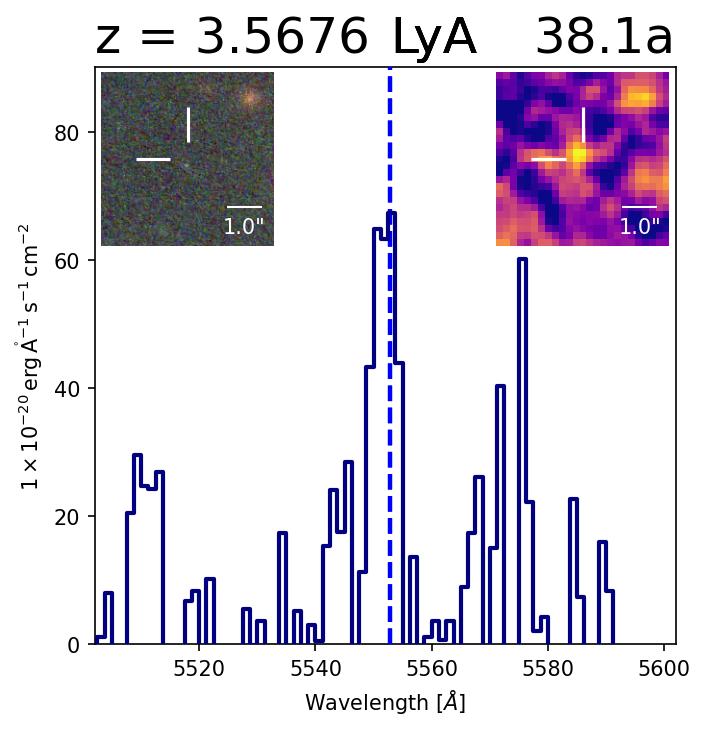}
        \end{subfigure}
        \begin{subfigure}{.25\textwidth}
            \centering
            \includegraphics[width=1.0\textwidth]{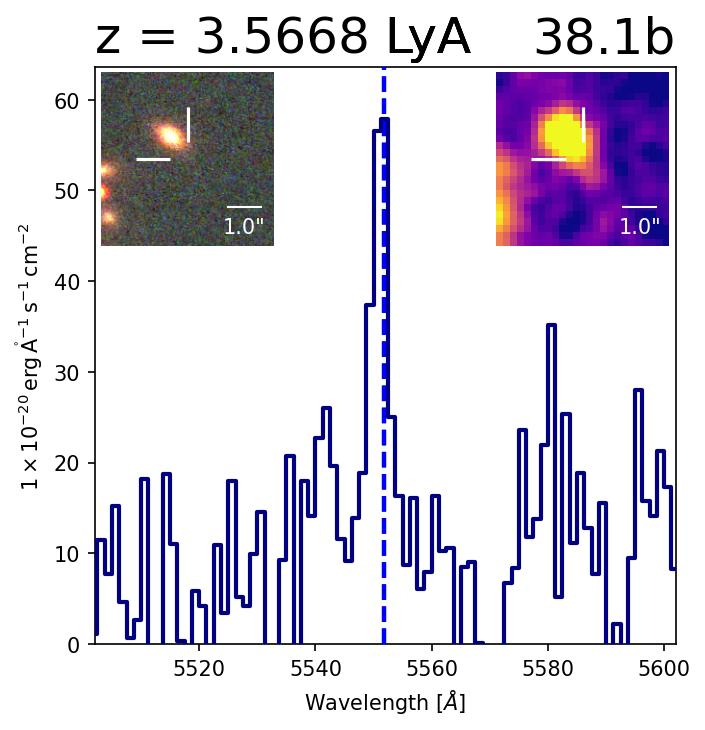}
        \end{subfigure}%
        \begin{subfigure}{.25\textwidth}
            \centering
            \includegraphics[width=1.0\textwidth]{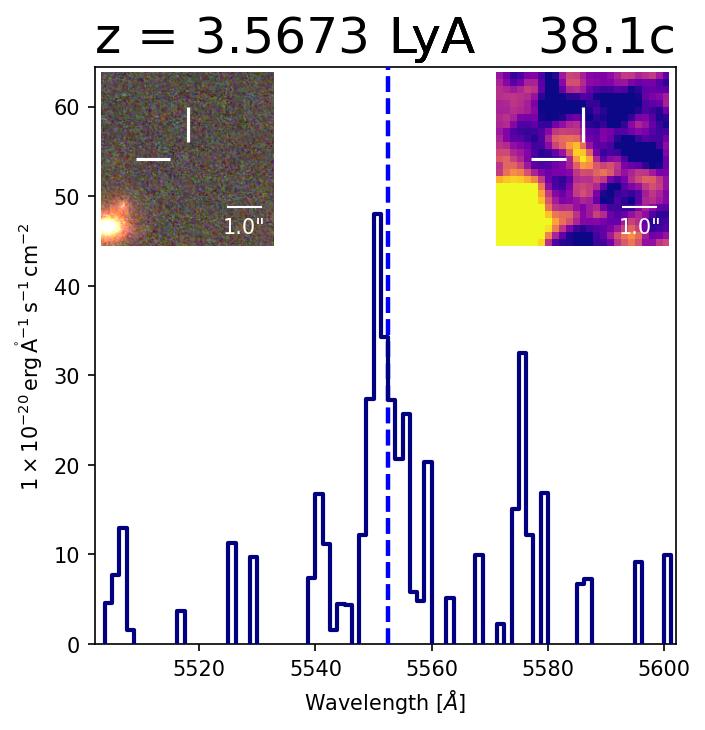}
        \end{subfigure}%
        \begin{subfigure}{.25\textwidth}
            \centering
            \includegraphics[width=1.0\textwidth]{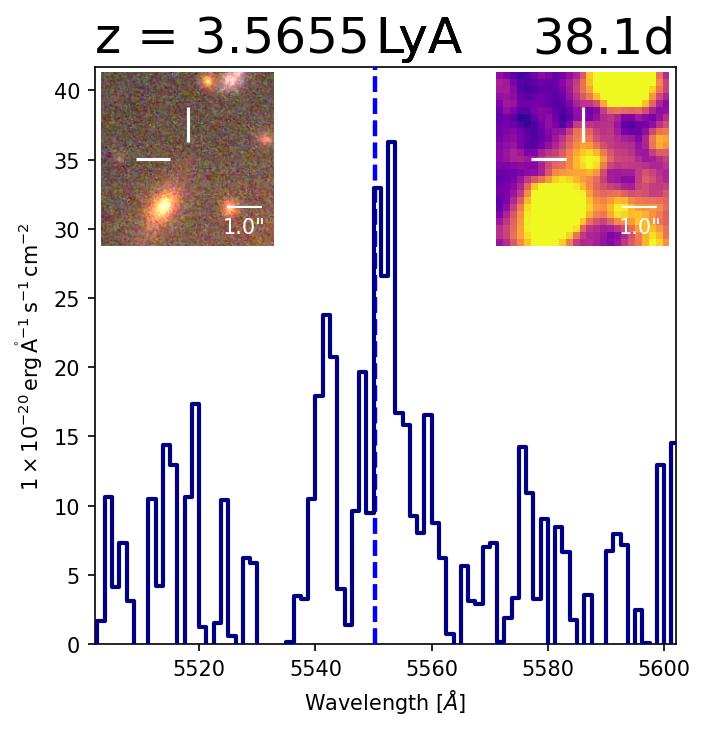}
        \end{subfigure}%
        \caption{\label{fig:cimg_sys_38}System 38}
    \end{figure}
    
    \begin{figure}[H]
        \raggedright{}
        \begin{subfigure}{.25\textwidth}
            \centering
            \includegraphics[width=1.0\textwidth]{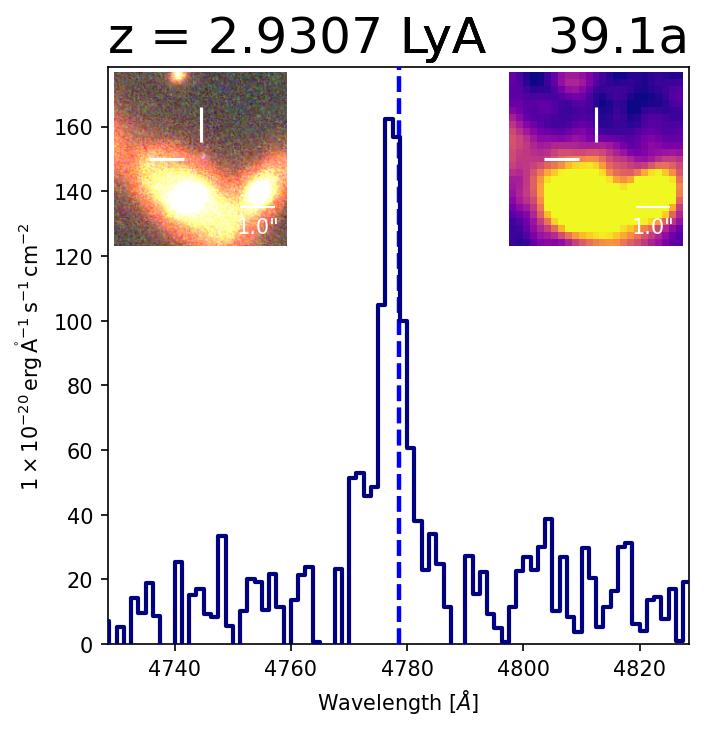}
        \end{subfigure}
        \begin{subfigure}{.25\textwidth}
            \centering
            \includegraphics[width=1.0\textwidth]{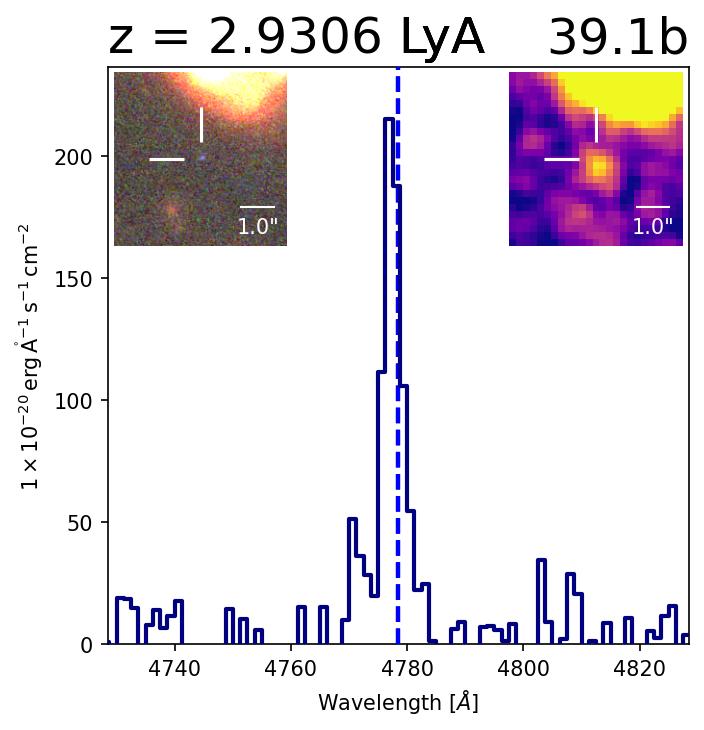}
        \end{subfigure}%
        \begin{subfigure}{.25\textwidth}
            \centering
            \includegraphics[width=1.0\textwidth]{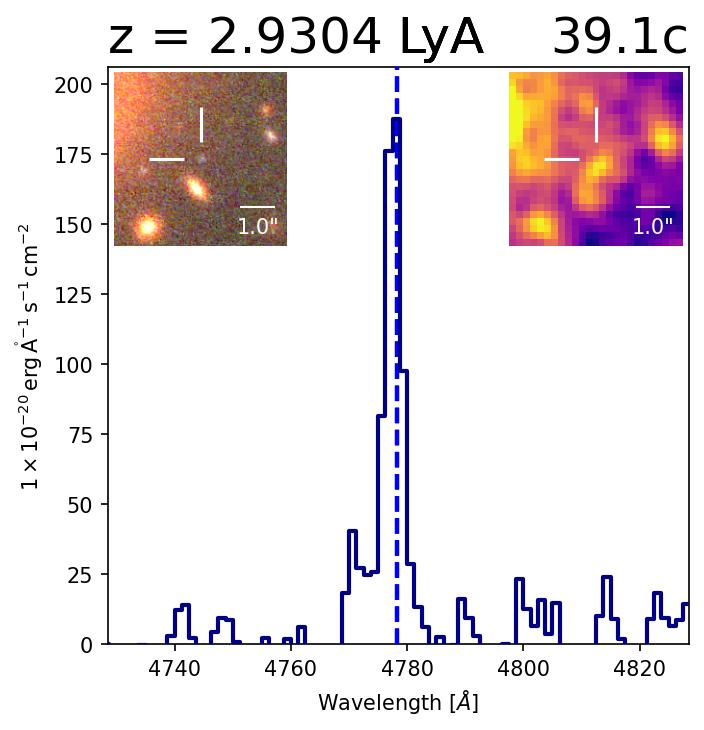}
        \end{subfigure}%
        \begin{subfigure}{.25\textwidth}
            \centering
            \includegraphics[width=1.0\textwidth]{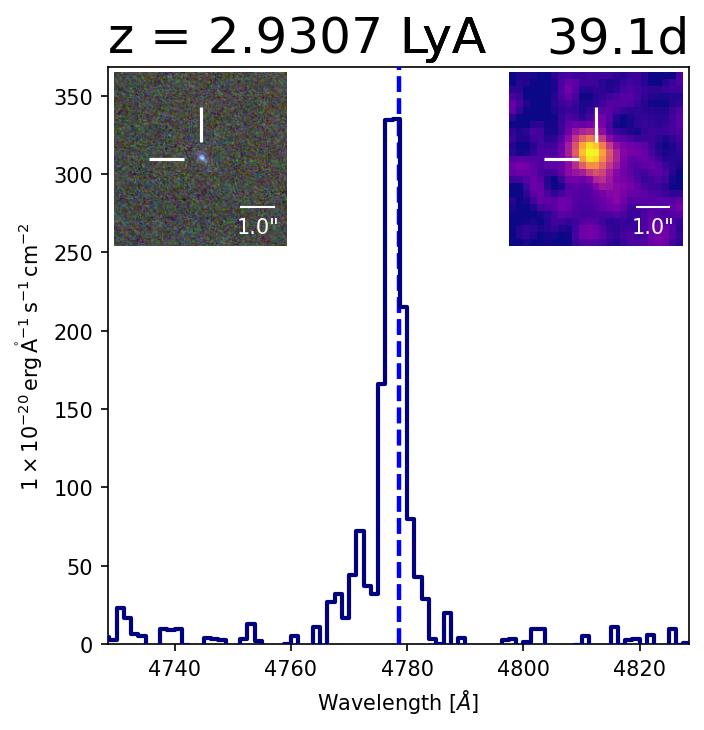}
        \end{subfigure}
        \begin{subfigure}{.25\textwidth}
            \centering
            \includegraphics[width=1.0\textwidth]{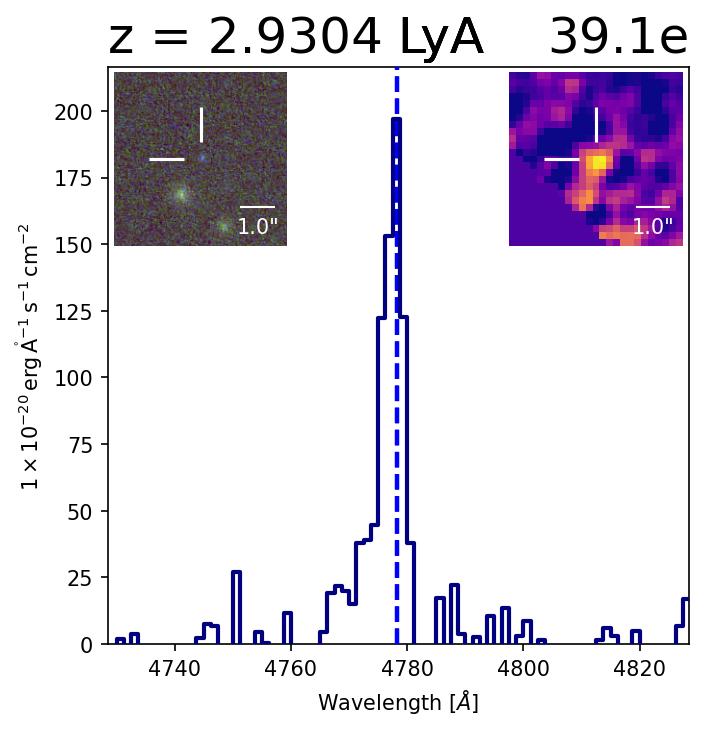}
        \end{subfigure}%
        \caption{\label{fig:cimg_sys_39}System 39}
    \end{figure}

    \begin{figure}[H]
        \raggedright{}
        \begin{subfigure}{.25\textwidth}
            \centering
            \includegraphics[width=1.0\textwidth]{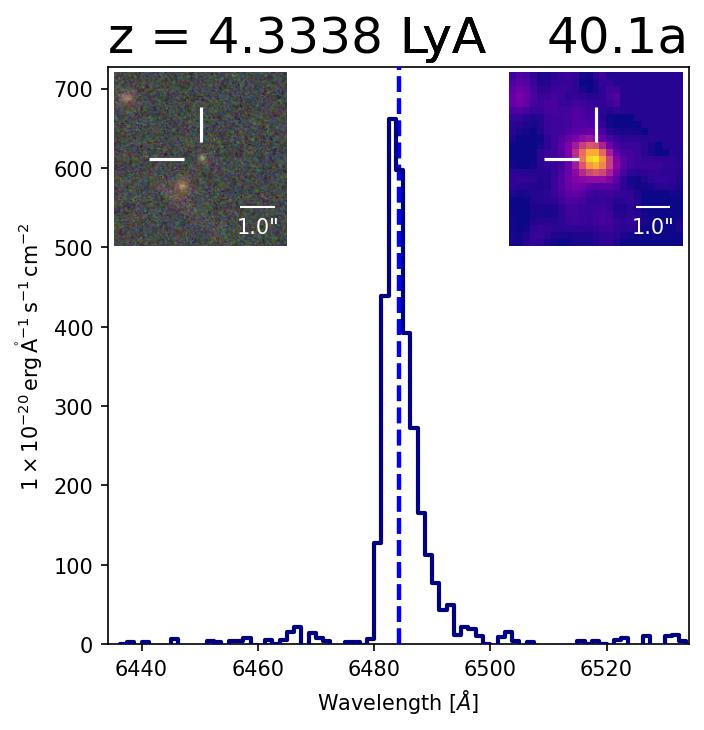}
        \end{subfigure}
        \begin{subfigure}{.25\textwidth}
            \centering
            \includegraphics[width=1.0\textwidth]{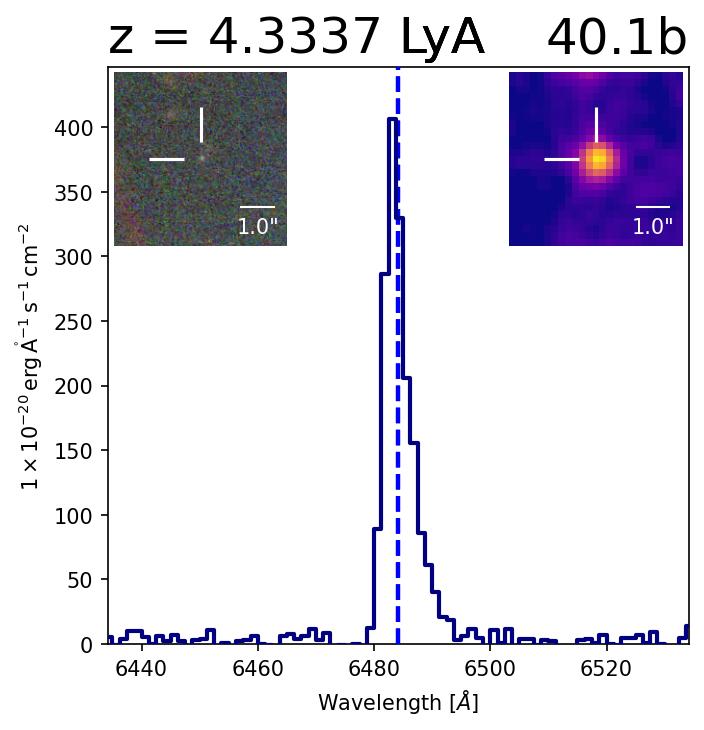}
        \end{subfigure}%
        \begin{subfigure}{.25\textwidth}
            \centering
            \includegraphics[width=1.0\textwidth]{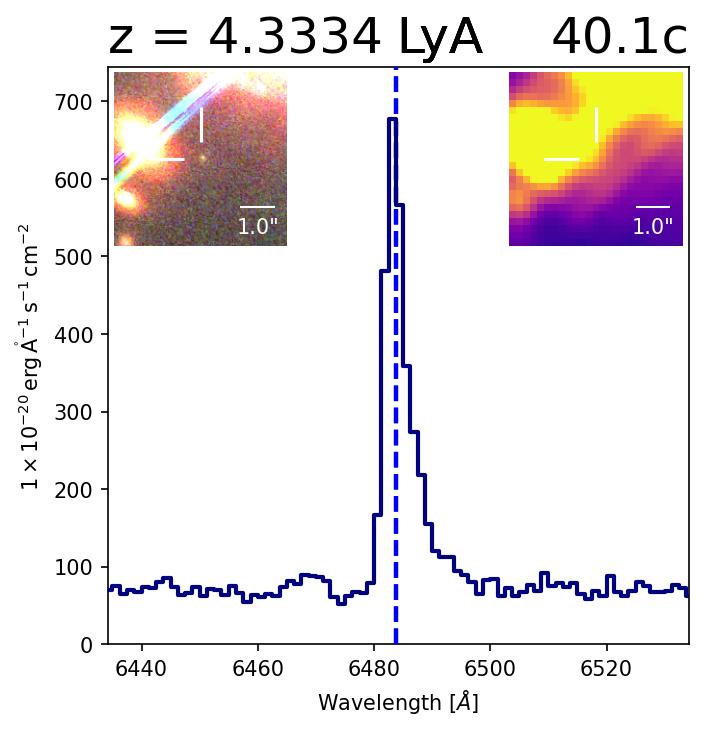}
        \end{subfigure}%
        \begin{subfigure}{.25\textwidth}
            \centering
            \includegraphics[width=1.0\textwidth]{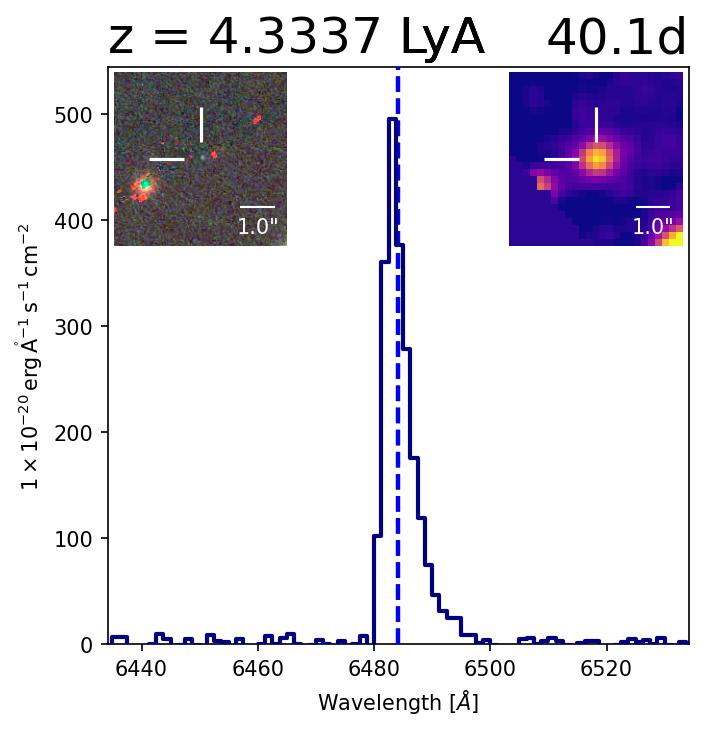}
        \end{subfigure}
        \begin{subfigure}{.25\textwidth}
            \centering
            \includegraphics[width=1.0\textwidth]{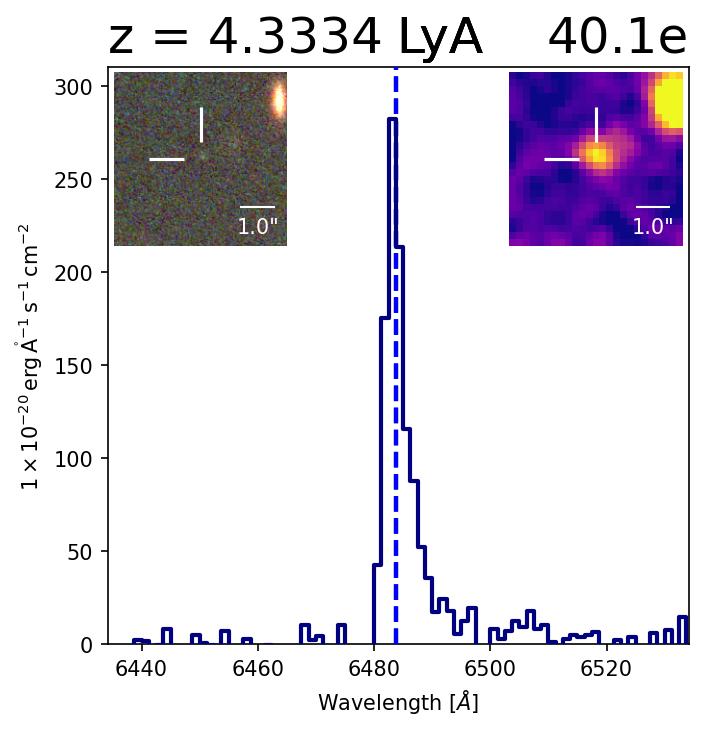}
        \end{subfigure}%
        \caption{\label{fig:cimg_sys_40}System 40}
    \end{figure}
    
    \begin{figure}[H]
        \raggedright{}
        \begin{subfigure}{.25\textwidth}
            \centering
            \includegraphics[width=1.0\textwidth]{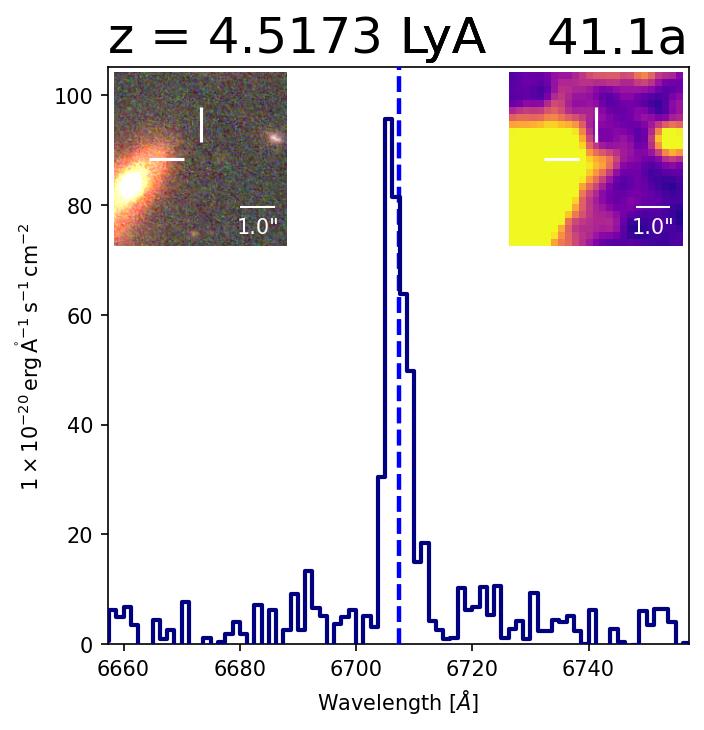}
        \end{subfigure}
        \begin{subfigure}{.25\textwidth}
            \centering
            \includegraphics[width=1.0\textwidth]{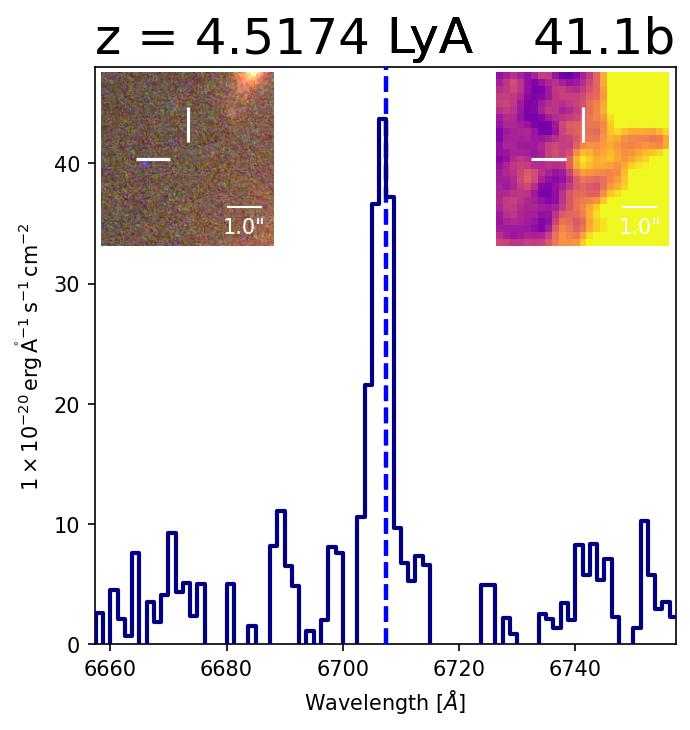}
        \end{subfigure}%
        \begin{subfigure}{.25\textwidth}
            \centering
            \includegraphics[width=1.0\textwidth]{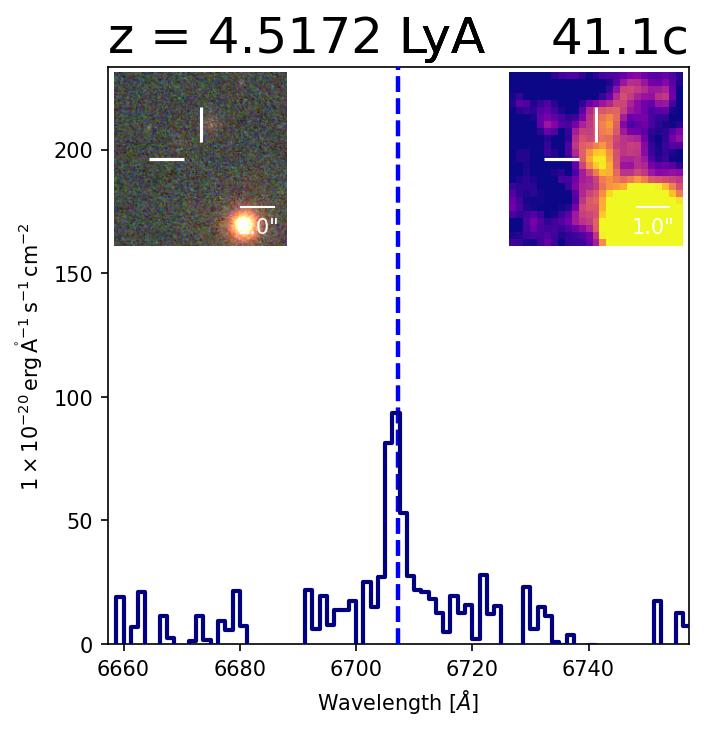}
        \end{subfigure}%
        \caption{\label{fig:cimg_sys_41}System 41}
    \end{figure}
    
    \begin{figure}[H]
        \raggedright{}
        \begin{subfigure}{.25\textwidth}
            \centering
            \includegraphics[width=1.0\textwidth]{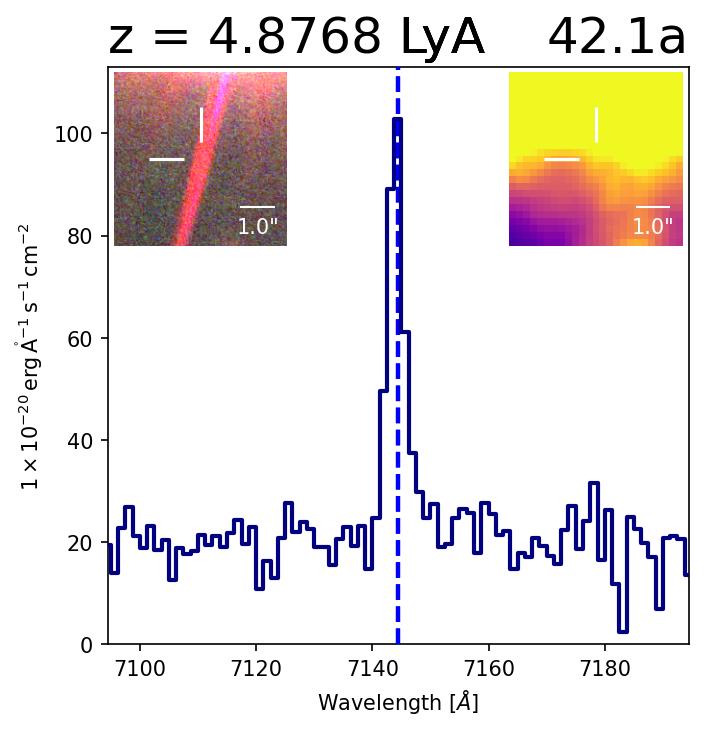}
        \end{subfigure}
        \begin{subfigure}{.25\textwidth}
            \centering
            \includegraphics[width=1.0\textwidth]{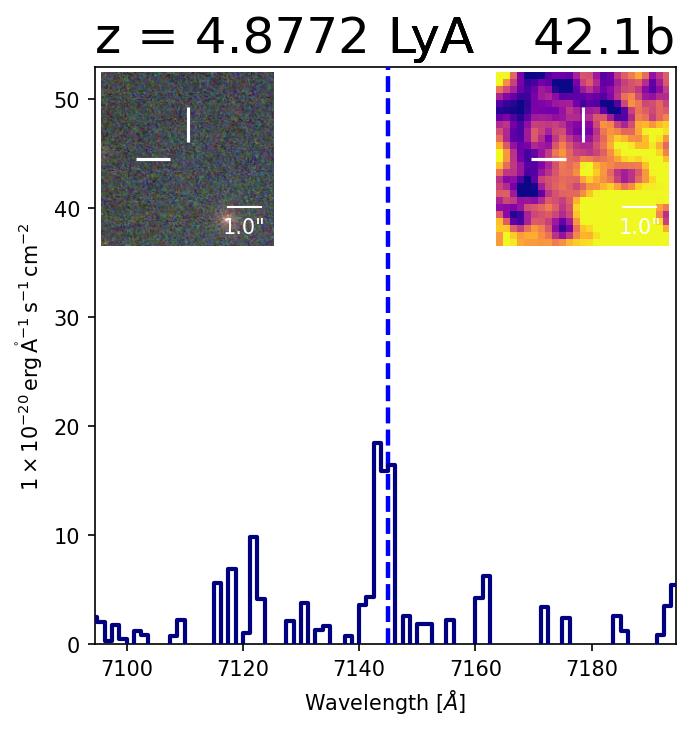}
        \end{subfigure}%
        \begin{subfigure}{.25\textwidth}
            \centering
            \includegraphics[width=1.0\textwidth]{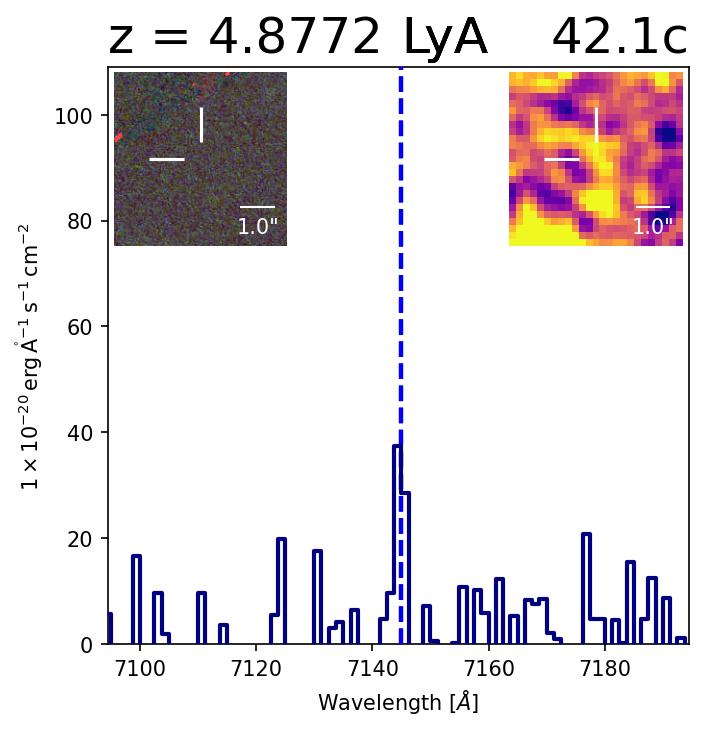}
        \end{subfigure}%
        \caption{\label{fig:cimg_sys_42}System 42}
    \end{figure}

    \begin{figure}[H]
        \raggedright{}
        \begin{subfigure}{.25\textwidth}
            \centering
            \includegraphics[width=1.0\textwidth]{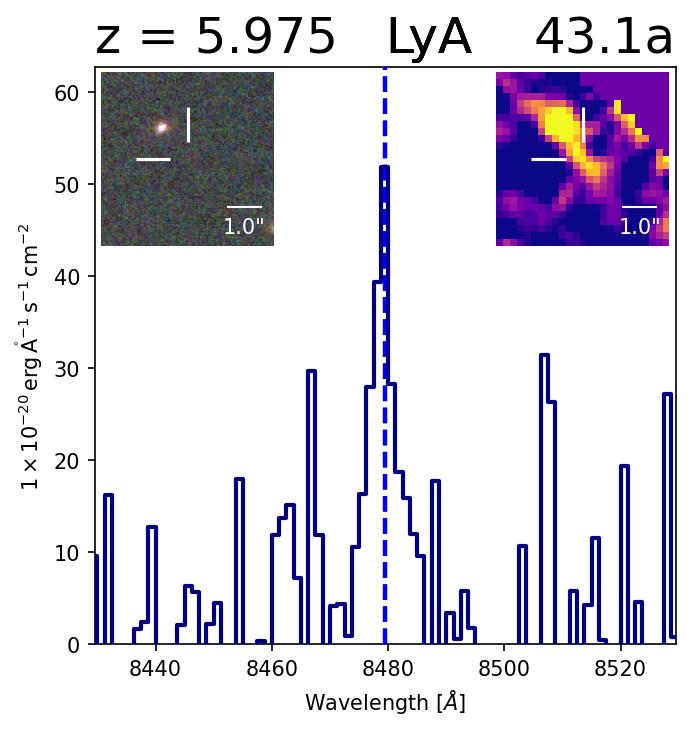}
        \end{subfigure}
        \begin{subfigure}{.25\textwidth}
            \centering
            \includegraphics[width=1.0\textwidth]{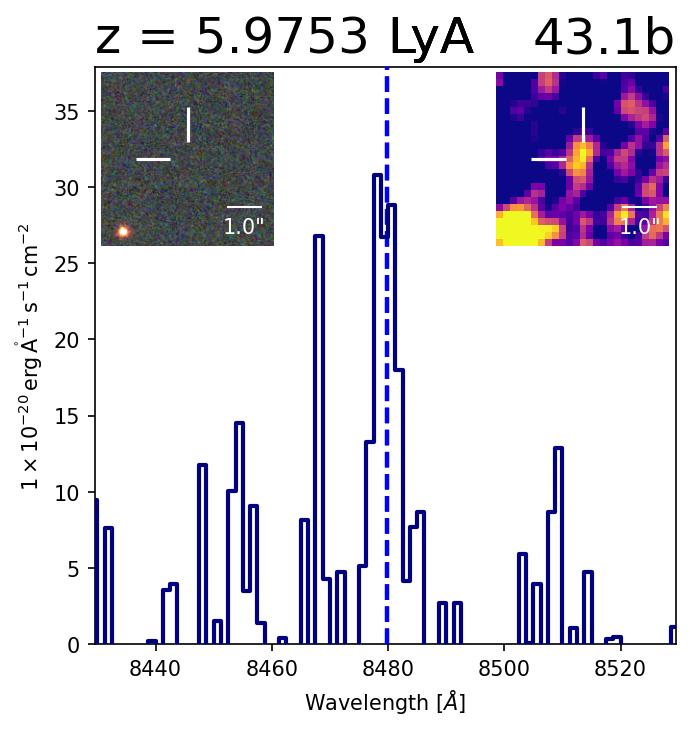}
        \end{subfigure}%
        \begin{subfigure}{.25\textwidth}
            \centering
            \includegraphics[width=1.0\textwidth]{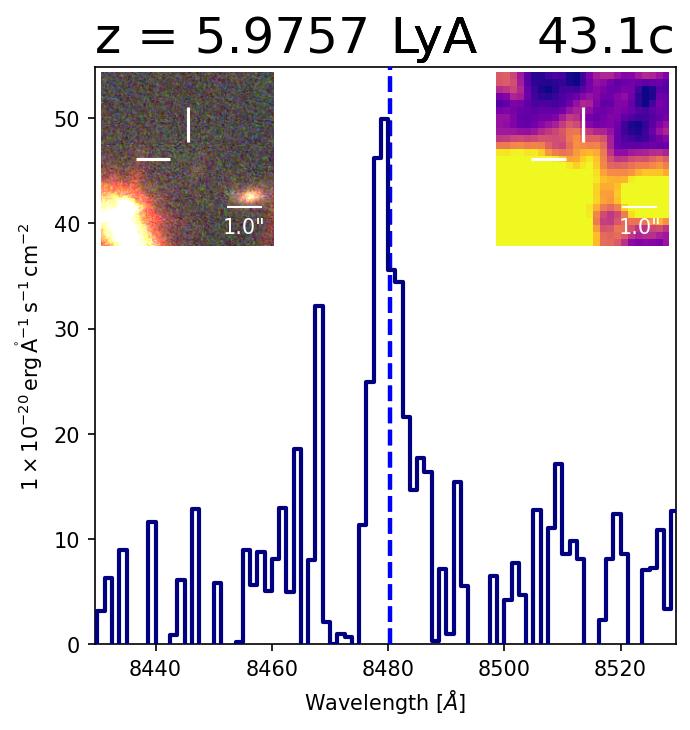}
        \end{subfigure}%
        \begin{subfigure}{.25\textwidth}
            \centering
            \includegraphics[width=1.0\textwidth]{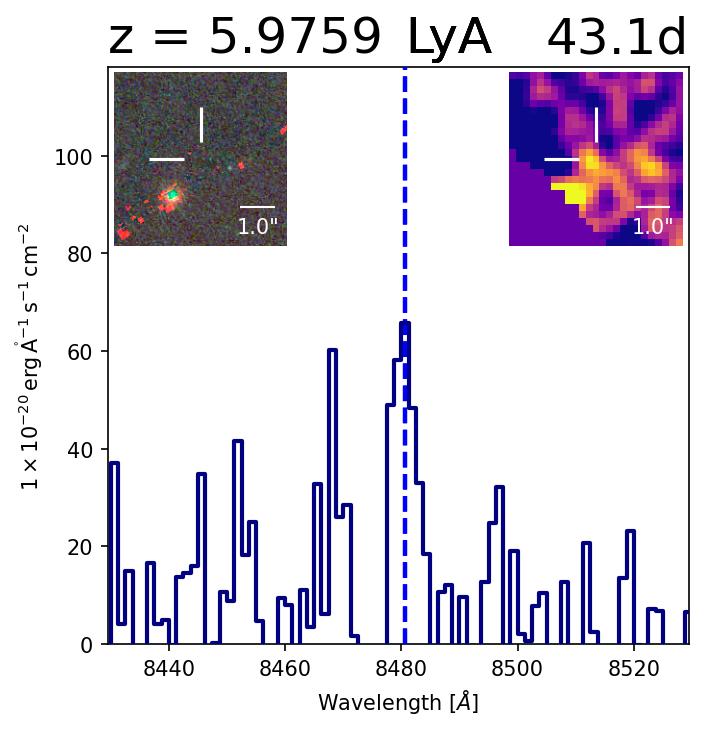}
        \end{subfigure}%
        \caption{\label{fig:cimg_sys_43}System 43}
    \end{figure}
    
    \begin{figure}[H]
        \raggedright{}
        \begin{subfigure}{.25\textwidth}
            \centering
            \includegraphics[width=1.0\textwidth]{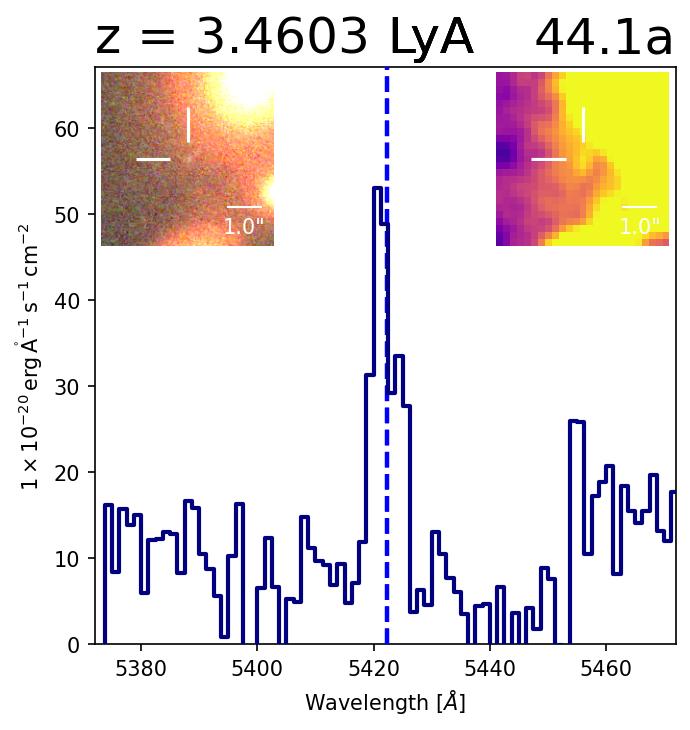}
        \end{subfigure}
        \begin{subfigure}{.25\textwidth}
            \centering
            \includegraphics[width=1.0\textwidth]{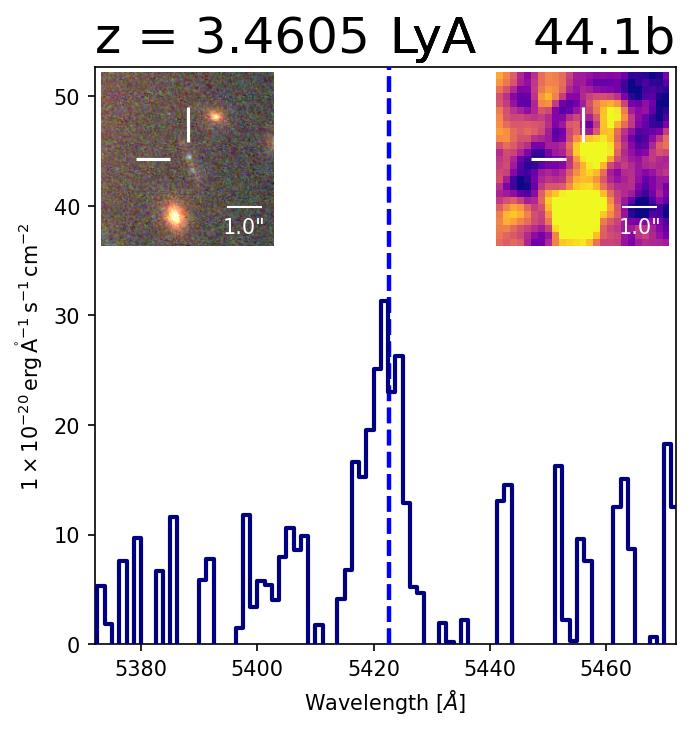}
        \end{subfigure}%
        \caption{\label{fig:cimg_sys_44}System 44}
    \end{figure}
    
    \begin{figure}[H]
        \raggedright{}
        \begin{subfigure}{.25\textwidth}
            \centering
            \includegraphics[width=1.0\textwidth]{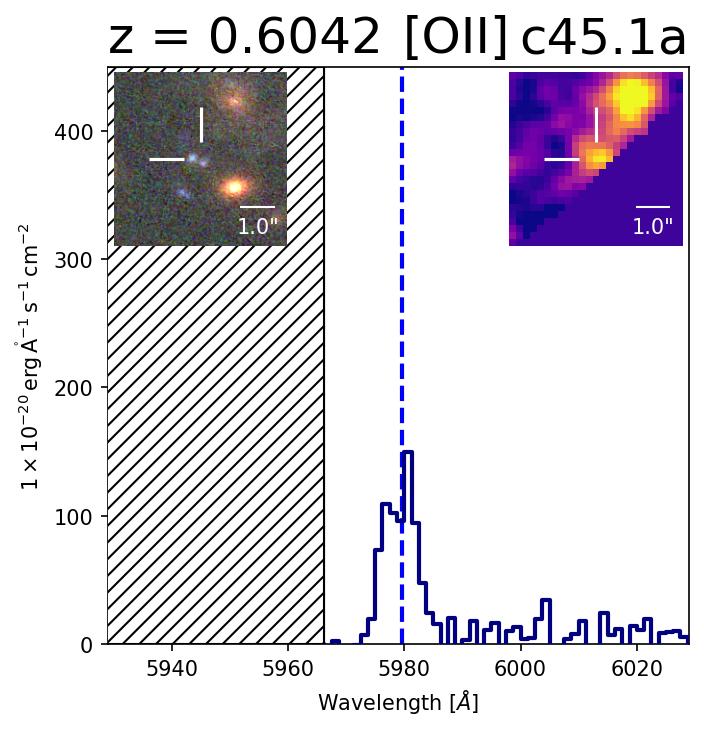}
        \end{subfigure}
        \begin{subfigure}{.25\textwidth}
            \centering
            \includegraphics[width=1.0\textwidth]{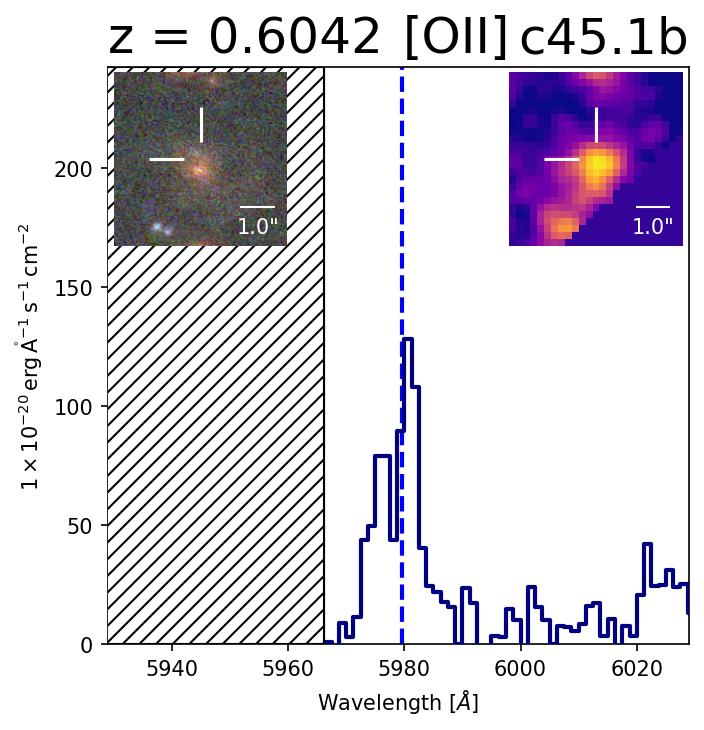}
        \end{subfigure}%
        \caption{\label{fig:cimg_sys_45}System 45}
    \end{figure}
    
    \begin{figure}[H]
        \raggedright{}
        \begin{subfigure}{.25\textwidth}
            \centering
            \includegraphics[width=1.0\textwidth]{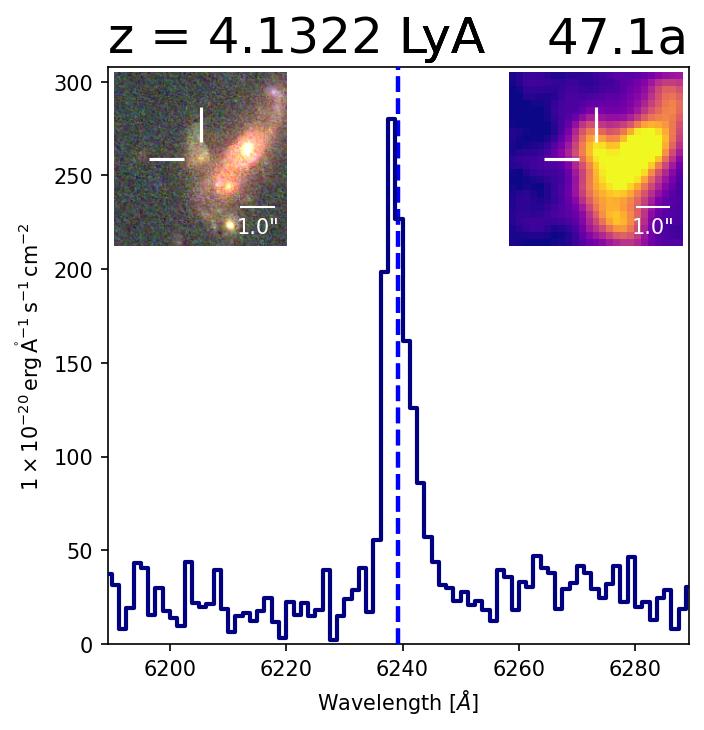}
        \end{subfigure}
        \begin{subfigure}{.25\textwidth}
            \centering
            \includegraphics[width=1.0\textwidth]{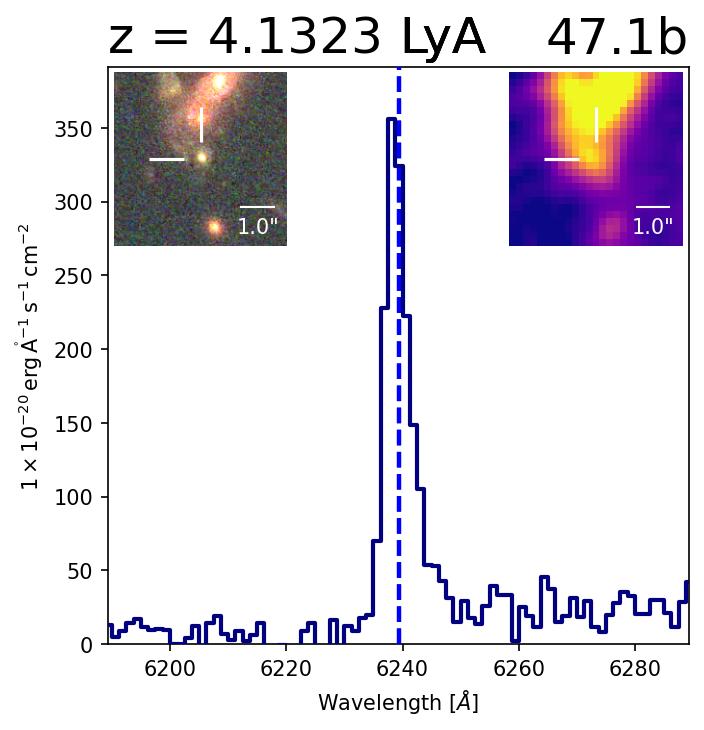}
        \end{subfigure}%
        \begin{subfigure}{.25\textwidth}
            \centering
            \includegraphics[width=1.0\textwidth]{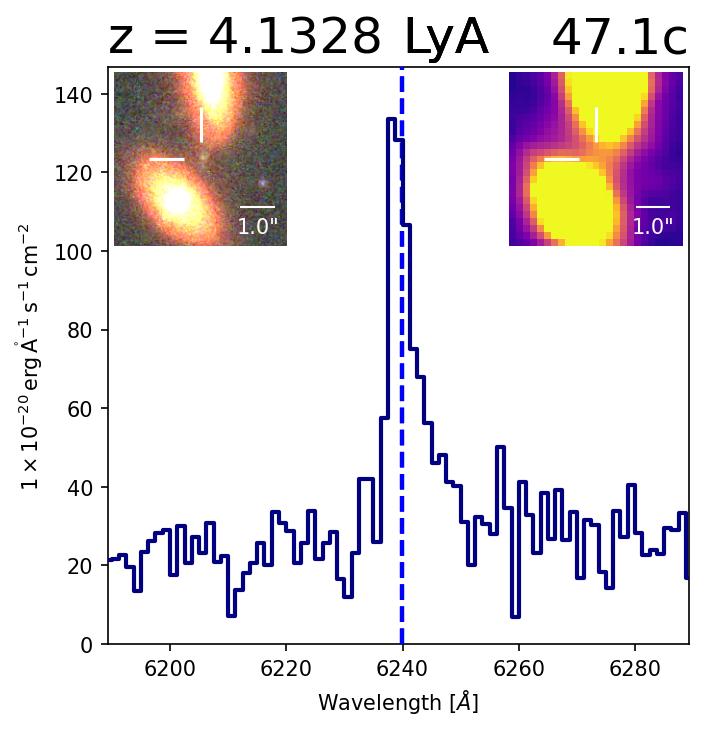}
        \end{subfigure}%
        \caption{\label{fig:cimg_sys_47}System 47}
    \end{figure}
    
    \begin{figure}[H]
        \raggedright{}
        \begin{subfigure}{.25\textwidth}
            \centering
            \includegraphics[width=1.0\textwidth]{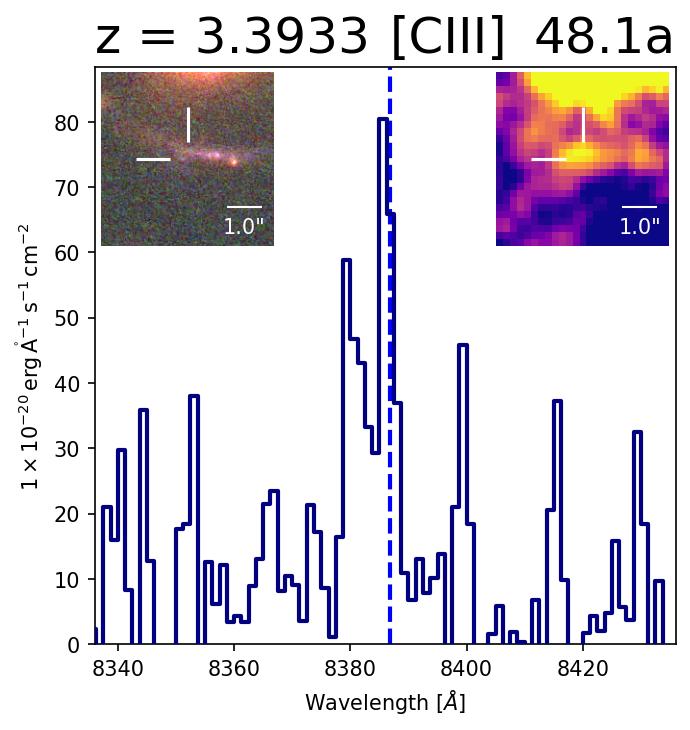}
        \end{subfigure}
        \begin{subfigure}{.25\textwidth}
            \centering
            \includegraphics[width=1.0\textwidth]{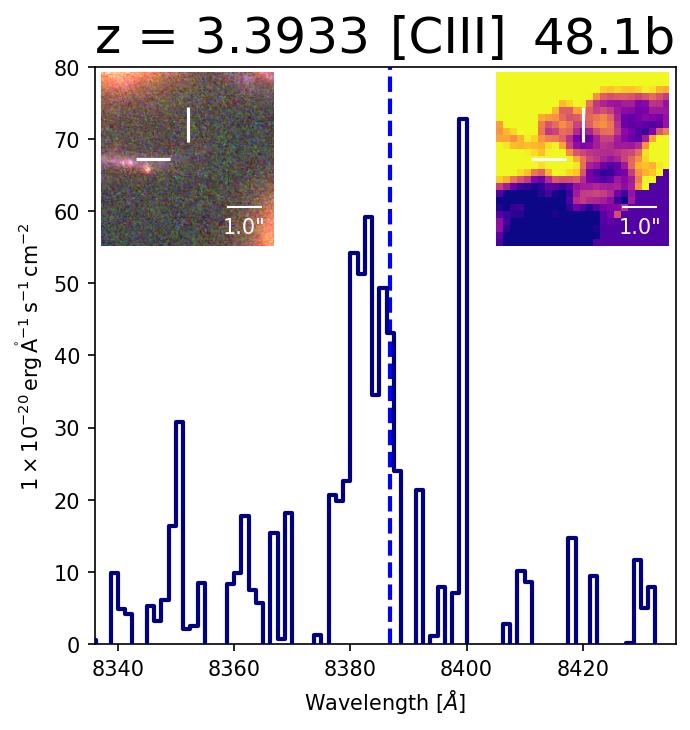}
        \end{subfigure}%
        \caption{\label{fig:cimg_sys_48}System 48}
    \end{figure}
    
    \begin{figure}[H]
        \raggedright{}
        \begin{subfigure}{.25\textwidth}
            \centering
            \includegraphics[width=1.0\textwidth]{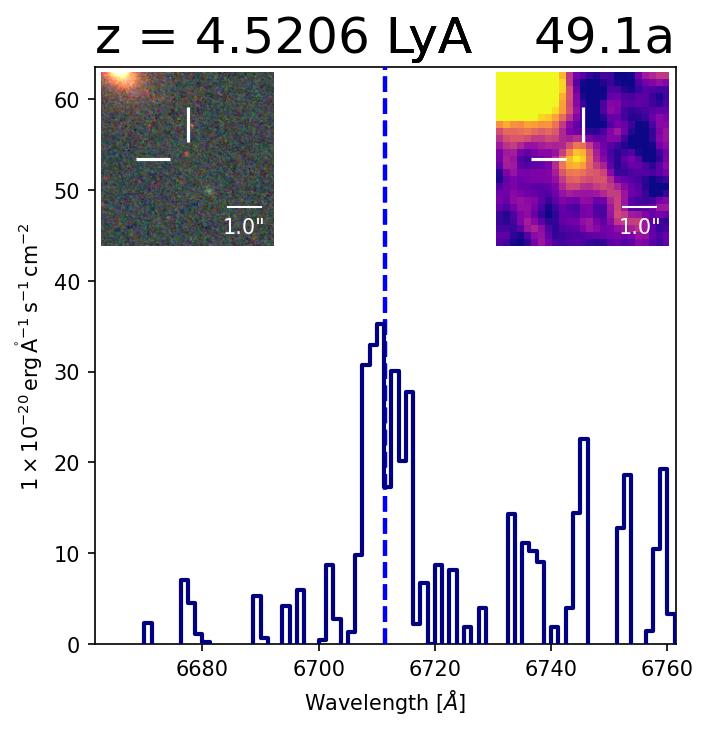}
        \end{subfigure}
        \begin{subfigure}{.25\textwidth}
            \centering
            \includegraphics[width=1.0\textwidth]{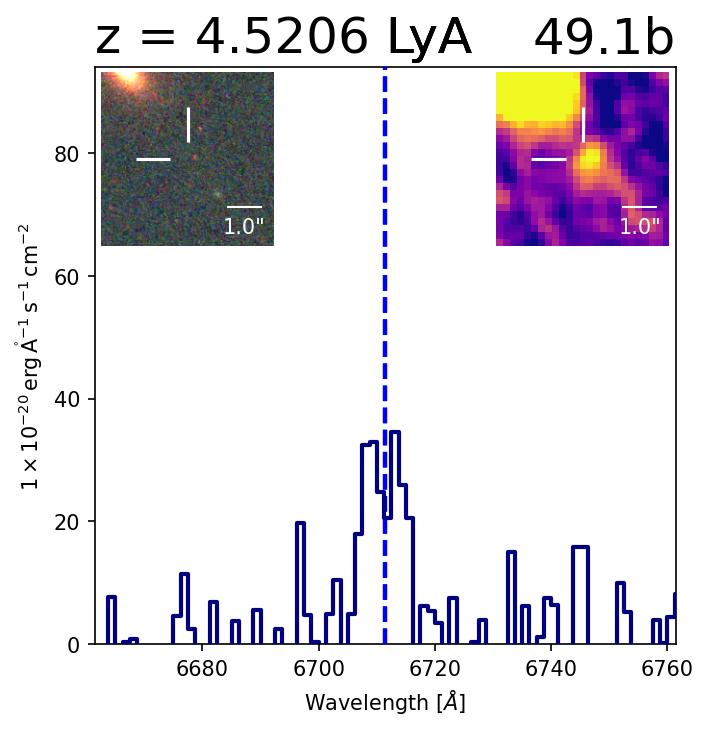}
        \end{subfigure}%
        \begin{subfigure}{.25\textwidth}
            \centering
            \includegraphics[width=1.0\textwidth]{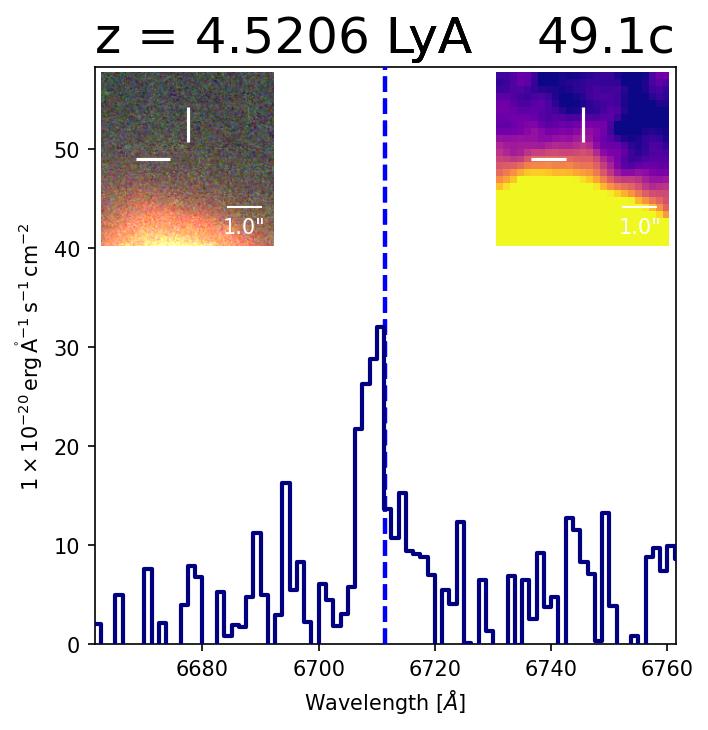}
        \end{subfigure}
    
        \begin{subfigure}{.25\textwidth}
            \centering
            \includegraphics[width=1.0\textwidth]{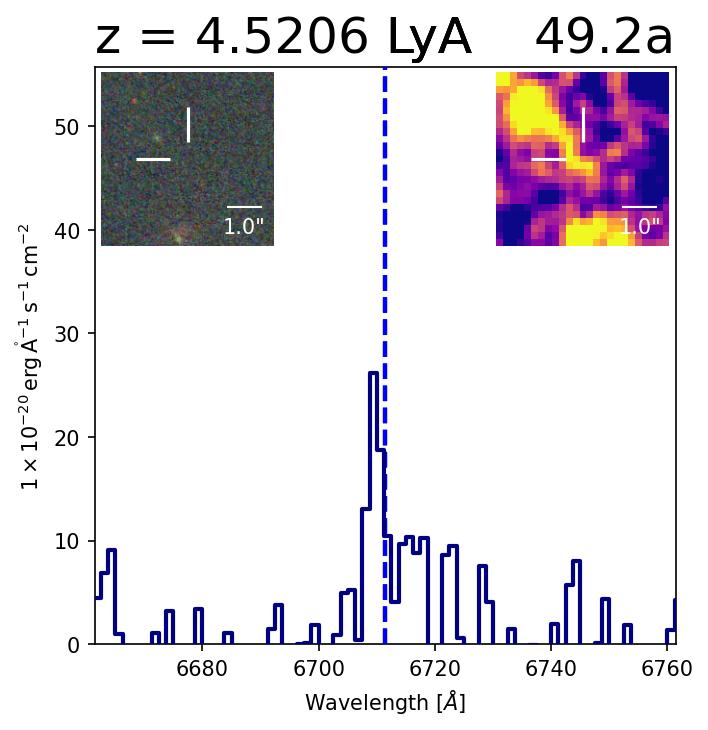}
        \end{subfigure}%
        \begin{subfigure}{.25\textwidth}
            \centering
            \includegraphics[width=1.0\textwidth]{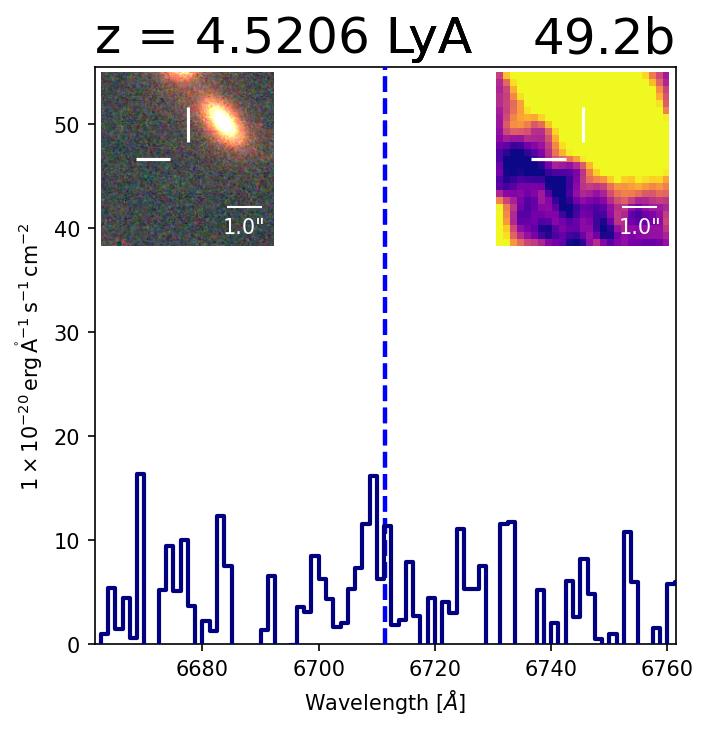}
        \end{subfigure}%
        \caption{\label{fig:cimg_sys_49}System 49}
    \end{figure}
    \onecolumn

    \section{Cluster members}
    \label{appendix:cluster_members}
    Table D.1 (only available in electronic form at the CDS) presents the selected cluster members used in the lens model, sorted by their increasing magnitude values in the band F160W. The first column (ID) reports the ID in the spectroscopic and photometric catalogs; the second and third columns show the J2000 ICRS right ascension and declination (R.A and Dec.); the fourth one (Source) reports the source of the redshift estimation for the spectroscopic members (m = MUSE; v = VIMOS; d = DEIMOS), while photometric members identified by the CNN are indicated by a c (see Sect. \ref{sub:model.members}); the fifth column ($z_{\rm{spec}}$) reports the measured spectroscopic redshift; the last column ($m_{\rm{F160W}}$ show the Kron magnitudes in the band F160W);

    \section{Spectroscopic catalog}
    \label{appendix:z_spec_catalog}

    Table E.1 (only available in electronic form at the CDS) presents the full spectroscopic catalog. The first column (ID) reports an increasing unique identification number; the second and third columns show the J2000 ICRS right ascension and declination (R.A and Dec.); the fourth one (Source) reports the source of the redshift estimation (m = MUSE; v = VIMOS; d = DEIMOS); the fifth and sixth columns ($z_{\rm{spec}}$ and QF) reports the measured spectroscopic redshift and its estimation quality flag  (1=insecure, 2=likely, 3=secure, and 9=based on one emission line); the last column (Comments) provides comments on the redshift estimation and/or the most prominent spectral features detected.\\

}
\end{appendix}
\end{document}